\documentclass[12pt]{article}
\pdfoutput=1
\usepackage[nosort]{cite}
\usepackage[height=8.85in,width=6.45in]{geometry}
\usepackage{pifont}

\usepackage{times}

\usepackage{comment}
\usepackage{booktabs}
\usepackage[utf8]{inputenc}
\usepackage{amsmath}
\usepackage{amssymb}
\usepackage{mathtools}
\numberwithin{equation}{section}
\usepackage{slashed}
\usepackage{braket}
\usepackage[svgnames,psnames]{xcolor}
\usepackage{url}
\usepackage[colorlinks,citecolor=DarkGreen,linkcolor=FireBrick,linktocpage,pagebackref]{hyperref}
\usepackage{cite}
\usepackage{graphicx}
\usepackage{tikz}
\usepackage{tikz-cd}
\usepackage{tikzit}

\tikzstyle{pointoperator}=[fill=black, draw=none, shape=circle, minimum size=.1 cm, inner sep=0 pt]

\tikzstyle{dashedline}=[dashed, -, thick]
\tikzstyle{blueline}=[-, draw=blue, thick]
\tikzstyle{arrowline}=[<-, thick]
\tikzstyle{normalline}=[-, thick]
\tikzstyle{fillline}=[-, thick, fill=red, fill opacity=.5]
\tikzstyle{bluefillline}=[-, fill opacity=.5, fill=blue, thick]
\tikzstyle{purplefillline}=[-, fill opacity=.5, fill=purple, thick]
\tikzstyle{bluedashedline}=[-, draw=blue, thick, dashed]

\usepackage{courier}
\usepackage{bm}

\usepackage{dashbox}
\usepackage{caption}
\usepackage{subcaption}
\usepackage{enumitem}

\usepackage{mdframed}

\usepackage{blkarray}
\usepackage{arydshln}
\usepackage{dsfont}
\usetikzlibrary{patterns}
\usetikzlibrary{arrows.meta}

\usepackage{calc}

\usepackage{accents}

\newcommand{\ie}{\begin{equation}\begin{aligned}}
\newcommand{\fe}{\end{aligned}\end{equation}}

\def\acts{\rotatebox[origin=c]{-90}{$\circlearrowright$}}

\renewcommand{\title}[1]{\vbox{\center\LARGE{#1}}\vspace{5mm}}
\renewcommand{\author}[1]{\vbox{\center#1}\vspace{5mm}}
\newcommand{\address}[1]{\vbox{\center\em#1}}

\makeatletter
\newsavebox{\@brx}
\newcommand{\llangle}[1][]{\savebox{\@brx}{\(\m@th{#1\langle}\)}%
  \mathopen{\copy\@brx\kern-0.5\wd\@brx\usebox{\@brx}}}
\newcommand{\rrangle}[1][]{\savebox{\@brx}{\(\m@th{#1\rangle}\)}%
  \mathclose{\copy\@brx\kern-0.5\wd\@brx\usebox{\@brx}}}
\makeatother

\begin{document}
    
\begin{titlepage}
    \hfill      YITP-SB-2023-10
    \\

\title{Remarks on Boundaries, Anomalies, and Noninvertible Symmetries}

\author{Yichul Choi${}^{1,2}$, Brandon C.\  Rayhaun${}^1$, Yaman Sanghavi${}^1$, and Shu-Heng Shao${}^1$}

        \address{${}^{1}$C.\ N.\ Yang Institute for Theoretical Physics, Stony Brook University\\
        ${}^{2}$Simons Center for Geometry and Physics, Stony Brook University}

\abstract
 
What does it mean for a boundary condition to be symmetric with respect to a non-invertible global symmetry? We discuss two possible definitions in   1+1d QFTs and lattice models. On the one hand, we call a boundary \emph{weakly symmetric} if the symmetry defects can terminate topologically on it, leading to conserved operators for the Hamiltonian on an interval (in the open string channel). 
On the other hand, we call a boundary \emph{strongly symmetric} if the corresponding boundary state is an eigenstate of the symmetry operators (in the closed string channel).
These two notions of symmetric boundaries are equivalent for invertible symmetries, but bifurcate for non-invertible symmetries. 
We discuss the relation to anomalies, where we observe that it is sometimes possible to gauge a non-invertible symmetry  in a generalized sense even though it is incompatible with a trivially gapped phase. 
The analysis of symmetric boundaries further leads to  constraints on bulk and boundary renormalization group flows.

In 2+1d, we study the action of  non-invertible condensation defects on the boundaries of $U(1)$ gauge theory and several TQFTs. 
Starting from the Dirichlet boundary of free Maxwell theory, the non-invertible symmetries  generate infinitely many boundary conditions that are neither Dirichlet nor Neumann.

\end{titlepage}

\eject

\tableofcontents

\section{Introduction}

Global symmetries serve as a powerful tool to study the non-perturbative dynamics of quantum field theories and lattice models, providing us with a general organizing principle for the physical observables.
A particularly fruitful application is the interplay between  global symmetries and boundary conditions.
Often times, boundary conditions are sensitive to the fine algebraic data associated with a global symmetry, such as its 't Hooft anomalies, and the spectrum of boundary conditions can correspondingly be greatly constrained.
For instance, nontrivial 't Hooft anomalies are known to obstruct the existence of a symmetric boundary condition (see for instance \cite{Wang:2013yta,Han:2017hdv,Jensen:2017eof,Numasawa:2017crf,Smith:2020rru,Smith:2020nuf,Thorngren:2020yht,Tong:2021phe,Li:2022drc,Zeng:2022grc,Wang:2022ucy}).

The notion of a global symmetry has been generalized in various directions in the past decade to include higher-form symmetries \cite{Gaiotto:2014kfa} and non-invertible symmetries.
See \cite{McGreevy:2022oyu,Cordova:2022ruw} for recent reviews. 
In 1+1d, non-invertible  symmetries are generated by topological defect lines \cite{Verlinde:1988sn,Moore:1988qv,Fuchs:2002cm,Frohlich:2004ef,Frohlich:2006ch,Frohlich:2009gb,Carqueville:2012dk,Brunner:2014lua,Aasen:2016dop,Bhardwaj:2017xup,Chang:2018iay,Thorngren:2019iar,Cordova:2019wpi,Lin:2019hks,Pal:2020wwd,Komargodski:2020mxz,Aasen:2020jwb,Thorngren:2021yso,Huang:2021zvu,Kaidi:2022cpf,Lin:2022dhv,Chang:2022hud,Lu:2022ver,Kaidi:2023maf,Zhang:2023wlu,Lin:2023uvm}, and are  
described by the mathematical theory  of fusion categories.

In this paper, we study the interplay between boundary conditions and finite non-invertible global symmetries in diverse spacetime dimensions, with a particular focus on 1+1d and 2+1d bulk field theories and lattice systems. 
In 1+1d, relationships between boundary conditions and generalized global symmetries have been discussed, for example, in  \cite{Fuchs:2001qc,Brunner:2007ur,Fredenhagen:2009tn,Kojita:2016jwe,Chang:2018iay,Konechny:2019wff,Konechny:2020jym,Fukusumi:2021zme,Collier:2021ngi}. Boundary conditions for non-invertible symmetries in 3+1d lattice gauge theory have recently been analyzed in \cite{Koide:2023rqd} based on the model in \cite{Koide:2021zxj}.

We begin by asking the most basic question: What does it mean for a boundary condition to be symmetric under a non-invertible global symmetry?
Interestingly, we find that such a simple question does not have a unique natural answer in the case of non-invertible symmetries.

We call a boundary condition \emph{weakly symmetric} if the corresponding defect lines can topologically terminate on the boundary.
Such a definition is natural in the open string channel, as it implies that the topological defect lines commute with the Hamiltonian on the interval with the boundary condition imposed at its two endpoints, leading to conserved quantities.
This definition applies equally well  to 1+1d lattice systems, where we call a boundary condition weakly symmetric if the Hamiltonian on an open chain commutes with   operators which locally look like the symmetry operators in the bulk.

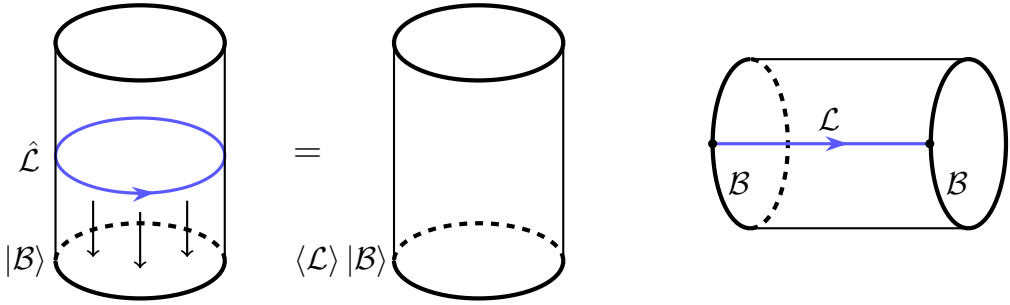
\begin{figure}
\begin{center}
\begin{tikzpicture}

\begin{scope}[shift={(0,-6)}]

\tikzset{
    partial ellipse/.style args={#1:#2:#3}{
        insert path={+ (#1:#3) arc (#1:#2:#3)}
    }
}

\draw[ultra thick, black!100, dashed] (-1.75,2.5) [partial ellipse= 0:180:2.25cm and 1cm];
\draw[ultra thick, black!100] (-1.75,2.5) [partial ellipse= 180:360:2.25cm and 1cm];
\draw[ultra thick, black!100] (-1.75,8.3) [partial ellipse= 0:360:2.25cm and 1cm];

\draw[very thick, blue!65] (-1.75,5.3) [partial ellipse= 0:360:2.25cm and 1cm];
\draw[very thick, blue!65,-{Stealth[length=3mm,width=2mm]}] (-1.75,5.3) [partial ellipse= 268:280:2.25cm and 1cm];

\node [style=none] (0) at (-4, 2.5) {};
\node [style=none] (1) at (0.5, 2.5) {};
\node [style=none] (2) at (-4, 8.275) {};
\node [style=none] (3) at (0.5, 8.275) {};

\draw [style=normalline, black] (2.center) to (0.center);
\draw [style=normalline, black] (3.center) to (1.center);

\draw [style=normalline, black, -to] (-3,4.1) to (-3,2.6);
\draw [style=normalline, black, -to] (-1.75,3.8) to (-1.75,2.3);
\draw [style=normalline, black, -to] (-0.5,4.1) to (-0.5,2.6);

\node [style=none] (6) at (-4.7, 5.3) {$\hat{\mathcal{L}}$};
\node [style=none] (7) at (-4.8, 2.5) {$|\mathcal{B}\rangle$};

\end{scope}

\node [style=none] (6) at (2.7, -0.7) {\large $=$};

\begin{scope}[shift={(9,-6)}]

\tikzset{
    partial ellipse/.style args={#1:#2:#3}{
        insert path={+ (#1:#3) arc (#1:#2:#3)}
    }
}

\draw[ultra thick, black!100, dashed] (-1.75,2.5) [partial ellipse= 0:180:2.25cm and 1cm];
\draw[ultra thick, black!100] (-1.75,2.5) [partial ellipse= 180:360:2.25cm and 1cm];
\draw[ultra thick, black!100] (-1.75,8.3) [partial ellipse= 0:360:2.25cm and 1cm];


\node [style=none] (0) at (-4, 2.5) {};
\node [style=none] (1) at (0.5, 2.5) {};
\node [style=none] (2) at (-4, 8.275) {};
\node [style=none] (3) at (0.5, 8.275) {};

\draw [style=normalline, black] (2.center) to (0.center);
\draw [style=normalline, black] (3.center) to (1.center);

\node [style=none] (7) at (-5.4, 2.5) {$\langle \mathcal{L} \rangle \ket{\mathcal{B}}$};

\end{scope}

\end{tikzpicture}\quad\quad\quad\quad
\raisebox{-0.5\height}{\begin{tikzpicture}[rotate=270]

\tikzset{
    partial ellipse/.style args={#1:#2:#3}{
        insert path={+ (#1:#3) arc (#1:#2:#3)}
    }
}

\draw[ultra thick, black!100, dashed] (-1.75,2.5) [partial ellipse= 0:180:2.25cm and 1cm];
\draw[ultra thick, black!100] (-1.75,2.5) [partial ellipse= 180:360:2.25cm and 1cm];
\draw[ultra thick, black!100] (-1.75,8.3) [partial ellipse= 0:360:2.25cm and 1cm];


\node [style=none] (0) at (-4, 2.5) {};
\node [style=none] (1) at (0.5, 2.5) {};
\node [style=none] (2) at (-4, 8.275) {};
\node [style=none] (3) at (0.5, 8.275) {};

\draw [style=normalline, black] (2.center) to (0.center);
\draw [style=normalline, black] (3.center) to (1.center);
\draw [very thick, blue!65] (-1.75, 7.275) to (-1.75, 1.5);
\draw [very thick, blue!65, -{Stealth[length=3mm,width=2mm]}] (-1.75, 5.0) to (-1.75, 5.1);
\filldraw [black] (-1.75, 7.275) circle (3pt);  
\filldraw [black] (-1.75, 1.5) circle (3pt);  
\node [style=none] at (-2.4, 4.6) {$\mathcal{L}$};
\node [style=none] at (-0.7, 2.2) {$\mathcal{B}$};
\node [style=none] at (-0.7, 8.0) {$\mathcal{B}$};
\end{tikzpicture}}
\end{center}
\caption{We call a boundary condition strongly symmetric if the corresponding boundary state is an eigenstate under the action of the symmetry, as shown in the figure on the left (closed string channel). Here $\langle {\cal L}\rangle\geq1$ is the quantum dimension  of the topological line. On the other hand, we call a boundary condition weakly symmetric if the topological defect lines can topologically end on the boundary, as shown in the figure on the right (open string channel). 
}
\label{fig:open_closed}
\end{figure}

There is an alternative way to define a symmetric boundary condition, which is more natural in the closed string channel.
In the context of 1+1d conformal field theory (CFT), boundary conditions \footnote{Throughout the paper, for 1+1d CFTs, we only consider compact boundary conditions of a compact CFT, i.e. those boundary conditions $\cal B$ for which the Hilbert space $\mathcal{H}_{\mathcal{B}\mathcal{B}}$, defined on an interval with a finite length $L$ with boundary conditions $\cal B$, has a discrete spectrum. For instance, the $c=1$ compact boson CFT at an irrational multiple of self-dual radius has a family of non-compact boundaries \cite{Janik:2001hb}. We thank P. Boyle Smith for the discussions on this point.} define a special class of quantum states, called boundary states, which are subject to consistency conditions known as the Cardy conditions \cite{Cardy:1989ir}.
We call a boundary condition \emph{strongly symmetric} if the corresponding boundary state is a simultaneous eigenstate under the action of the topological defect lines.
This definition may also be extended to lattice systems by requiring that the boundary condition is invariant (up to a positive number) with respect to parallel fusion with the symmetry defects.\footnote{See, for example,  \cite{Aasen:2020jwb} for such fusion processes of defects and boundaries on the lattice.}  
See Figure \ref{fig:open_closed} for an illustration of these two notions.

 The two definitions of what a symmetric boundary is turn out to be equivalent when the symmetry is invertible.
However, we find that this is not the case for general non-invertible symmetries described by fusion categories, and the very notion of a symmetric boundary condition bifurcates.
In particular, as the names suggest, a strongly symmetric boundary condition is always also weakly symmetric, but not necessarily the other way around.  
Fusion categories therefore fall into three types, based on the kinds of boundary conditions with which they are kinematically compatible.\footnote{As is common in high energy physics, we refer to theory-independent consequences of symmetries and anomalies as ``kinematics," and to effects that depend on the specific choice of a Hamiltonian or Lagrangian as ``dynamics." This is to be contrasted with terminology in other fields, where ``dynamics" often refers to evolution under real time.  }
See Table \ref{table:fusioncategoryexamples}. In the main text, we discuss CFTs and lattice models which realize these symmetries and their corresponding boundaries. The specific lattice models   we will study   are the golden chain and other anyonic chains.\footnote{For more general lattice models, the relation between  non-invertible symmetries and boundary conditions  can be more intricate since the structure of non-invertible symmetries on the lattice can be fundamentally different from the corresponding symmetries in the continuum. For instance, in some cases, the non-invertible symmetry on a tensor product Hilbert space can mix with the lattice translation symmetry \cite{Seiberg:2023cdc,Seifnashri:2023dpa}. We leave the analysis for such non-invertible lattice translations  for the future.}

 \begin{table}[h!]
\begin{center}
\begin{tabular}{ c|c } 
Type of Boundary & Examples of Fusion Categories \\
 \toprule
 Strongly Symmetric Boundary &  $\text{Vec}_G$, $\text{Rep}(G)$, $\text{TY}(\mathbb{Z}_2 \times \mathbb{Z}_2, \chi, +)$  \\
 \midrule
 Weakly Symmetric Boundary & $\text{Fib}$, $(A_1,k)_{\frac12}$ for $k>4$\\
 \midrule
 No Symmetric Boundary & $\text{Vec}_G^{\omega \neq 0}$, $\text{TY}_{\pm}(\mathbb{Z}_2)$  \\
 \bottomrule
\end{tabular}
\end{center}
\caption{Various fusion category symmetries and boundary conditions.
The fusion categories on the first row are compatible with a trivially gapped phase, and  admit a strongly symmetric boundary condition, whereas the ones on the second row only admit a weakly symmetric boundary condition but not a strongly symmetric one. The fusion  categories on the third row do not admit any symmetric boundary. Here $\text{Vec}_G^\omega$ stands for an ordinary global symmetry based on a finite group $G$ with an 't Hooft anomaly $\omega\in H^3(G,U(1))$, while $\text{Vec}_G= \text{Vec}_G^{\omega=0}$. Fib and TY stand for the unitary Fibonacci and Tambara-Yamagami fusion categories. $(A_1,k)$ is the fusion category realized by the Verlinde lines of the diagonal $SU(2)_k$ WZW model, and $(A_1,k)_{\frac12}$ is its fusion subcategory which is generated by lines corresponding to primary operators with integer $SU(2)$ spin. Finally, Rep$(G)$ is the (untwisted) fusion category whose fusion rule is the representation ring of $G$. }
\label{table:fusioncategoryexamples}
\end{table}

We further discuss the relation to anomalies. 
For invertible symmetries, 't Hooft anomalies can be equivalently defined either in terms of (1) the incompatibility with a symmetry-preserving, trivially gapped phase \footnote{A symmetry-preserving  trivially gapped phase is a gapped phase with a unique ground state which is invariant under the symmetry action. It is the low energy limit of an SPT phase.}  (which will be our definition of an anomaly throughout), or (2) the obstruction to gauging. 
For non-invertible symmetries, we point out that these two notions of anomaly again bifurcate. 
It is possible that a fusion category $\cal C$ can be ``gauged" in a generalized sense by inserting a mesh of all the topological defects of $\cal C$ \cite{Frohlich:2009gb,Carqueville:2012dk,Brunner:2014lua,Bhardwaj:2017xup}, and yet is incompatible with a trivially gapped phase. 
The simplest such example is the Fibonacci fusion category. 
We show that a fusion category kinematically admits a strongly symmetric boundary if and only if it is free of anomaly (in the sense that it is compatible with a trivially gapped phase), while it admits a weakly symmetric boundary if and only if it can be ``gauged."

By employing the relation between  renormalization group (RG) boundaries and relevant deformations, the analysis of  symmetric boundary conditions leads to new constraints on  bulk RG flows beyond those from anomaly matching. 
In particular, we find that the tetracritical Ising CFT cannot be trivially gapped by any relevant deformation while preserving the $\mathrm{Rep}(S_3)$ fusion category, even though that symmetry is non-anomalous in the sense that it is kinematically compatible with a trivially gapped phase. 

In higher spacetime dimensions, generalized global symmetries act on boundary conditions as a matrix representation with ``coefficients" valued in topological quantum field theories (TQFTs). 
This generalizes the non-negative integer matrix representation in 1+1d \cite{Fuchs:2001qc}. 
We demonstrate this general structure by focusing on a special kind of non-invertible global symmetry, known as a \textit{condensation defect} \cite{Kong:2013aya,Kong:2014qka,Else:2017yqj,Gaiotto:2019xmp,Roumpedakis:2022aik} (see also \cite{Choi:2021kmx,Kaidi:2021xfk,Bhardwaj:2022yxj,Choi:2022zal,Bhardwaj:2022lsg,Lin:2022xod,Bartsch:2022mpm,Freed:2022qnc,Bhardwaj:2022kot,Bhardwaj:2022maz,Bartsch:2022ytj,Delcamp:2023kew,Inamura:2023qzl}). 
More specifically, we study the action of non-invertible condensation defects on boundary conditions in 2+1d $U(1)$ gauge theory, $\mathbb{Z}_2$ gauge theory (i.e., the low energy limit of the toric code), and other TQFTs.  
Starting from the ordinary Dirichlet boundary of the 2+1d $U(1)$ Maxwell gauge theory, the non-invertible symmetry surfaces lead to infinitely many boundary conditions ${\cal B}_N$ (indexed by a positive integer $N$) that are neither Dirichlet nor Neumann. 
We call them \textit{partially Dirichlet boundaries}, because while a minimally charged Wilson line cannot terminate on ${\cal B}_N$, a charge $N$ Wilson line can.

The rest of this paper is organized as follows.
In Section \ref{sec:symm_boundaries}, we explain in more detail the two notions of symmetric boundary conditions.
In Section \ref{sec:anomaly}, we discuss the relation between the existence of strongly/weakly symmetric boundary conditions, 't Hooft anomalies, and gauging in the case of non-invertible symmetries. 
In Section \ref{sec:example} we provide several examples of strongly and weakly symmetric boundaries in 1+1d CFTs and anyonic chains. In Section \ref{sec:app}, we  discuss applications to bulk and boundary RG flows. 
Intriguingly, we find that the analysis of strongly symmetric boundary conditions can sometimes lead to non-trivial constraints on bulk RG flows.
In Section \ref{sec:2+1d}, we discuss the interplay between non-invertible symmetries and boundary conditions in 2+1d, and provide examples in $U(1)$ and $\mathbb{Z}_2$ gauge theories. 
In Appendix \ref{app:categories}, we review some concepts in category theory which are used in the main text.
Appendix \ref{NIMrepG} proves a basic fact about non-negative integer-valued matrix representations of finite groups.

\section{Symmetric Boundaries in 1+1d} \label{sec:symm_boundaries}

\subsection{Simple Boundaries}

We start with a discussion on  the basic structure of the space of conformal boundary conditions of a 1+1d CFT.

First, we note that there is a well-defined notion of taking a direct sum of boundaries. Indeed, the direct sum ${\cal B}_1\oplus{\cal B}_2$ of two boundary conditions ${\cal B}_1, {\cal B}_2$ is defined so that every Euclidean correlation function in the presence of ${\cal B}_1\oplus{\cal B}_2$ is a sum of those in the presence of ${\cal B}_1$ and ${\cal B}_2$. Given a pair of boundary conditions $(\mathcal{B}_1,\mathcal{B}_2)$, we can quantize the theory on an interval, placing ${\cal B}_1$ and ${\cal B}_2$ on its two ends, to obtain a Hilbert space of states ${\cal H}_{\mathcal{B}_1,\mathcal{B}_2}$. 
The Hilbert space associated with the pair $({\cal B}_1\oplus{\cal B}_1',{\cal B}_2)$ is then ${\cal H}_{{\cal B}_1\oplus {\cal B}_1',{\cal B}_2}= {\cal H}_{{\cal B}_1,{\cal B}_2}\oplus{\cal H}_{{\cal B}_1',{\cal B}_2}$. 
We can define any non-negative integer linear combination of a finite set of valid boundary conditions analogously.

In a 1+1d CFT, any conformal boundary condition $\cal B$ can be mapped to a (non-normalizable) state $|{\cal B}\rangle$ in the Hilbert space of the theory on a circle. This is known as the boundary state corresponding to $\mathcal{B}$. 
However,  general linear combinations of boundary states $\sum_i c_i |{\cal B}_i\rangle$ with $c_i\in \mathbb{C}$ are generally not valid boundary conditions because there is no well-defined Hilbert space associated with them.\footnote{For instance, $\frac17|{\cal B}\rangle$ is generally not a valid boundary state because we cannot divide a Hilbert space by 7. Similarly, $|{\cal B}_1\rangle-|{\cal B}_2\rangle$ is also generally not a boundary condition.} 
In other words, the set of all conformal boundary states does not form a vector space over the complex numbers.  
 Rather, it forms a set which is closed under taking linear combinations with coefficients in $\mathbb{Z}_{\geq 0}$. 

An alternative way to understand this is to note that a generic complex number is not a local counterterm that one can add along a conformal boundary. On the other hand, multiplying a boundary state by a non-negative integer $n$ is equivalent to stacking a decoupled $n$-state quantum mechanics, i.e., a free qunit,  along the boundary, which respects locality. 
(For boundaries that are not necessarily conformal, there exist more general local counterterms, such as the boundary cosmological constant term, i.e., the length of the boundary. They are forbidden by conformal invariance in the case of conformal boundaries.)
The constraint of locality plays a crucial role in Cardy's construction of boundary states in rational CFTs \cite{Cardy:1989ir}.

A boundary condition $\cal B$ is said to be \textit{simple} (also called \textit{elementary}) if the interval Hilbert space ${\cal H}_{\cal B,  B}$ has a unique ground state.\footnote{Throughout the paper, we assume that the bulk theory has a unique vacuum on the $S^1$ Hilbert space. In particular, there is no nontrivial bulk topological local operator except for the identity operator.}   
By a conformal  map (or equivalently, the operator-state correspondence), this is equivalent to the condition that there is a unique topological point operator (which can be thought of as the restriction of the bulk identity operator) on the boundary $\cal B$. 
A simple boundary cannot be written as a sum of other boundary conditions with non-negative integer coefficients. 
A more detailed mathematical description is given in Section \ref{SWSB}.

\subsection{Review of Invertible   Symmetries and Boundaries}

Suppose a bulk 1+1d CFT has an ordinary, finite global symmetry group $G$. What kind of representations do conformal boundary states form? 
Since only non-negative integer linear combinations of boundary states are valid,  boundary states cannot be in a general representation of $G$. Rather, they are necessarily in matrix representations with non-negative integer entries. These are known as \textit{non-negative integer matrix representations}, or \textit{NIM-reps} for short \cite{Fuchs:2001qc}. 
In fact, all finite-dimensional NIM-reps of a group are permutation representations (see Appendix \ref{NIMrepG}). 
In contrast, the local operators of a CFT can be in general irreducible representations of $G$.

There are two possible definitions of a $G$-symmetric boundary. 
One definition is that a conformal boundary $\cal B$ is $G$-symmetric if  
\ie
\hat{\cal L}_g|{\cal B}\rangle = |{\cal B}\rangle\,,~~~\forall ~g\in G\,,
\fe
where $\hat{\cal L}_g$ is the unitary operator that implements the symmetry transformation labeled by $g\in G$.\footnote{We use $\cal L$ for a topological line in Euclidean spacetime, and $\hat {\cal L}$ for the corresponding operator acting on the Hilbert space. }
This definition is natural in the closed string channel. 
Another definition involves requiring the existence   of a topological junction between the symmetry line ${\cal L}_g$ and the boundary $\cal B$. 
This junction leads to a conserved operator that commutes with the interval Hamiltonian. 
This is more natural in the open string channel. 
  These two definitions are equivalent for invertible finite symmetries. We will elaborate more on this point in Section \ref{SWSB}.

Given a CFT with a finite global symmetry $G$, it is natural to ask if it admits a $G$-symmetric boundary. 
However, given any boundary $|\cal B\rangle$, one can always add to it the images of the $G$ action to construct a  symmetric but non-simple boundary $\sum_{g\in G}{\hat{\cal L}}_g |{\cal B}\rangle$. 
Therefore, the more interesting question is whether   there is a \textit{simple} $G$-symmetric boundary condition. 

Below we illustrate some of the main ideas and themes that we explore throughout the rest of the paper in the context of familiar models with basic invertible symmetry groups.

\subsubsection{Ising Model and its $\mathbb{Z}_2$ Spin-Flip Symmetry}\label{subsubsec:Isingmodel}

As an invitation, we begin  with the 1+1d Ising CFT.

The Ising model has a non-anomalous $\mathbb{Z}_2$ symmetry, whose generator we denote by $\eta$.
There are three primary operators in the bulk theory: the identity operator $\mathds{1}$, the spin field $\sigma$ of dimension $(h,\bar h)=(\frac{1}{16},\frac{1}{16})$, and the energy/thermal operator  $\epsilon$ of dimension $(h,\bar h)=(\frac{1}{2},\frac{1}{2})$.  
The $\mathbb{Z}_2$ symmetry  acts on the spin field and its descendants as $\sigma \to -\sigma$, and leaves the descendants of the identity and $\epsilon$ invariant.

First, we study the conformal boundary states and ask whether they are eigenstates of $\eta$, a question which is natural from the perspective of the closed-string channel.  Recall that the
Ising CFT has three simple conformal boundary conditions \cite{Cardy:1989ir}, in correspondence with the three bulk primary operators, $\mathds{1},\epsilon,\sigma$. We denote the corresponding Cardy states as $\ket{\uparrow}$, $\ket{\downarrow}$, $\ket{f}$, respectively. From the point of view of the lattice,  $\ket{\uparrow}$ corresponds to the boundary condition in which the boundary spin is pinned so that it points up, and similarly for $\ket{\downarrow}$; on the other hand, $\ket{f}$ corresponds to the free boundary condition in which the boundary spin is allowed to fluctuate. The action of the $\mathbb{Z}_2$ symmetry on the Cardy states is as follows,
\begin{equation}
\hat{\eta} \ket{\uparrow} = \ket{\downarrow}, \quad \; \; \hat{\eta} \ket{\downarrow} = \ket{\uparrow}, \quad \; \; \hat{\eta} \ket{f} = \ket{f}.
\end{equation}
We find that $\eta$ exchanges the two fixed boundaries $\ket{\uparrow} ,\ket{\downarrow}$ and leaves the free boundary $\ket{f}$ invariant. In particular, $\ket{\uparrow}$ and $\ket{\downarrow}$ form a 2-dimensional NIM-rep (the smallest non-trivial permutation representation of $\mathbb{Z}_2$) and the free boundary $\ket{f}$ forms the trivial representation. We conclude that $\ket{f}$ is a $\mathbb{Z}_2$-symmetric boundary, while $\ket{\uparrow}$ and $\ket{\downarrow}$ are not.\footnote{Note that it is inconsistent to have a  boundary state transforming  in the sign representation of $\mathbb{Z}_2$, i.e.\ $\eta|{\cal B}\rangle=-|{\cal B}\rangle$, since it is not a NIM-rep.}

Let us reproduce this result in another way, which is natural in the open string channel. It is convenient to work directly on the lattice, where a Hamiltonian realization of the transverse field Ising model with its free boundary condition can be obtained as follows, 
\begin{align}
    H_f=-J\left(\sum_{i=1}^{N-1} Z_iZ_{i+1}+\sum_{i=1}^N X_i\right).
\end{align}
In the above expression, we have placed a qubit $\mathbb{C}^2$ on the $N$ sites of an open spin chain, and $X_i$/$Z_i$ denote the Pauli-X/Z operators supported at site $i$. We note that the above Hamiltonian commutes with the standard $\mathbb{Z}_2$ spin-flip operator, 
\begin{align}
    [H_f,U]=0, \ \ \ \ \ U = \prod_{i=1}^N X_i.
\end{align}
This is a lattice manifestation of the continuum fact that the $\mathbb{Z}_2$ line $\eta$ is capable of ending topologically on the free boundary $\ket{f}$. For this reason, we again conclude that $\ket{f}$ is $\mathbb{Z}_2$-symmetric.

On the other hand, the  boundary conditions $\ket{\uparrow}$ and $\ket{\downarrow}$ can be implemented on the lattice by adding a boundary pinning field to the Hamiltonian (see, for instance, \cite{Balaska:2011tw}), 
\begin{align}
    H_{\uparrow} = H_f - \alpha (Z_1+Z_N), \ \ \ \ \ H_{\downarrow}=H_f + \alpha(Z_1+Z_N),
\end{align}
where $\alpha$ is an arbitrary positive coupling constant.
Alternatively, one could implement the pinning as a constraint on the Hilbert space, as opposed to energetically as we have done above. It is straight-forward to see that the spin-flip operator $U$ does not commute with $H_{\uparrow}$ nor $H_{\downarrow}$. In fact, no modification of $U$ close to the boundary will remedy the situation. This is a lattice manifestation of the continuum fact that the $\mathbb{Z}_2$ line cannot end topologically on the pinned boundaries, and so we conclude again that they are not $\mathbb{Z}_2$-symmetric.

To summarize, whether we use open string reasoning or closed string reasoning, we are led to the same conclusions about the structure of $\mathbb{Z}_2$-symmetric boundaries in the Ising model. In Section \ref{SWSB}, we will see that the equivalence of these two criteria  holds in general for invertible finite symmetries.

\subsubsection{Compact Free Boson and $U(1)$ Momentum and Winding Symmetries}

Consider a compact free boson $\phi$ with radius $R$, i.e., $\phi\sim \phi+2\pi R$. At generic radius, the theory possesses a $U(1)$ Kac--Moody chiral algebra, and the boundary conditions which preserve this chiral algebra are  the Dirichlet and Neumann boundary conditions. The Dirichlet boundary conditions come in a continuous family parametrized by $S^1$,
\begin{align}
    \phi\vert=\theta R,
\end{align}
where $\theta\in [0,2\pi)$. 
In the string theory context, $\theta$ 
parametrizes  the position of the corresponding Dirichlet brane in the target space circle. 
Here $|$ means the restriction to the boundary.
We label the corresponding boundary states as $|\theta\rangle_D$.

On the other hand, $\phi$ at a Neumann boundary condition obeys $\partial_n \phi\vert = 0$, 
where $\partial_n$ is the partial derivative in the direction normal to the boundary. 
Since $\phi$ is unconstrained on a Neumann boundary, one can add a boundary theta angle term,
\ie
{i\tilde\theta\over 2\pi R} \oint d\phi\,.
\fe
Therefore, the Neumann boundary conditions also come in an $S^1$ family, which we denote by $|\tilde \theta\rangle_N$ with $\tilde\theta \in [0,2 \pi)$.
Neumann boundaries for $\phi$ are equivalent to  Dirichlet boundaries of the T-dual compact scalar field $\tilde\phi\sim \tilde\phi+{2\pi\over R }$, defined by $- i d\phi = \star d\tilde\phi$,
\ie
\tilde\phi| =  {\tilde \theta\over R}\,.
\fe
 In the string theory picture, there is a Dirichlet brane wrapping the target space circle which supports a $U(1)$ gauge field on its world-volume, and $\tilde\theta$ is the corresponding holonomy of this gauge field.  

The free boson at generic radius has a $(U(1)^\text{(m)}\times U(1)^\text{(w)})\rtimes \mathbb{Z}_2$ global symmetry, and we may contemplate how it acts on the boundary conditions that we have identified. 
The $U(1)^\text{(m)}$ momentum symmetry shifts the $\phi$ field while the $U(1)^\text{(w)}$ winding symmetry shifts the $\tilde\phi$ field. 
Therefore, $U(1)^\text{(m)}$ acts on the two sets of boundaries as 
\ie
U_\alpha^\text{(m)} |\theta\rangle_D = |\theta+\alpha\rangle_D  \,,~~~
U_\alpha^\text{(m)} |\tilde\theta\rangle_N = |\tilde\theta\rangle_N\,.
\fe
In particular, we see that the Neumann boundaries are invariant under the $U(1)^\text{(m)}$ symmetry. 
Similarly, the Dirichlet boundaries are invariant under the $U(1)^\text{(w)}$ symmetry. 
If however we consider the diagonal $U(1)^\text{(D)}$ subgroup of $U(1)^\text{(m)}\times U(1)^\text{(w)}$, we see that there is no symmetric boundary.

The discussion above exemplifies a general connection between symmetric boundaries and 't Hooft anomalies \cite{Han:2017hdv,Jensen:2017eof,Numasawa:2017crf,Thorngren:2020yht,Li:2022drc}. 
The $\mathbb{Z}_2$ symmetry of the Ising CFT,  and the $U(1)^\text{(m)}$ (or $U(1)^\text{(w)}$) symmetry of a compact boson are both free of 't Hooft anomaly. 
As we describe in more detail in Section \ref{sec:anomaly}, this means that these non-anomalous symmetries kinematically admit simple, symmetric boundaries, and in the above examples they are furthermore dynamically realized in these CFTs. 
In contrast, the diagonal $U(1)^\text{(D)}$ symmetry is anomalous and it is kinematically incompatible with a simple, symmetric boundary. 
See, for example, \cite{Lin:2019hks,Lin:2021udi} for a review of this anomaly and for a related discussion on its implications for the operator spectrum.

\subsection{Strongly and Weakly Symmetric Boundaries}\label{SWSB}

We now extend the discussion of symmetric boundary conditions to non-invertible global symmetries in 1+1d CFTs.

We denote the set of simple boundary conditions as $\{\mathcal{B}_a \}_{a\in\mathcal{J}}$ where $\mathcal{J} = \{a,b,c, \cdots \}$ is a set of labels for these boundaries.\footnote{Generally, there are infinitely many, or even continuous families of,  simple conformal boundary conditions in a 1+1d CFT. However, given a finite symmetry category $\mathcal{C}$, we are always free to restrict attention to a set of finitely many simple boundaries  which transform into one another under the action of $\mathcal{C}$. Thus, for ease of exposition, we take the indexing set $\mathcal{J}$ to be finite. }
The corresponding boundary states are denoted as $\{\ket{\mathcal{B}_a} \}_{a\in\mathcal{J}}$.
We focus only on conformal boundary conditions.

A finite generalized  global symmetry of a 1+1d CFT is characterized by a \textit{fusion category} ${\cal C}$ of topological defect lines \cite{Fuchs:2002cm,Frohlich:2009gb,Bhardwaj:2017xup,Chang:2018iay}, encompassing both  the invertible and non-invertible cases.
As with boundary conditions, we call a topological defect line \emph{simple} if the only topological point operator on the line is the identity operator \cite{Chang:2018iay}.
The set of simple lines is denoted as $\{\mathcal{L}_i\}_{i\in\mathcal{I}}$ where $\mathcal{I} = \{i,j,k,\cdots \}$ is a set of labels.
The fusion algebra  takes the following form,
\begin{equation} \label{eq:fusion_lines}
    \mathcal{L}_i \otimes \mathcal{L}_j = \bigoplus_{k\in\mathcal{I}} N_{ij}^{k} \mathcal{L}_k \,,
\end{equation}
where $N_{ij}^{k} \in \mathbb{Z}_{\geq 0}$ are non-negative integer-valued fusion coefficients.
We review some basic facts (including $F$-symbols) about fusion categories in Appendix \ref{app:categories}. 

\begin{figure}
    \centering
    \raisebox{-0.5\height}{\begin{tikzpicture}

\draw[blue!65, very thick] (-3,-4) -- (0,0);
\draw[blue!65, very thick,-{Stealth[length=3mm,width=2mm]}] (-3,-4) -- 
 (-1.5,-2);
\node [style=none] at (-1,-2.3) {$\mathcal{L}_i$};

\draw[blue!65, very thick] (3,-4) -- (0,0);
\draw[blue!65, very thick,-{Stealth[length=3mm,width=2mm]}] (3,-4) -- 
 (1.5,-2);
\node [style=none] at (1,-2.3) {$\mathcal{L}_j$};

\draw[blue!65, very thick] (0,0) -- (0,4.5);
\draw[blue!65, very thick,-{Stealth[length=3mm,width=2mm]}] (0,0) -- (0,2.5);
\node [style=none] at (0.8,2) {$\mathcal{L}_k$};

\filldraw [black] (0,0) circle (4pt);  
\node [style=none] at (0.6,0) {$v$};
\end{tikzpicture} }\hspace{4cm}\raisebox{-0.65\height}{\begin{tikzpicture}
\begin{scope}[shift={(0,-4)}]
\draw[very thick, blue!65] (-1.75,3.5) -- (-1.75,7.2);
\draw[very thick, blue!65] (-1.75,7.4) -- (-1.75,9.3);
\draw[very thick, blue!65, -{Stealth[length=3mm,width=2mm]}] (-1.75,6.3) -- (-1.75,6.5);

\tikzset{
    partial ellipse/.style args={#1:#2:#3}{
        insert path={+ (#1:#3) arc (#1:#2:#3)}
    }
}
\draw[thick, black!100, dashed] (-1.75,2.5) [partial ellipse= 0:180:2.25cm and 1cm];
\draw[thick, black!100] (-1.75,2.5) [partial ellipse= 180:360:2.25cm and 1cm];
\draw[thick, black!100] (-1.75,8.3) [partial ellipse= 0:360:2.25cm and 1cm];
\draw[very thick, blue!65] (-0.5,1.7) -- (-0.5,7.5);
\draw[very thick, blue!65, -{Stealth[length=3mm,width=2mm]}]  (-0.5,5) -- (-0.5,4.5);

\draw[very thick, blue!65] (-3,1.7) -- (-3,7.5);
\draw[very thick, blue!65, -{Stealth[length=3mm,width=2mm]}] (-3,5) --  (-3,4.5);

	\begin{pgfonlayer}{nodelayer}
		\node [style=none] (0) at (-4, 2.5) {};
		\node [style=none] (1) at (0.5, 2.5) {};
		\node [style=none] (2) at (-4, 8.275) {};
		\node [style=none] (3) at (0.5, 8.275) {};
		\node [style=none] (8) at (0, 4) {$\mathcal{L}_j$};
		\node [style=none] (13) at (-1.2, 5.5) {$\mathcal{L}_k$};
		\node [style=none] (14) at (-3.5, 4) {$\mathcal{L}_i$};
	\end{pgfonlayer}
	\begin{pgfonlayer}{edgelayer}
		\draw [style=normalline] (2.center) to (0.center);
		\draw [style=normalline] (3.center) to (1.center);
  
	\end{pgfonlayer}

\end{scope}
\end{tikzpicture}}
    \caption{Left: A topological junction operator $v$ in $\mathrm{Hom}_{\mathcal{C}}(\mathcal{L}_i\otimes\mathcal{L}_j,\mathcal{L}_k)$. Right: By a conformal map, it is mapped to a dimension-zero state in the $S^1$ Hilbert space twisted by $\mathcal{L}_i$, $\mathcal{L}_j$, and $\mathcal{L}_k$.}
    \label{fig:trivalent}
\end{figure}
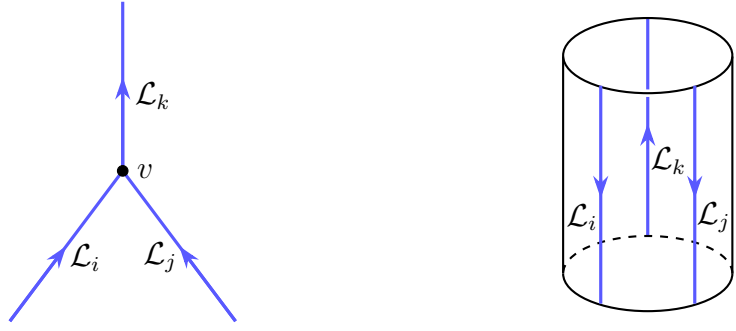

The bulk topological defect lines naturally act on boundary conditions by parallel fusion, as in the left of Figure \ref{fig:open_closed}.
Such a parallel fusion of a topological line with a conformal boundary yields another conformal boundary condition, which in general can be written as a linear combination of simple conformal boundaries with non-negative integer coefficients.
We denote the action of a topological line $\mathcal{L}$ on a conformal boundary $\mathcal{B}$ by  parallel fusion as $\mathcal{L} \otimes \mathcal{B}$, using the same product symbol as for the fusion of lines. 
In the closed string channel, this corresponds to acting with $\hat{\cal L}$ as an operator on the boundary state $|{\cal B}\rangle$, which yields another boundary state $\hat{\cal L}|{\cal B}\rangle$. 
We will use ${\cal L}\otimes {\cal B}$ and $\hat{\cal L}|{\cal B}\rangle$ interchangeably.

When a simple line $\mathcal{L}_i$ acts on a simple boundary $\mathcal{B}_a$, we can decompose the result as
\begin{equation}
    \mathcal{L}_i \otimes \mathcal{B}_a = \bigoplus_{b\in\mathcal{J}} \widetilde{N}_{ia}^{b} \mathcal{B}_b \,,
\end{equation}
where $\widetilde{N}_{ia}^{b} \in \mathbb{Z}_{\geq 0}$.
In other words, the set of simple boundary conditions forms a non-negative integer-valued matrix representation (NIM-rep) of the fusion algebra \eqref{eq:fusion_lines} of lines, and $\widetilde{N}_{ia}^{b}$ are the NIM-rep coefficients. 
Namely,
\begin{equation}
    \sum_{b \in \mathcal{J}} \widetilde{N}^{c}_{ib} \widetilde{N}^{b}_{ja} =
    \sum_{k \in \mathcal{I}} N_{ij}^k \widetilde{N}_{ka}^c \,.
\end{equation}
On top of the data of the NIM-rep, there are various consistency conditions that the set of boundary conditions must satisfy with respect to the action of the bulk topological lines.
These conditions are packaged into the  mathematical theory of \textit{module categories}. 
The objects of a module category are the conformal boundary conditions that are related by the action of the fusion category $\mathcal{C}$ of topological lines. As a simple example, the two fixed boundary conditions $\ket{\uparrow}$ and $\ket{\downarrow}$ of the Ising model form a module category of the $\mathbb{Z}_2$ global symmetry. 
We review the concept of a module category in Appendix \ref{app:categories}.
We denote the module category of boundary conditions as $\mathcal{M}$.

It is well-known that the fusion coefficients $N_{ij}^k$ themselves furnish a NIM-rep, i.e.,
\begin{equation}
    \sum_{p \in \mathcal{I}} N_{ij}^p N_{pk}^s =
    \sum_{q \in \mathcal{I}} N_{jk}^q N_{iq}^s \,.
\end{equation}
In particular, this NIM-rep is realized by the \emph{regular module category}, whose simple objects are those of $\mathcal{C}$ (i.e., $\mathcal{J}=\mathcal{I}$) and whose NIM-rep coefficients are given by the bulk fusion coefficients (i.e., $\widetilde{N}_{ij}^k=N_{ij}^k$).
More details are given in Appendix \ref{app:categories}.
Regular module categories are familiar in the study of RCFTs.
Indeed, the Cardy boundary conditions in any diagonal RCFT arrange themselves into the regular module category over the fusion category formed by the Verlinde lines \cite{Cardy:1989ir,Fuchs:2002cm}.
Relatedly, both the Cardy boundary conditions as well as the Verlinde lines are labeled by the bulk primary operators (that is, $\cal I = \cal J$).

In the case of topological line defects, the fusion coefficient $N_{ij}^k$ measures the dimension of the vector space of topological point operators at the trivalent junction where the three simple lines $\mathcal{L}_i$, $\mathcal{L}_j$, and $\mathcal{L}_k$ meet, as in Figure \ref{fig:trivalent}.
Such a vector space corresponds to the Hom space $\text{Hom}_{\mathcal{C}}(\mathcal{L}_i \otimes \mathcal{L}_j,\mathcal{L}_k)$ in the abstract fusion category $\mathcal{C}$, and the statement is that
\begin{equation}
    N_{ij}^k = \text{dim}_{\mathbb{C}}\,\text{Hom}_{\mathcal{C}}(\mathcal{L}_i \otimes \mathcal{L}_j,\mathcal{L}_k) \,.
\end{equation}
Using a conformal map, the topological point operators at a trivalent junction correspond to dimension-zero states in the $S^1$ Hilbert space twisted by the three lines $\mathcal{L}_i$, $\mathcal{L}_j$, and $\mathcal{L}_k$.  See Figure \ref{fig:trivalent}.

\begin{figure}
\begin{center}
\begin{tikzpicture}

\filldraw [gray!20] (6,0) rectangle (8,7);

\draw[black, ultra thick] (6,0) -- (6,7);

\draw[black, ultra thick, -{Stealth[length=3mm,width=2mm]}] (6,0) -- (6,2);
\draw[black, ultra thick, -{Stealth[length=3mm,width=2mm]}] (6,3.5) -- (6,5.8);
\draw[blue!65, very thick] (0,3.5) -- (6,3.5);
\draw[blue!65, very thick, -{Stealth[length=3mm,width=2mm]}] (0,3.5) -- (3.4,3.5);

\filldraw [black] (6,3.5) circle (4pt);  
\node [style=none] at (3, 2.8) {$\mathcal{L}_i$};

\node [style=none] at (6.7, 3.5) {$\tilde{v}$};

\node [style=none] at (6.8, 1.5) {$\mathcal{B}_a$};
\node [style=none] at (6.8, 5.5) {$\mathcal{B}_b$};

\end{tikzpicture}\hspace{3cm}\begin{tikzpicture}

\filldraw [gray!20] (0,0) rectangle (-2,7);
\filldraw [gray!20] (6,0) rectangle (8,7);
\draw[black, ultra thick] (0,0) -- (0,7);
\draw[black, ultra thick] (6,0) -- (6,7);
\draw[black, ultra thick, -{Stealth[length=3mm,width=2mm]}] (0,7) -- (0,3);
\draw[black, ultra thick, -{Stealth[length=3mm,width=2mm]}] (6,0) -- (6,4);
\draw[blue!65, very thick] (3,0) -- (3,7);
\draw[blue!65, very thick, -{Stealth[length=3mm,width=2mm]}] (3,7) -- (3,3);
 
\node [style=none] at (3.7, 3.3) {$\mathcal{L}_i$};

\node [style=none] at (-1, 3.5) {$\mathcal{B}_a$};
\node [style=none] at (7, 3.5) {$\mathcal{B}_b$};

\end{tikzpicture}
\end{center}
\caption {Left: A topological junction operator $\tilde{v}$ in $\mathrm{Hom}_{\mathcal{M}}(\mathcal{L}_i\otimes \mathcal{B}_a,\mathcal{B}_b)$. Right: By a conformal transformation, it is mapped to  a dimension zero state in the interval Hilbert space  with the boundary conditions $\mathcal{B}_a$ and $\mathcal{B}_b$ imposed at the two ends, and twisted by the line $\mathcal{L}_i$.}\label{fig:boundaryjunction}
\end{figure}
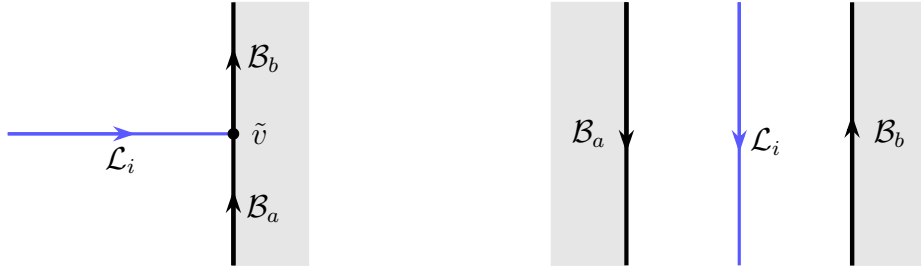

Similarly, the NIM-rep coefficient $\widetilde{N}_{ia}^b$ corresponds to the dimension of the vector space of topological point operators at the junction where the two simple boundaries $\mathcal{B}_a$ and $\mathcal{B}_b$ meet with the simple line $\mathcal{L}_i$, as in the left of Figure \ref{fig:boundaryjunction}.
This vector space corresponds to the Hom space $\mathrm{Hom}_{\mathcal{M}} (\mathcal{L}_i \otimes \mathcal{B}_a, \mathcal{B}_b)$ in the module category $\mathcal{M}$.
Thus, we have
\begin{equation}
    \widetilde{N}_{ia}^b = \text{dim}_{\mathbb{C}}\,\text{Hom}_{\mathcal{M}} (\mathcal{L}_i \otimes \mathcal{B}_a, \mathcal{B}_b) \,.
\end{equation}
Upon a conformal  map, the topological point operators at the junction of $\mathcal{B}_a$, $\mathcal{B}_b$, and $\mathcal{L}_i$ correspond to dimension-zero states in the interval Hilbert space twisted by the line $\mathcal{L}_i$ and with the  boundary conditions $\mathcal{B}_a$ and $\mathcal{B}_b$ imposed at the two ends.\footnote{The boundary conditions on the right end of the interval forms the \emph{left} module category $\cal M$ over $\cal C$, since the topological lines act from the left. On the other hand, boundary conditions on the left end of the interval forms the dual category ${\cal M}^{\vee}$ (which has the same set of objects as $\cal M$ but with the direction of all the morphisms reversed) which is a \emph{right} module category over $\cal C$, since the topological lines act from the right. See \cite[Remark~7.1.5]{etingof2016tensor}.}
 See the right of Figure \ref{fig:boundaryjunction}.
We denote such a Hilbert space as $\mathcal{H}_{ab}^{i}$.
When $i=\mathds{1}$ is the identity line, we simply write $\mathcal{H}_{ab} \equiv \mathcal{H}_{ab}^{\mathds{1}}$.

\subsubsection*{Weakly Symmetric Boundaries}

When $\widetilde{N}_{ia}^a \ge1$ for a boundary $\mathcal{B}_a$ and a topological line $\mathcal{L}_i$, the line $\mathcal{L}_i$ can \emph{topologically} end on the boundary $\mathcal{B}_a$.
The topological endpoint of the line at the boundary is generally not unique. The choice of junction operator lives in $\text{Hom}_{\mathcal{M}} (\mathcal{L}_i \otimes \mathcal{B}_a, \mathcal{B}_a)$ in the case that the line is oriented pointing into the boundary; 
similarly, the choice of junction operator lives in $\text{Hom}_{\mathcal{M}} (\bar{\mathcal{L}}_i \otimes \mathcal{B}_a, \mathcal{B}_a)$ in the case that the line is oriented pointing out of the boundary.

\begin{figure}
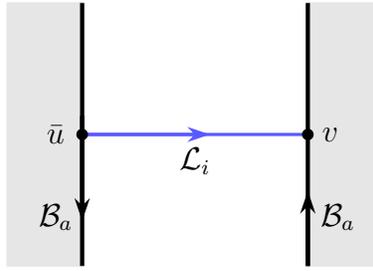

    \centering
    \ctikzfig{figures/weaklysymmetrictriple}
    \caption{A topological line ${\cal L}_i$ ends topologically on a weakly symmetric boundary ${\cal B}_a$, giving a conserved operator that commutes with the Hamiltonian on an interval.}
    \label{fig:weaklysymmetrictriple}
\end{figure}

Consider the theory defined on an interval with the $\mathcal{B}_a$ boundary condition imposed at the two ends.
Because the line $\mathcal{L}_i$ is able to end topologically on $\mathcal{B}_a$, it is free to move up and down along the (Euclidean) time direction, and defines a conserved quantity on the interval. 
To be more precise, every triple\footnote{More generally, every triple $(\mathcal{L}_i, v, \bar{u})$ of a topological line $\mathcal{L}_i$ and topological junction operators $v \in \text{Hom}_{\mathcal{M}} (\mathcal{L}_i \otimes \mathcal{B}_a, \mathcal{B}_a)$ and  $\bar{u} \in \text{Hom}_{\mathcal{M}^\vee} (\mathcal{B}_b, \mathcal{B}_b \otimes \mathcal{L}_i) \cong \text{Hom}_{\mathcal{M}} (\overline{\mathcal{L}}_i \otimes \mathcal{B}_b, \mathcal{B}_b)$ gives rise to an operator $\hat{\mathcal{L}}_{i;v}^{\bar{u}}$ which commutes with the Hamiltonian $H_{ab}$ on the interval Hilbert space $\mathcal{H}_{ab}$.} $(\mathcal{L}_i, v,\bar{u})$ of a topological line $\mathcal{L}_i$ and topological junction operators\footnote{More precisely, the junction $\bar u$ on the left boundary is an element of $\text{Hom}_{\mathcal{M}^\vee} (\mathcal{B}_a, \mathcal{B}_a \otimes \mathcal{L}_i)$ which is isomorphic to $\text{Hom}_{\mathcal{M}} (\overline{\mathcal{L}}_i \otimes \mathcal{B}_a, \mathcal{B}_a)$.} $v \in \text{Hom}_{\mathcal{M}} (\mathcal{L}_i \otimes \mathcal{B}_a, \mathcal{B}_a)$ and $\bar{u} \in \text{Hom}_{\mathcal{M}} (\overline{\mathcal{L}}_i \otimes \mathcal{B}_a, \mathcal{B}_a)$ at the boundaries gives rise to an operator $\hat{\mathcal{L}}_{i ; v}^{\bar{u}}$ acting on the interval Hilbert space  $\mathcal{H}_{aa}$, which commutes with the corresponding Hamiltonian $H_{aa}$,
\begin{equation}
[\hat{\mathcal{L}}_{i,v}^{\bar{u}}, H_{aa} ] = 0 \,.
\end{equation}
See Figure \ref{fig:weaklysymmetrictriple}.

Motivated by this, we make the following definition.
We say that \emph{a simple conformal boundary condition $\mathcal{B}_a$ is weakly symmetric under the fusion category symmetry $\mathcal{C}$ if every topological line in $\mathcal{C}$ can  end topologically on $\mathcal{B}_a$.}
In other words, $\mathcal{B}_a$ is weakly symmetric under $\mathcal{C}$ if $\widetilde{N}_{ia}^a \ge1$ for every simple line $\mathcal{L}_i$ in $\mathcal{C}$. 
Equivalently, $\mathcal{B}_a$ is weakly symmetric if
\begin{equation}
    \hat{\mathcal{L}}_i \ket{\mathcal{B}_a} = \ket{\mathcal{B}_a} + \cdots 
\end{equation}
for all $i\in \mathcal{I}$ (where $\cdots$ may contain more copies of $\ket{\mathcal{B}_a}$, if $\widetilde{N}_{ia}^a > 1$).
If a boundary condition $\mathcal{B}_a$ is weakly symmetric under $\mathcal{C}$, then every topological line in $\mathcal{C}$ gives rise to at least one conserved operator acting on the interval Hilbert space $\mathcal{H}_{aa}$.\footnote{A single topological defect line $\mathcal{L}_i$ may give rise to multiple conserved operators if the choice of the topological junction vector $v$ is not unique (up to rescaling).}
One may regard this as an open string channel definition of a symmetric boundary condition.
The definition of a weakly symmetric boundary condition is especially natural in the context of an open quantum chain, such as the golden chain \cite{Feiguin:2006ydp}. We will discuss this example in detail   in Section \ref{sec:goldenchain}.

\begin{figure}
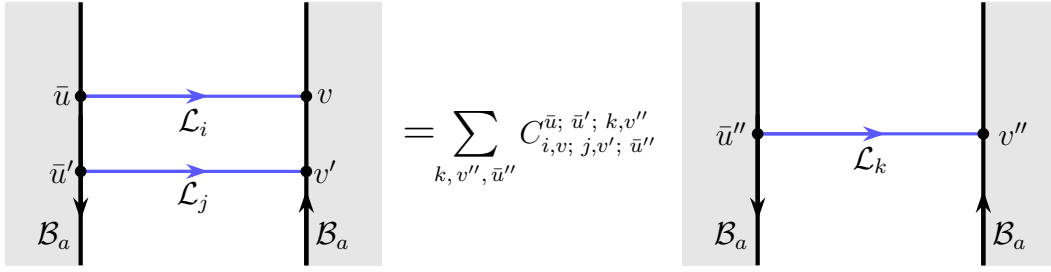

    \centering
\ctikzfig{figures/projectivealgebra}
    \caption{Fusion of two topological lines $\mathcal{L}_i$ and $\mathcal{L}_j$ on the interval with boundary conditions $\mathcal{B}_a$ at both ends depends on the junction vectors $v$, $\bar{u}$, $v'$, and $\bar{u}'$. It is written as a linear combination of topological lines $\mathcal{L}_k$ on the interval with junction vectors $v''$ and $\bar{u}''$ with, possibly, non-integer coefficients $C_{i,v;~j,v';~\bar{u}''}^{\bar{u};~ \bar{u}';~ k,v''}$.}
    \label{fig:projectivealgebra}
\end{figure}
Given a boundary $\mathcal{B}_a$ which is weakly symmetric with respect to a fusion category $\mathcal{C}$, we can consider the fusion algebra of the operators $\hat{\mathcal{L}}_{i,v}^{\bar{u}}$. In other words, we are looking for the fusion algebra of the lines $\mathcal{L}_i$ on an interval with the boundary condition $\mathcal{B}_a$ at both ends, as shown in Figure \ref{fig:projectivealgebra}. This algebra generally depends on the choice of topological junction operators $v$ and $\bar{u}$. A natural question to ask is how this fusion algebra compares to the fusion algebra of lines $\mathcal{L}_i \in \mathcal{C}$ in the bulk. In particular, one may ask if there is always a choice of topological junction vectors $v$ and $\bar{u}$ such that the fusion algebra in the presence of the boundary $\mathcal{B}_a$ agrees with the fusion algebra in the bulk. It turns out that this is not always possible. In particular, when the fusion of two operators $\hat{\mathcal{L}}_{i,v}^{\bar{u}}$ and $\hat{\mathcal{L}}_{j,v'}^{\bar{u}'}$  is expressed as a linear combination of the operators $\{\hat{\mathcal{L}}_{i,v''}^{\bar{u}''}\}$, there may be some (possibly non-integer) coefficients   $C_{i,v;~j,v';~\bar{u}''}^{\bar{u};~ \bar{u}';~ k,v''}$ \cite{Konechny:2019wff, Kojita:2016jwe, Barter_2022},
\begin{equation}\label{projective algebra}
 \hat{\mathcal{L}}_{i,v}^{\bar{u}} \times \hat{\mathcal{L}}_{j,v'}^{\bar{u}'} = 
 \sum_{k,v'',\bar{u}''}
 C_{i,v;~j,v';~\bar{u}''}^{\bar{u};~ \bar{u}';~ k,v''} \; \hat{\mathcal{L}}_{k,v''}^{\bar{u}''}.\end{equation}
This is in contrast with the bulk fusion of lines \eqref{eq:fusion_lines}, where the coefficients are always non-negative integers. This is reminiscent of projective representations of a group, where the usual group multiplication law is modified by phase factors which cannot always be removed by a group element redefinition.\footnote{See \cite{Li:2023mmw} for a related lattice discussion.}

A simple example of such a ``projective'' algebra on the interval can be found, for instance, for the Fibonacci fusion category symmetry, which we denote as ${\cal C} = \text{Fib}$.
The simple objects in this category are the identity line $\mathds{1}$ and the Fibonacci line $W$, satisfying the fusion algebra $W \otimes W = \mathds{1} \oplus W$.
This Fibonacci fusion category admits a weakly symmetric boundary condition.
To see this, consider the regular module category of Fib, whose simple objects (i.e., boundary states) we write as $\ket{\mathds{1}}$ and $\ket{W}$.
It is straightforward to see that $\ket{W}$ is weakly symmetric, since $\hat{\mathds{1}}\ket{W} =\ket{W}$ and $\hat{W} \ket{W} = \ket{W} \oplus \ket{\mathds{1}}$.
As will be discussed in Section \ref{subsec:examplesWeaklySymmetric}, this weakly symmetric boundary of Fib is realized in many models, including the open golden chain, the tricritical Ising model, and the diagonal WZW models based on $(G_2)_1$, $(F_4)_1$, and $SU(2)_3$.
We can define a symmetry operator $\hat{W}_v^{\bar u}$ acting on the interval Hilbert space ${\cal H}_{WW}$ with the weakly symmetric boundary condition imposed on both ends, where $v$ and $\bar u$ are junction operators which are unique up to an overall scale in this case.
With an appropriate choice of the normalization of the junction operators $v$ and $\bar u$, one can show that
\begin{equation}
    \hat{W}_v^{\bar u} \times \hat{W}_v^{\bar u} = \mathds{1} + \varphi^{-3/2}  \hat{W}_v^{\bar u} \,,
\end{equation}
where $\varphi = (1+\sqrt{5})/2$ is the golden ratio.
This can be dervied from the $F$-symbols following \cite{Konechny:2019wff, Kojita:2016jwe}.
It is easy to see that no rescaling of the junction opeartors $v$ and $\bar u$ can recover the bulk fusion algebra $W \otimes W = \mathds{1} \oplus W$.

\subsubsection*{Strongly Symmetric Boundaries}

Another definition of a symmetric boundary is in terms of  boundary states. 
We say  \emph{a simple conformal  boundary condition $\mathcal{B}$ is strongly symmetric with respect to the fusion category symmetry $\mathcal{C}$ if the corresponding boundary state $\ket{\mathcal{B}}$ is an eigenstate under the action of $\mathcal{C}$ with eigenvalues given by the quantum dimensions $\langle\mathcal{L}_i\rangle$,
\begin{equation} \label{eq:strong_NIM}
    \hat{\mathcal{L}}_i \ket{\mathcal{B}} = \langle \mathcal{L}_i \rangle \ket{\mathcal{B}} \quad \text{for all $i\in\mathcal{I}$} \,.
\end{equation}
}To be more precise, a strongly symmetric boundary condition $\mathcal{B}$ forms an indecomposable module category $\mathcal{M} = \text{Vec}$ over $\mathcal{C}$ by itself.\footnote{A module category is called indecomposable if it cannot be written as a direct sum of two module categories. See \cite{ostrik2003module} for more details.} That is, the only simple object in $\mathcal{M}$ is $\mathcal{B}$, and all the other objects are direct sums of several copies of $\mathcal{B}$. In particular, a strongly symmetric boundary condition $\mathcal{B}$ forms a one-dimensional NIM-rep under the fusion algebra \eqref{eq:fusion_lines} of topological lines. 
In contrast with the previous definition, one may regard this as a closed string channel definition of a symmetric boundary condition.

One immediate observation is that a strongly symmetric boundary condition cannot exist unless every topological line in the fusion category $\mathcal{C}$ has an integer quantum dimension, since otherwise \eqref{eq:strong_NIM} does not define a NIM-rep.  

\begin{figure}
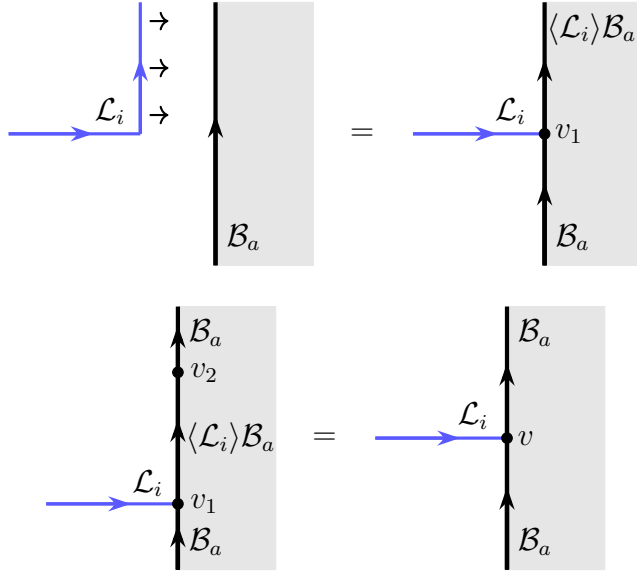

    \centering
    \ctikzfig{figures/stronglyimpliesweakly1}
    \ctikzfig{figures/stronglyimpliesweakly2}
    \caption{A topological line $\mathcal{L}_i$ can be bent into a right angle and fused with a strongly symmetric boundary $\mathcal{B}_a$ to produce a boundary junction $v_1$ (top figure). This junction may then be fused with another junction $v_2$ from $\langle \mathcal{L}_i\rangle\mathcal{B}_a$ to $\mathcal{B}_a$ (which always exists) to produce a topological endpoint $v$ for the line $\mathcal{L}_i$ on the boundary $\mathcal{B}_a$ (bottom figure). }
    \label{fig:stronglyimpliesweakly}
\end{figure}

 Let us compare these two notions of symmetric boundaries. 
For a general fusion category, we  note  that if a boundary   is strongly symmetric, then it is also weakly symmetric.
Indeed, for a strongly symmetric $\mathcal{B}_a$, we have 
\begin{equation}
    \widetilde{N}_{ia}^a = \langle \mathcal{L}_i \rangle \ge 1\,,~~~\forall~~i\in\mathcal{I}
\end{equation}
which implies that the defect line $\mathcal{L}_i$ can terminate topologically on $\mathcal{B}_a$ in $\langle \mathcal{L}_i\rangle$ linearly independent ways.\footnote{We focus on unitary conformal field theories, where the fusion category $\mathcal{C}$ is unitary. In a unitary fusion category, the quantum dimension of a simple line is always greater than or equal to 1.} One can reach the same conclusion pictorially. Indeed, by bending the line $\mathcal{L}_i$ into a right angle and pushing it onto the boundary $\mathcal{B}_a$ as in Figure \ref{fig:stronglyimpliesweakly}, one produces a trivalent topological junction $v_1$ between $\mathcal{L}_i$, $\mathcal{B}_a$, and $\langle \mathcal{L}_i\rangle \mathcal{B}_a$. Furthermore, calling $v_2$ any of the $\langle \mathcal{L}_i\rangle$ topological junctions between $\langle\mathcal{L}_i\rangle\mathcal{B}_a$ and $\mathcal{B}_a$, we may obtain the desired topological endpoint $v$ of $\mathcal{L}_i$ on the boundary $\mathcal{B}_a$ by fusing $v_1$ with $v_2$, thereby proving that $\mathcal{B}_a$ is weakly symmetric.

The converse is not true in general because there are examples of weakly symmetric boundaries (for non-invertible fusion categories) that are not strongly symmetric.
This happens if ${\cal L}_i \otimes {\cal B}_a = n_{ia} {\cal B}_a\oplus \cdots$ contains some copies of ${\cal B}_a$, but there are also other boundaries on the right-hand side. 
In Section \ref{subsec:examplesStronglySymmetric} and Section \ref{subsec:examplesWeaklySymmetric}, we will provide examples of symmetric boundaries of both kinds.

\begin{figure}
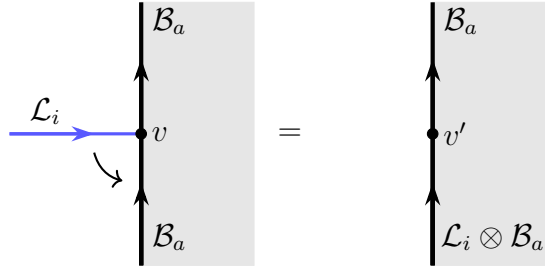

    \centering
\ctikzfig{figures/weaklyimpliesstronglyinvertible}
    \caption{An invertible line $\mathcal{L}_i$ which terminates on a weakly symmetric boundary $\mathcal{B}_a$ can be fused with half of the boundary to produce a topological junction between $\mathcal{L}_i$ and $\mathcal{L}_i\otimes\mathcal{B}_a$, which implies that $\mathcal{L}_i\otimes\mathcal{B}_a=\mathcal{B}_a$.}
    \label{fig:weaklyimpliesstronglyinvertible}
\end{figure}

However, if we restrict attention to \emph{invertible} symmetries, the converse does become true.  
  For a finite group symmetry $G$, 
 every irreducible NIM-rep of $G$ is given by a set of permutation matrices (see Appendix \ref{NIMrepG}).
Furthermore, the quantum dimension of an invertible line is always equal to 1.
This implies that, for the case of invertible symmetries, a weakly symmetric  boundary is also strongly symmetric, and the two notions of symmetric boundary conditions coincide. Again, it is also possible to see this pictorially. Indeed, given a junction between an invertible line $\mathcal{L}_i$ and a simple boundary $\mathcal{B}_a$, one may deform the line until it fuses with half of the boundary, producing a topological junction between $\mathcal{L}_i\otimes \mathcal{B}_a$ and $\mathcal{L}_i$. 
See Figure \ref{fig:weaklyimpliesstronglyinvertible}. 
The invertibility of $\mathcal{L}_i$ implies that $\mathcal{L}_i\otimes\mathcal{B}_a$ is simple,\footnote{Assume counterfactually that $\mathcal{L}_i\otimes\mathcal{B}_a$ is not simple. This means that it can be decomposed (not necessarily uniquely) into the sum of two  not necessarily simple boundaries, $\mathcal{L}_i\otimes \mathcal{B}_a=\mathcal{B}\oplus \mathcal{B}'$. Applying the inverse line $\overline{\mathcal{L}}_i$ to the left-hand side of this equation recovers the simple boundary $\mathcal{B}_a$, but applying it to the right-hand side recovers the non-simple boundary $(\overline{\mathcal{L}}_i \otimes \mathcal{B})\oplus (\overline{\mathcal{L}}_i \otimes\mathcal{B}')$, a contradiction.} and by Schur's lemma there can only exist a topological junction between two simple boundaries if the boundaries are in fact equal, so we conclude that $\mathcal{L}_i\otimes\mathcal{B}_a=\mathcal{B}_a$. Thus, if $\mathcal{B}_a$ is weakly symmetric with respect to an invertible $\mathcal{L}_i$, then it is an eigenstate of $\mathcal{L}_i$ and hence strongly symmetric as well. 
 
\section{Relation to Anomalies and Gauging}\label{sec:anomaly}

For internal, invertible symmetries, 't Hooft anomalies admit two equivalent definitions. 
One is as an obstruction to a symmetry-preserving, trivially gapped phase, and the other is as an obstruction to gauging. 
In this section, we argue that these two notions bifurcate for non-invertible symmetries:
sometimes it is  possible to ``gauge'' a fusion category $\mathcal{C}$ (by inserting a mesh of lines) even if it is not compatible with a trivially gapped phase. 
Correspondingly, we will see that the first notion is related to the kinematic existence of a strongly symmetric boundary, while the second notion is related to the kinematic existence of a weakly symmetric boundary.

\subsection{Strongly Symmetric Boundaries and Anomalies}

In this paper, we adopt the definition that a fusion category symmetry $\mathcal{C}$ is anomalous if $\mathcal{C}$ is incompatible with a trivially gapped phase. 

This physical definition can be translated to a mathematical condition on $\mathcal{C}$. We recall that there is a one-to-one correspondence between 1+1d $\mathcal{C}$-symmetric TQFTs and  module categories for $\mathcal{C}$ \cite{Thorngren:2019iar,Huang:2021zvu}. In this correspondence, the module category arises as the category of boundary conditions for the TQFT, and the content of \cite{Huang:2021zvu} is that the entire structure of the theory, including its bulk correlators, can be reconstructed just from this initial data. In particular, the dimension of the $S^1$ Hilbert space is equal to the number of simple objects of the corresponding module category. In a trivially gapped phase, this Hilbert space should be one-dimensional, and so we conclude that the trivially-gapped $\mathcal{C}$-symmetric phases are in correspondence with indecomposable module categories for $\mathcal{C}$ with \emph{one} simple object, which in turn are in correspondence with fiber functors $\mathcal{C}\to\mathrm{Vec}$ \cite[Example 7.4.6]{etingof2016tensor}. If $\mathcal{C}$ does not admit a fiber functor, we say that $\mathcal{C}$ has an anomaly \cite{Thorngren:2019iar}.

Recall that a strongly symmetric boundary corresponds to the unique simple object of an indecomposable module category ${\cal M}$. 
We therefore conclude that \textit{a fusion category is non-anomalous (i.e., admits a fiber functor) if and only if it admits a strongly symmetric boundary condition.}
This generalizes the familiar relation between 't Hooft anomalies and symmetric boundary conditions to the case of general non-invertible symmetries in 1+1d.

On the other hand, the anomaly of $\mathcal{C}$ does not necessarily forbid the existence of a weakly symmetric boundary condition.
For instance, the Fibonacci fusion category Fib, whose simple objects are $\mathds{1}$ and $W$ satisfying the fusion algebra $W \otimes W = \mathds{1} \oplus W$, is known to be anomalous \cite{Chang:2018iay}, and yet it admits a weakly symmetric boundary as we briefly explained in Section \ref{sec:symm_boundaries} and discuss more in Section \ref{subsec:examplesWeaklySymmetric}.
What is special about fusion categories that admit a weakly symmetric boundary? 
For such a category $\cal C$, we will see that 
 one can ``gauge $\mathcal{C}$'' in a generalized sense, even though it is incompatible with a trivially gapped phase.

 \subsection{Weakly Symmetric Boundaries and Gauging}

We briefly review how to gauge finite (non-invertible) symmetries in 1+1d, following \cite{Fuchs:2002cm,Frohlich:2009gb,Carqueville:2012dk,Brunner:2014lua,Bhardwaj:2017xup}. 
Gauging an invertible finite group symmetry  $G$ is  defined as summing over flat $G$-connections on the spacetime manifold. 
This is equivalent to inserting a network of $G$ symmetry lines along the dual triangulation of the manifold. There is also a freedom in assigning phases at the trivalent junctions associated with a choice of discrete torsion in $H^2(G,U(1))$. 
The gauging is consistent  if the resulting partition function does not depend on the choice of the triangulation; the obstruction is called a 't Hooft anomaly, and is labeled by an element of $H^3(G,U(1))$.  
This definition of gauging can be extended to a general   fusion category $\mathcal{C}$ by inserting a mesh of topological lines. 
The generalization to the non-invertible setting involves the choice of an algebra object in $\mathcal{C}$, as we explain below.

An \textit{algebra object} consists of a triple $(\mathcal{A}, \mu, u)$. 
First, 
$\mathcal{A}$ is an  object in $\mathcal{C}$  which is not necessarily simple. 
Every object of $\cal C$ can be expressed as a non-negative integer linear combination of the simple lines of $\cal C$, and we denote these coefficients for $\cal A$ as $\langle {\cal L}_i ,{\cal A}\rangle$:
\begin{equation} \label{eq:algebra}
\mathcal{A} = \bigoplus_{i\in \mathcal{I}} \langle{\cal L}_i ,{\cal A}\rangle  \mathcal{L}_i \,,~~~\langle{\cal L}_i ,{\cal A}\rangle\in\mathbb{Z}_{\ge 0}\,,
\end{equation}
where $\mathcal{I}$ denotes the set of labels for the simple lines in $\mathcal{C}$.
The coefficient $\langle{\cal L}_i ,{\cal A}\rangle$ is called the multiplicity of $\mathcal{L}_i$ in $\mathcal{A}$, and
we say $\mathcal{L}_i$ is a subobject of $\mathcal{A}$ if $\langle{\cal L}_i,{\cal A}\rangle\ge1 $.
Next, $\mu \in \text{Hom}_{\mathcal{C}}(\mathcal{A} \otimes \mathcal{A}, \mathcal{A})$ is called the \emph{multiplication} morphism, and $u \in \text{Hom}_{\mathcal{C}}(\mathds{1},\mathcal{A})$ is called the \emph{unit} morphism.
Together, the triple $(\mathcal{A},\mu,u)$ must satisfy the consistency conditions which are diagrammatically depicted in Figure \ref{fig:consistencyA}.
We explain this in more detail in Appendix \ref{app:categories}. 

\begin{figure}[t]
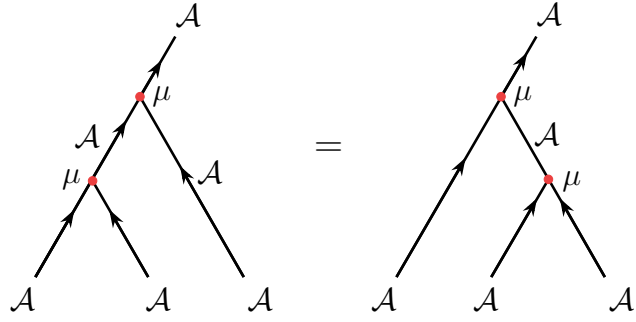
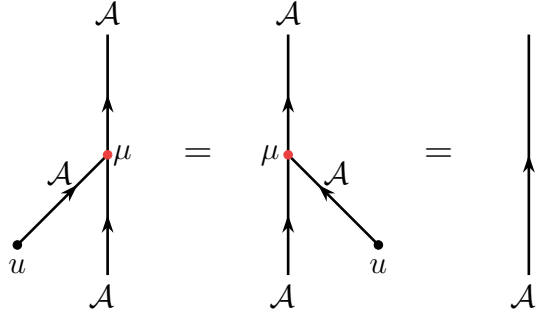

 \centering
    \begin{subfigure}[h]{1\textwidth}
    \centering
    \ctikzfig{figures/associativityAappa}
    \caption{Associativity condition for an algebra object.}
    \label{fig:associativityA}
    \end{subfigure}

    \begin{subfigure}[h]{1\textwidth}
  \centering
    \ctikzfig{figures/unitmorphismA}
    \caption{Compatibility with the unit morphism.}
    \label{fig:unitmorphismA}
    \end{subfigure}
    \caption{Consistency conditions satisfied by an algebra object.}
    \label{fig:consistencyA}
\end{figure}

We restrict attention to algebra objects $\mathcal{A}$ for which the multiplicity of the identity line in $\mathcal{A}$ is equal to 1, that is, $\langle\mathds{1},{\cal A}\rangle =1$.
An algebra object satisfying $\langle\mathds{1},{\cal A}\rangle = 1$ is called \emph{haploid} (or \emph{connected}).\footnote{Every semisimple indecomposable algebra object is Morita equivalent to a haploid algebra object \cite{ostrik2003module}.}
A haploid algebra object obeys the property \cite{Fuchs:2004dz}
\ie
\langle{\cal L}_i,{\cal A}\rangle\le \langle {\cal L}_i\rangle \,,
\fe
where $\langle {\cal L}_i\rangle$ is the quantum dimension of ${\cal L}_i$.
We briefly explain this in Appendix \ref{app:categories} as well.
Finally, for a haploid algebra object, the choice of the unit morphism $u \in \text{Hom}_{\mathcal{C}}(\mathds{1},\mathcal{A})$ is unique up to rescaling, since $\text{dim}_{\mathbb{C}}\,\text{Hom}_{\mathcal{C}}(\mathds{1},\mathcal{A})=1$.

A haploid algebra object is the non-invertible generalization of an anomaly-free subgroup  with a specific choice of discrete torsion. 
In the invertible case, a subgroup $G$ of $H$ is free of 't Hooft anomalies if $({\cal A}, \mu,u)$ is an algebra object, where
\begin{equation} \label{eq:algebra_group}
    \mathcal{A} = \bigoplus_{g\in G} \mathcal{L}_g
\end{equation}
and $\mu$ is given by an element of $H^2(G,U(1))$. 
The consistency conditions obeyed by $({\cal A},\mu ,u)$ guarantee that when we insert a mesh of $\cal A$ on any spacetime manifold with $\mu$ at the trivalent junctions, the resulting correlation functions are independent of the choice of the triangulation of the manifold.\footnote{To be more precise, to be able to consistently gauge an algebra object, the algebra object must be equipped with an additional structure of a coalgebra such that it becomes a \emph{symmetric $\Delta$-separable Frobenius algebra object} \cite{Fuchs:2002cm,Carqueville:2018sld}. However, every semisimple haploid algebra object admits a unique coalgebra structure such that it automatically becomes symmetric, $\Delta$-separable, and Frobenius (this can be seen by combining Proposition 2.(ii) of \cite{ostrik2003module} with the discussions in Section 3 of  \cite{Fuchs:2002cm}). Thus, we will not worry about these additional structures.}

Interestingly,  the subobjects of $\mathcal{A}$ might not be closed under fusion, i.e., they need not form a fusion subcategory of $\mathcal{C}$. 
In other words, when it comes to gauging, the non-invertible generalization of a subgroup is not a subcategory, and one must truly contend with algebra objects.

Now, we claim that \textit{if a fusion category symmetry $\mathcal{C}$  admits a weakly symmetric boundary condition, then there exists an algebra object $(\mathcal{A},\mu,u)$  that contains all the simple lines in $\mathcal{C}$.} 
Namely, there exists an algebra obejct  $\mathcal{A} = \bigoplus_{i \in \mathcal{I}} \langle{\cal L}_i,{\cal A}\rangle \mathcal{L}_i$ with $\langle{\cal L}_i,{\cal A}\rangle \ge1$ for all $i\in \mathcal{I}$.
Since $\mathcal{A}$ contains all the simple lines in $\mathcal{C}$, gauging $\mathcal{A}$ may be thought of as ``gauging $\mathcal{C}$." 
In this sense we say ${\cal C}$ is ``gaugeable," becuase we can consistently insert a mesh of lines in which every object of $\cal C$ participates. 
To prove this claim, an essential ingredient is the concept of the \emph{internal Hom} \cite{ostrik2003module}.\footnote{Many of the category theory concepts we use here, including internal Hom, are reviewed in \cite{Bhardwaj:2017xup}. We also briefly review them in Appendix \ref{app:categories}.}

Recall that the set of boundary conditions in a 1+1d theory with a fusion category symmetry $\mathcal{C}$ forms a module category $\mathcal{M}$ over $\mathcal{C}$.
Given an object $\mathcal{B}$ in $\mathcal{M}$ (that is, a boundary condition), the internal Hom from $\mathcal{B}$ to $\mathcal{B}$, denoted as $\underline{\text{Hom}}(\mathcal{B},\mathcal{B})$, is an object in $\mathcal{C}$ (that is, a topological line) with the property that
\begin{equation}
    \text{Hom}_{\mathcal{M}}(\mathcal{L} \otimes \mathcal{B}, \mathcal{B}) \cong \text{Hom}_{\mathcal{C}}(\mathcal{L},\underline{\text{Hom}}(\mathcal{B},\mathcal{B}))
\end{equation}
for every  $\mathcal{L}\in {\cal C}$. 
In other words, the topological line $\mathcal{A} \equiv \underline{\text{Hom}}(\mathcal{B},\mathcal{B}) $ satisfies
\begin{equation}\label{intHomZ}
   \langle{\cal L}_i,{\cal A}\rangle = \text{dim}_{\mathbb{C}}\, \text{Hom}_{\mathcal{M}}(\mathcal{L}_i \otimes \mathcal{B}, \mathcal{B})\,,~~~\forall i \in {\cal I}\,.
\end{equation}
Importantly, such an internal Hom $\mathcal{A}$ is always equipped with the structure of an algebra object for any choice of the object $\mathcal{B}$ in $\mathcal{M}$.
Moreover, if $\mathcal{B}$ is a simple boundary condition, then
\begin{align}
    \langle \mathds{1} , \mathcal{A} \rangle = \text{dim}_\mathbb{C}\, \text{Hom}_{\mathcal{M}}(\mathds{1} \otimes \mathcal{B}, \mathcal{B}) = \text{dim}_\mathbb{C}\, \text{Hom}_{\mathcal{M}}(\mathcal{B}, \mathcal{B})= 1.
\end{align} 
In particular, when $\mathcal{B}$ is simple, $\mathcal{A} \equiv \underline{\text{Hom}}(\mathcal{B},\mathcal{B}) $ is a haploid algebra object.
For more details on the explicit construction of the internal Hom algebra object, see Appendix \ref{app:categories}.

Now, assume that $\cal C$ admits a weakly symmetric boundary condition.
This means that there exists a module category $\mathcal{M}$ over $\mathcal{C}$ which contains a simple object $\mathcal{B}$ for which 
\begin{equation}
    \text{dim}_{\mathbb{C}}\, \text{Hom}_{\mathcal{M}}(\mathcal{L}_i \otimes \mathcal{B}, \mathcal{B}) \geq 1\,, ~~\forall i\in\mathcal{I}\,.
\end{equation}
It follows from \eqref{intHomZ} that the internal Hom $\mathcal{A}=\underline{\text{Hom}}(\mathcal{B},\mathcal{B}) $ is a haploid algebra object that contains all the simple lines of $\cal C$.\footnote{We will assume that the module category $\cal M$ arising from boundary conditions is semisimple. Physically, this means that every boundary condition is a direct sum (superposition) of simple boundary conditions. This then implies $\cal A$ is semisimple.}
This proves the claim that if $\cal C$ admits a weakly symmetric boundary, then it is ``gaugeable."

\begin{figure}
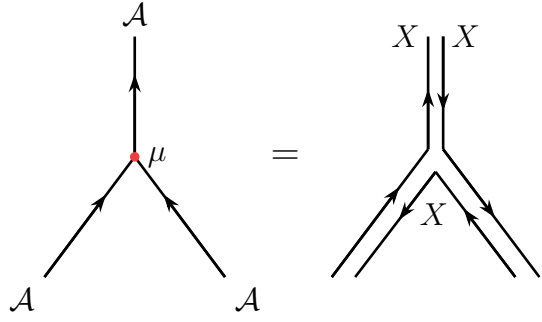

    \centering
    \ctikzfig{figures/moritatrivialfib}
    \caption{An object of the form ${\cal A}=X\otimes \overline{X}$ can be lifted to an algebra object $({\cal A},\mu,u)$  that is Morita equivalent to a trivial object \cite[Example~7.8.4]{etingof2016tensor}. Physically, it means that gauging such an algebra object leaves the CFT invariant. Intuitively, this is because a mesh of ${\cal A}$ can be desolved into the double lines of $X$, which can be shrunk to nothing. Examples include ${\cal A}=\mathds{1}\oplus W=W\otimes W$ of Fib and ${\cal A}=\mathds{1}\oplus\eta 
 ={\cal N}\otimes {\cal N}$ of TY$_\pm(\mathbb{Z}_2)$.}
    \label{fig:moritatrivialfib}
\end{figure}

Some remarks are in order.
First, recall that physically distinct gaugings correspond to  Morita equivalence classes of algebra objects in $\mathcal{C}$ (see, for instance, \cite{Bhardwaj:2017xup}). 
An algebra object  obtained from the internal Hom of a weakly symmetric boundary condition is generally Morita equivalent to another algebra object which does not contain all the simple lines in $\mathcal{C}$. 
For instance, as mentioned previously, the Fibonacci fusion category admits a weakly symmetric boundary condition $\ket{W}$ which belongs to the regular module category, and
the internal Hom of $\ket{W}$ produces
\begin{equation}
    {\cal A} = \mathds{1} \oplus W \,,
\end{equation}
which is equipped with a structure of an algebra object. 
Therefore, the Fibonacci fusion category is ``gaugeable" even though it is incompatible with a trivially gapped phase because of the non-integral quantum dimension of $W$. 
However, this algebra object is Morita trivial.
Indeed, as explained in Appendix \ref{app:categories}, the internal Hom based on the regular module category always produces a Morita trivial algebra object.
More intuitively, the fact that ${\cal A} = \mathds{1} \oplus W=W\otimes W$ admits a Morita trivial algebra structure is because a mesh of $\cal A$ can be dissolved into a mesh of double lines of $W$, which can then be shrunk to nothing. See Figure \ref{fig:moritatrivialfib}.\footnote{It is easy to give examples of Morita non-trivial algebra objects that contain all the simple lines for a fusion category that doesn't admit a fiber functor. For instance, consider the tensor product of  $\text{Fib}\boxtimes\text{Vec}_{\mathbb{Z}_2}$, then the algebra object $(\mathds{1}\oplus W)\otimes (\mathds{1}\oplus \eta)$ is one such an example.} 
Physically, it means that any CFT with a Fibonacci category is invariant under gauging  ${\cal A}=\mathds{1}\oplus W$.\footnote{This is analogous to the self-duality under gauging a $\mathbb{Z}_2$ symmetry in any CFT with a $\mathrm{TY}_\pm(\mathbb{Z}_2)$ category, in which case there is an algebra object ${\cal A}=\mathds{1}\oplus\eta = {\cal N}\otimes {\cal N}$ which is Morita trivial (see \cite[Example~7.8.18]{etingof2016tensor}). 
Here $\eta$ is the $\mathbb{Z}_2$ line and $\cal N$ is the Kramers-Wannier duality line.}

Second, one can similarly relate the existence of a strongly symmetric boundary condition to gauging using the internal Hom construction.
If the fusion category $\mathcal{C}$ kinematically allows   a strongly symmetric boundary condition, then there is a module category $\mathcal{M}$ over $\mathcal{C}$ which contains a unique simple object $\mathcal{B}$, satisfying
\begin{equation}
\text{dim}_{\mathbb{C}}\, \text{Hom}_{\mathcal{M}}(\mathcal{L}_i \otimes \mathcal{B}, \mathcal{B}) = \langle \mathcal{L}_i \rangle\,,~~\forall ~i\in\mathcal{I}\,.
\end{equation}
This implies that 
\begin{equation} \label{eq:max_algebra}
\mathcal{A} = \underline{\text{Hom}}(\mathcal{B},\mathcal{B}) = \bigoplus\limits_{i \in \mathcal{I}} \langle \mathcal{L}_i \rangle  \mathcal{L}_i
\end{equation}
is an algebra object.
Since $\langle \mathcal{L}_i , \mathcal{A} \rangle \leq \langle \mathcal{L}_i \rangle$ for all $i \in \mathcal{I}$ \cite{Fuchs:2004dz} (see also Appendix \ref{app:categories}), the object \eqref{eq:max_algebra} has the maximum quantum dimension among all the haploid algebra objects in $\mathcal{C}$.
More specifically, the quantum dimension of $\mathcal{A}$ is equal to the total dimension of the fusion category $\mathcal{C}$, i.e.\ $\langle \mathcal{A} \rangle = \sum_{i\in \mathcal{I}} \langle \mathcal{L}_i \rangle^2$.
Intuitively, one may say that the fusion category $\mathcal{C}$ can be ``maximally gauged'' if it admits a strongly symmetric boundary condition, or equivalently, if it is anomaly-free.
Furthermore, in the case of a strongly symmetric boundary condition, the algebra object \eqref{eq:max_algebra} is never Morita equivalent to another algebra object which does not contain all the simple lines of $\mathcal{C}$.

To conclude, we find that sometimes it is  possible to ``gauge'' a fusion category $\mathcal{C}$ (by inserting a mesh of lines) even if $\cal C$ is not compatible with a trivially gapped phase.
This happens precisely for those fusion categories which only admit  a weakly but not strongly  symmetric boundary.
One may thus call a fusion category $\mathcal{C}$ strongly anomaly-free if it admits a fiber functor (i.e., if it is compatible with a trivially gapped phase), and  weakly anomaly-free if there exists an algebra object containing all the simple objects of $\mathcal{C}$ as subobjects (i.e., if $\mathcal{C}$ can be ``gauged'').
In this terminology,   a fusion category symmetry $\mathcal{C}$ kinematically allows the existence of a weakly/strongly symmetric boundary condition if and only if it is weakly/strongly anomaly-free.\footnote{We have not shown that if $\mathcal{C}$ is weakly anomaly-free then it  admits the existence of a weakly symmetric boundary condition. Suppose $\mathcal{C}$ is weakly anomaly-free, so that there exists an algebra object $\mathcal{A}$ containing all the simple objects in $\mathcal{C}$ as subobjects. The algebra object $\mathcal{A}$ defines a simple object in a module category $\mathcal{M}$ over $\mathcal{C}$ such that $\mathcal{A} = \underline{\text{Hom}}(\mathcal{A},\mathcal{A})$ \cite[Exercise~7.10.6]{etingof2016tensor}.
This implies that $\mathcal{A}$ as an object in $\mathcal{M}$  defines a weakly symmetric boundary condition under the action of $\mathcal{C}$.}

\section{Examples}\label{sec:example}

In this section, we provide several examples of strongly and weakly symmetric boundaries in 1+1d CFTs and lattice models.

\subsection{Strongly Symmetric Boundaries}\label{subsec:examplesStronglySymmetric}

We begin with strongly symmetric boundary conditions, where we offer two examples. The first involves a certain Tambara--Yamagami category in a  $c=1$ orbifold model. The second involves $G$ and $\mathrm{Rep}(G)$ symmetries in holomorphically factorized CFTs constructed from permutation orbifolds.

\subsubsection{Ising$^2$ and a $\mathbb{Z}_2\times \mathbb{Z}_2$ TY Category} \label{sec:ising2}

Let us begin with the first example one might think to consider, the Ising model, and show that it does not admit a strongly symmetric boundary with respect to its full fusion category of Verlinde lines, which includes a non-invertible symmetry. Recall that the topological line defects of the Ising model are in one-to-one correspondence with its three primary operators and obey the same fusion rules, i.e.,
\begin{align}\label{eqn:Isingfusionrules}
    \eta\otimes \eta =\mathds{1}\,, \ \ \  \eta\otimes \mathcal{N} = \mathcal{N}\otimes \eta = \mathcal{N}\,, \ \ \ \mathcal{N}\otimes\mathcal{N} = 1\oplus\eta \,.
\end{align}
Here, $\eta$ is an invertible  $\mathbb{Z}_2$ line and $\cal N$ is the non-invertible Kramers-Wannier duality line \cite{Frohlich:2004ef}.
The fusion category generated by these lines is $\mathrm{TY}_{+}(\mathbb{Z}_2)$, also known as the Ising category.\footnote{The Ising fusion category generated by these topological line defects can be obtained  from   the modular tensor category (MTC) of the $c=\frac12$ Virasoro algebra (often called the Ising MTC) by forgetting its braiding and twist. } 
Here $\mathrm{TY}_\pm (\mathbb{Z}_2)$ are the two Tambara--Yamagami (TY) fusion categories \cite{TAMBARA1998692} based on $\mathbb{Z}_2$. 
 These two TY categories  have the same fusion rules but are distinguished by the sign $\pm$ of the Frobenius-Schur indicator of the line $\mathcal{N}$.\footnote{Incidentally, the other $\mathbb{Z}_2$ Tamabara-Yamagami category $\mathrm{TY}_-$ is realized in the $SU(2)_2$ WZW model.}

The boundary conditions $\ket{\uparrow}$, $\ket{\downarrow}$, and $|f\rangle$ (described in Section \ref{subsubsec:Isingmodel}) generate a module category of this $\mathrm{TY}_+(\mathbb{Z}_2)$ fusion category. In fact, because the Ising model is a diagonal RCFT, this module category is  the regular module category of $\mathrm{TY}_+(\mathbb{Z}_2)$. 
This means that there is a one-to-one correspondence between the  boundary conditions and the topological lines (and also with the local primaries), i.e.\ $\ket{\uparrow}\leftrightarrow \mathds{1},\ket{\downarrow} \leftrightarrow \eta, |f\rangle\leftrightarrow {\cal N}$. 
The action of the lines on the boundaries is the same as the fusion of the corresponding lines (see \eqref{VonC} below for generalizations to minimal models):
\begin{align}\label{eqn:Isinglinesactionboundary}
\begin{split}
   &\hspace{.3in} \hat{\eta} \ket{\uparrow} = \ket{\downarrow}\,, \ \ \ \hat{\eta}\ket{\downarrow} = \ket{\uparrow}\,, \ \ \ \hat{\eta}|f\rangle  = |f\rangle\,, \\ 
   & \hat{\mathcal{N}}\ket{\uparrow} = |f\rangle\,, \ \ \ \hat{\mathcal{N}}\ket{\downarrow} = |f\rangle\,, \ \ \ \hat{\mathcal{N}}|f\rangle = \ket{\uparrow}\oplus\ket{\downarrow}\,.
\end{split}
\end{align}
We discover that there are neither strongly symmetric boundary conditions, nor weakly symmetric ones. Indeed, the absence of strongly symmetric boundary conditions could have been anticipated even on kinematic grounds due to the fact that the category is not integral and thus anomalous (the duality line has a non-integer quantum dimension, $\langle \mathcal{N}\rangle=\sqrt{2}$). Furthermore, it is straightforward to see from  \eqref{eqn:Isinglinesactionboundary} that $\mathcal{N}$ cannot end topologically on any of the simple boundary conditions either.

Instead, let us consider the simplest TY category which \emph{is} kinematically compatible with a strongly symmetric boundary condition. Take the 1+1d CFT which, in the bulk, is described by stacking two decoupled Ising models. We refer to this CFT as Ising$^2$. Because it has central charge $c=1$, it should arise somewhere in the (conjecturally completely classified \cite{Ginsparg:1987eb}) conformal manifold of $c=1$ CFTs. In fact, it turns out to correspond to the theory obtained by performing an $\phi\to -\phi$ orbifold of the compact free boson with radius $R=\sqrt{2}$, in conventions where $R=1$ is the radius which is self-dual under T-duality.

This theory admits at least a $\mathrm{TY}_+(\mathbb{Z}_2)\boxtimes \mathrm{TY}_+(\mathbb{Z}_2)$ fusion category of topological line defects, one factor coming from each of its two Ising constituents. We work with the subcategory generated by $\mathds{1},~\eta_1,~\eta_2,~\eta_1\otimes\eta_2$, and $\mathcal{N}'\equiv \mathcal{N}_1\otimes \mathcal{N}_2$, where $\eta_i$ and $\mathcal{N}_i$ are the $\mathbb{Z}_2$ symmetry and duality lines respectively of the $i$th Ising factor.  
Note that the non-invertible line ${\cal N}'$ has integer quantum dimension, $\langle {\cal N}'\rangle =2$. 
 This subcategory is equivalent to the Tambara-Yamagami category $\mathrm{TY}(\mathbb{Z}_2\times\mathbb{Z}_2,\chi,+)$ based on $\mathbb{Z}_2\times \mathbb{Z}_2$ with the following choice of the bicharacter $\chi$,
\begin{align}
    \chi(a_1,a_2;b_1,b_2) = (-1)^{a_1 b_1 + a_2b_2}\,,
\end{align}
where we use $(a_1,a_2) ,(b_1,b_2)$ with $a_i,b_i=0,1$ to denote the elements of $\mathbb{Z}_2\times\mathbb{Z}_2$. The $+$ stands for the choice of the Frobenius-Schur indicator of the line $\mathcal{N}'$.
Its fusion rules are correspondingly
\begin{align}
g\otimes\mathcal{N}'=\mathcal{N}'\otimes g = \mathcal{N}', \ \ \ \mathcal{N}'\otimes\mathcal{N}' = \bigoplus_{g\in \mathbb{Z}_2\times\mathbb{Z}_2} g.
\end{align}
Here $g$  runs over the lines $\mathds{1}$, $\eta_1$, $\eta_2$, and $\eta_1\otimes \eta_2$, which furnish the fusion rules of $\mathbb{Z}_2\times \mathbb{Z}_2$.
Another way to describe this subcategory is that it is equivalent to $\mathrm{Rep}(H_8)$, where $H_8$ is the non-grouplike Hopf algebra of dimension 8 known as the Kac-Paljutkin algebra \cite{kats1966finite}. 

We would like to show that --- unlike for a single copy of the Ising model, which does not admit a strongly or weakly symmetric boundary condition with respect to its non-invertible symmetry --- this theory \emph{does} admit a strongly symmetric boundary condition with respect to $\mathrm{Rep}(H_8)$. 
\begin{figure}[t]
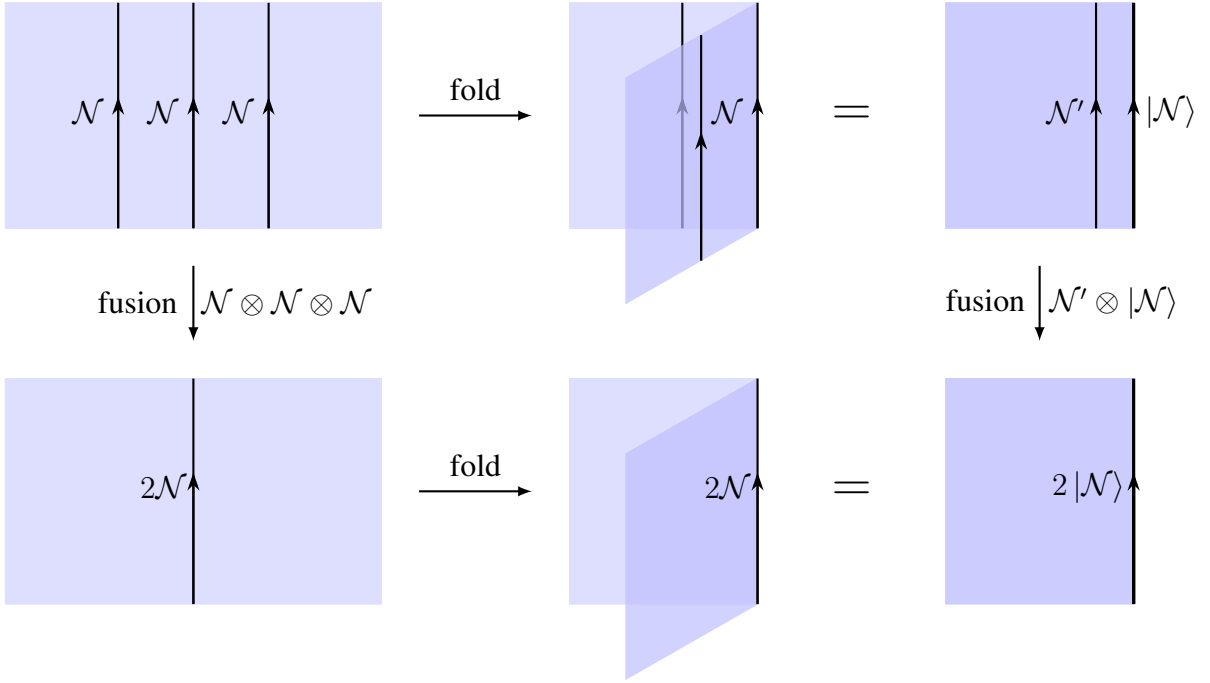

    \centering
    \ctikzfig{figures/ising2folding}
    \caption{Folding the Ising CFT with the Kramers-Wannier duality line $\mathcal{N}$ inserted at the crease creates a  conformal boundary in the $\text{Ising}^2$ CFT, denoted  $\ket{\mathcal{N}}$. 
    This boundary $\ket{\cal N}$ is strongly symmetric under the non-invertible  $\mathrm{Rep}(H_8)$ symmetry of the Ising$^2$ CFT. In particular, the diagram illustrates that $\hat{\mathcal{N}}'\ket{\mathcal{N}}=2\ket{\mathcal{N}}$, where $\mathcal{N}'=\mathcal{N}_1\otimes\mathcal{N}_2$ is the fusion of the two duality lines in the two copies of Ising.
    }
    \label{fig:ising2folding}
\end{figure}
The boundary condition we study can be obtained as follows. Consider the duality line $\mathcal{N}$ in \emph{one} copy of the Ising model. By the folding trick, this becomes a boundary condition in Ising$^2$, which we denote by $|\mathcal{N}\rangle$. We can calculate the action of $\mathrm{Rep}(H_8)$ on $|\mathcal{N}\rangle$ by unfolding to one copy of the Ising model as illustrated in Figure \ref{fig:ising2folding}. For example, the fusion of $\eta_1$ with $|\mathcal{N}\rangle$ in Ising$^2$ corresponds in the unfolded theory to computing the fusion rule of $\eta$ (say, from the left) with $\mathcal{N}$. In particular, we conclude from  \eqref{eqn:Isingfusionrules} that 
\begin{align}
    \eta\otimes\mathcal{N}  =\mathcal{N} \text{ in Ising} \implies \hat{\eta}_1|\mathcal{N}\rangle = \ket{\cal N} \text{ in Ising}^2.
\end{align}
Similarly, the fusion rule of $\eta_2$ with $|\mathcal{N}\rangle$ in Ising$^2$ can be computed in the unfolded theory by fusing $\eta$ with $\mathcal{N}$ again, but this time from the right, and the same argument shows that $\hat{\eta}_2|\mathcal{N}\rangle=|\mathcal{N}\rangle$. Finally, because $\mathcal{N}_1$ goes over in the unfolded theory to $\mathcal{N}$ to the left of the interface, while $\mathcal{N}_2$ goes over to $\mathcal{N}$ to the right of the interface, we can deduce the action of $\mathcal{N}'$ on $|\mathcal{N}\rangle$ by calculating 
\begin{align}
    \mathcal{N}\otimes\mathcal{N}\otimes\mathcal{N} = \mathcal{N}\otimes (1\oplus\eta) = 2\mathcal{N} \text{ in Ising} \implies \hat{\mathcal{N}}'|\mathcal{N}\rangle = 2|\mathcal{N} \rangle = \langle \mathcal{N}'\rangle |\mathcal{N}\rangle \text{ in Ising}^2.
\end{align}
Thus, we conclude that $|\mathcal{N}\rangle$ is a strongly symmetric boundary condition  for the non-invertible symmetry $\mathrm{Rep}(H_8)$  in the Ising$^2$ CFT.

\subsubsection{CFTs from Permutation Orbifolds and $\mathrm{Rep}(G)$ Symmetry}\label{subsubsec:repG}

Recall that a holomorphic vertex operator algebra (VOA) $\mathcal{V}$ is a VOA with exactly one irreducible representation. The unique character
\begin{align}
    \mathrm{ch}_{\mathcal{V}}(\tau) = \mathrm{Tr}_{\mathcal{V}}q^{L_0-c/24}
\end{align}
of a holomorphic VOA is modular invariant up to a phase, 
\begin{align}
    \mathrm{ch}_{\mathcal{V}}(-1/\tau) = \mathrm{ch}_{\mathcal{V}}(\tau), \ \ \ \ \mathrm{ch}_{\mathcal{V}}(\tau+1) = e^{-\pi i c/12} \mathrm{ch}_{\mathcal{V}}(\tau).
\end{align}
When $c$ is a multiple of $24$, this character is completely modular invariant, and we may think of $\mathcal{V}$ as a consistent chiral CFT in its own right, and $\mathrm{ch}_{\mathcal{V}}(\tau)$ as its torus partition function. Examples of holomorphic VOAs include the monster CFT $V^\natural$ \cite{frenkel1989vertex}, the 70 Schellekens CFTs \cite{Schellekens:1992db}, and any chiral boson theory based on an even unimodular lattice.

Now, let $\mathcal{V}$ be a holomorphic VOA and $G$ a solvable finite group.\footnote{A long-standing conjecture in the mathematics community is that if $\mathcal{V}$ is a completely rational VOA with an action of $G$ by symmetries, then so is its corresponding fixed-point subalgebra $\mathcal{V}^G=\{v\in \mathcal{V}\mid g v = v \text{ for all }g\in G\}$. This has been proven when $G$ is a solvable finite group \cite{Carnahan:2016guf}. If the conjecture is true for general finite groups, then we can relax our assumption of solvability as well.\label{footnote:regularityconjecture}} Using Cayley's theorem, we can realize $G$ as a subgroup of the symmetric group $S_n$ for some positive integer $n$. Then, $G$ acts by permutation symmetries on the holomorphic VOA 
\begin{align}\label{eqn:tensorpowerV}
    \mathcal{T}_0\equiv \mathcal{V}^{\otimes n}.
\end{align} If the central charge of $\mathcal{V}$ is a multiple of $24$ (so that it is a chiral CFT), then it is guaranteed that this action is non-anomalous \cite{Evans:2018qgz} and we may contemplate the gauged theory $\mathcal{T}_0\big/G$, which has a quantum $\mathrm{Rep}(G)$ symmetry \cite{Vafa:1989ih,Bhardwaj:2017xup}.

\begin{figure}
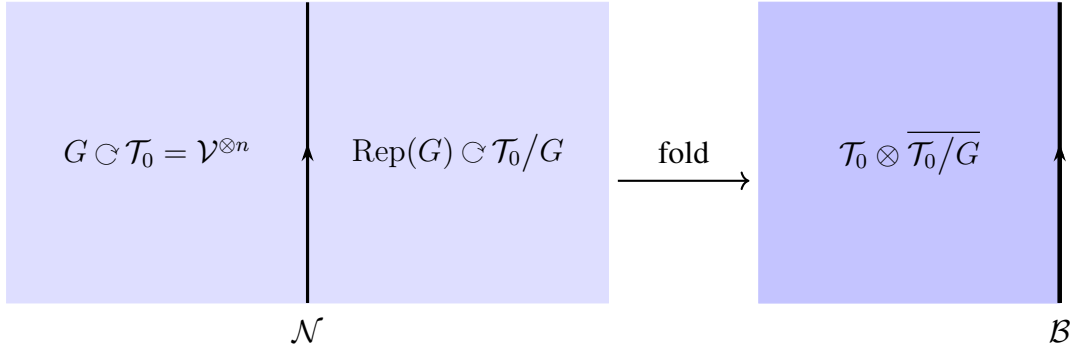

    \centering
\ctikzfig{figures/repGbc}
    \caption{The holomorphic CFT $\mathcal{T}_0={\cal V}^{\otimes n}$ and its gauged theory $\mathcal{T}_0/G$ are folded around an interface $\mathcal{N}$ which is obtained by gauging $\mathcal{T}_0$ by a group $G$ in half of the  spacetime. The resulting boundary $\mathcal{B}$ in the folded theory $\mathcal{T}_0 \otimes \overline{\mathcal{T}_0/G}$ is strongly symmetric under the action of $G\boxtimes\text{Rep}(G)$. }
    \label{fig:repGbc}
\end{figure}

Next, we form a topological interface $\mathcal{N}$ between $\mathcal{T}_0$ and $\mathcal{T}_0\big/ G$ by considering the theory $\mathcal{T}_0$ and gauging $G$ in half of spacetime, taking Dirichlet boundary conditions for the gauge field at the interface \cite{Choi:2021kmx}. The Dirichlet boundary conditions guarantee that the fusion rules of the $G$ topological lines in $\mathcal{T}_0$ with the interface $\mathcal{N}$ are 
\begin{align}\label{eqn:GfusionN}
    g\otimes  \mathcal{N} = \mathcal{N} \text{ for all }g\in G.
\end{align}
Similarly, the fusion rules of the $\mathrm{Rep}(G)$ topological lines in $\mathcal{T}_0\big/ G$ with the interface are
\begin{align}\label{eqn:RepGfusionN}
    \mathcal{N} \otimes \rho = \langle \rho\rangle \mathcal{N} \text{ for all irreducible representations }\rho \in \mathrm{Rep}(G),
\end{align}
where $\langle\rho\rangle=\dim(\rho)$. By the folding trick, the interface $\mathcal{N}$ can be thought of as a conformal boundary condition in the theory $\mathcal{T} = \mathcal{T}_0 \otimes \overline{\mathcal{T}_0\big/G}$ as shown in Figure \ref{fig:repGbc}. By the fusion rules in  \eqref{eqn:GfusionN} and \eqref{eqn:RepGfusionN}, this boundary condition is strongly symmetric with respect to both $G$ and $\mathrm{Rep}(G)$ in the  CFT $\cal T$.

Hence, this construction furnishes an infinite family of examples which realize strongly symmetric boundary conditions for both invertible and non-invertible global symmetries. See Section \ref{subsec:arbitraryModuleCategory} for a generalization of this construction which produces boundary conditions in an arbitrary module category of a fusion category.

\subsection{Weakly Symmetric Boundaries}\label{subsec:examplesWeaklySymmetric}

Next, we turn to weakly symmetric boundary conditions in a variety of models, including minimal models in Section \ref{minimal model}, $SU(2)_k$ WZW models in Section \ref{subsubsec:SU2WZW}, general diagonal RCFTs in Section \ref{subsubsec:diagonalRCFTs}, and the golden anyon chain in Section \ref{sec:goldenchain}.

\subsubsection{Minimal Models and Their Topological Lines}\label{minimal model}

There are many examples of weakly symmetric boundaries in Virasoro minimal models. We restrict to unitary diagonal minimal models for simplicity. We first review some basic facts. The unitary diagonal minimal models $\mathcal{M}(m+1,m)$ are labeled by an integer $m\geq 3$. Their central charge is $c = 1 - \frac{6}{m(m+1)}$ and they contain $\frac{1}{2}m(m-1)$ primary operators. We denote the primary operators as $\phi_{(r,s)}$, where $(r,s)$ is a pair of integers obeying $1\leq r<m$, $1\leq s<m+1$, and $sm<r(m+1)$. 
We use the index $i$ to denote such a pair $(r,s)$, and write $\phi_i$ for the corresponding primary operator. 
In particular, $i=0$ corresponds to the identity operator with $(r,s)=(1,1)$.

The simple topological lines (which are also the Verlinde lines\footnote{In an RCFT, the topological lines that commute with the extended chiral algebra are known as the Verlinde lines. Since the chiral algebra of a diagonal Virasoro minimal model is just the Virasoro algebra, all of the topological lines are Verlinde lines.}) are in one-to-one correspondence with  the local  primary operators \cite{Petkova:2000ip}  and we denote them by ${\cal L}_i$.  
They act on the $j$th primary $\ket{\phi_j}$ (or any of its descendants)  as
\begin{equation}\label{eqn:verlindelineaction}
    \hat{\mathcal{L}}_i \ket{\phi_j} = \frac{S_{ij}}{S_{0j}} \ket{\phi_j}
\end{equation}
where $S_{ij}$ is the modular S-matrix. 
Equation \eqref{eqn:verlindelineaction} shows that all of the topological lines can be simultaneously diagonalized, and hence commute with one another. 
The fusion of two  lines is given by
\begin{equation}
    \mathcal{L}_i \otimes \mathcal{L}_j = \bigoplus_{k} N_{ij\;}^k \mathcal{L}_k
\end{equation}
where $N_{ij}^k$ are the fusion coefficients of primary operators.

The simple boundary states in the diagonal minimal models (also known as Cardy states) are also in one-to-one correspondence with the local primary operators  \cite{Cardy:1989ir}. 
We label the $i$th Cardy state  $\ket{\mathcal{L}_i}$ using the corresponding topological line. 
This is natural in light of the fact that they form the regular module category over the fusion category of Verlinde lines, i.e.,
\begin{equation}\label{VonC}
    \hat{\mathcal{L}}_i  \ket{\mathcal{L}_j} = \bigoplus_k N_{ij}^k \ket{\mathcal{L}_k},
\end{equation}
as is the case for any diagonal RCFT.

A structural fact about the diagonal minimal models is that they do not admit simple boundary conditions which are strongly symmetric with respect to any of their non-invertible symmetries. 
Indeed, consider a non-invertible line $\mathcal{L}_i$, which necessarily has quantum dimension $\langle \mathcal{L}_i\rangle>1$. By definition (\ref{eq:strong_NIM}), a strongly symmetric boundary must be an eigenstate of $\mathcal{L}_i$ with eigenvalue $\langle \mathcal{L}_i\rangle$. 
However, the fusion coefficients  $N^k_{ij}$ are all either 0 or 1 in the Virasoro minimal models, so we see from \eqref{VonC} that no simple boundary $|\mathcal{L}_j\rangle$ will ever be strongly symmetric with respect to a non-invertible $\mathcal{L}_i$.

As an example,  consider the diagonal minimal model $\mathcal{M}(5,4)$, i.e.\ the tricritical Ising model. The theory has 6 primary operators, and therefore also 6 Verlinde lines and 6 Cardy states. The topological lines are 
\begin{equation}
    \mathds{1}, \quad \quad \eta, \quad \quad {\cal N}, \quad \quad W, \quad \quad \eta W, \quad \quad {\cal N}W
\end{equation}
and their fusion rules are
\begin{equation}
    \eta \otimes \eta = \mathds{1},     \quad \;\;
    \eta \otimes {\cal N} = {\cal N}\otimes \eta= {\cal N},
    \quad \;\; {\cal N} \otimes {\cal N} = \mathds{1} \oplus \eta, \quad \;\; W \otimes W = \mathds{1} \oplus W \,.
\end{equation}
The quantum dimensions of the lines $\eta$, $\cal N$, and $W$ are $1$, $\sqrt{2}$, and $\frac{1+\sqrt{5}}{2}$, respectively. The pair $\{\mathds{1},W \}$ generates a subcategory equivalent to the Fibonacci category $\mathrm{Fib}$, the unique non-invertible unitary fusion category with two simple lines. Using  \eqref{VonC}, we see that  the Cardy states $\ket{W}$, $\ket{\eta W}$, and $\ket{{\cal N}W}$ are weakly symmetric with respect to this Fibonacci symmetry,
\begin{equation} \label{eq:wbc_tri}
    \hat{W} \ket{W} = \ket{W} \oplus \ket{\mathds{1}}, \quad \;\;\; \hat{W} \ket{\eta W} = \ket{\eta W} \oplus\ket{\eta}, \quad \;\;\; \hat{W} \ket{{\cal N} W} = \ket{{\cal N} W} \oplus \ket{\cal N} ,
\end{equation}
but not strongly symmetric. Similarly, one can check that none of the boundaries are strongly nor even weakly symmetric with respect to the $\mathbb{Z}_2$ Tambara-Yamagami category generated by $\{\mathds{1},\eta,{\cal N}\}$.
On the other hand, invertible symmetries evade the arguments of the previous paragraph. Indeed, the boundaries $\ket{\cal N}$ and $\ket{{\cal N}W}$ are both strongly and weakly symmetric with respect to the invertible $\mathbb{Z}_2$ generated by $\eta$.

As another example, consider the  tetracritical Ising model, $\mathcal{M}(6,5)$. It contains 10 primary operators, and hence 10 Verlinde lines and 10 Cardy states. The Verlinde lines are denoted as
\begin{equation}
    \mathds{1}, \quad \;\; \eta, \quad \;\; M, \quad \;\; N, \quad \;\; \eta N, \quad \;\; W, \quad \;\; \eta W, \quad \;\; MW, \quad \;\; NW, \quad \;\; \eta NW
\end{equation}
and their fusion rules are given by
\begin{align}
\begin{split}
    \eta \otimes \eta &= \mathds{1}, \quad \quad \quad \;\;\; M \otimes M = \mathds{1} \oplus \eta \oplus  M, \quad \quad\;\;\; \quad\quad\quad\;
    \eta \otimes M = M\otimes \eta=M,
     \\
    N \otimes N &= \mathds{1} \oplus  M,
 \;\; \quad M \otimes N = N\otimes M= N \oplus \eta N, \;\;\quad\quad \quad
 W \otimes W = \mathds{1} \oplus  W  \,.
\end{split}
\end{align}
The quantum dimensions of the lines $\eta$, $M$, $N$, and $W$ are $1$, $2$, $\sqrt{3}$, and $\frac{1+\sqrt{5}}{2}$, respectively.
The lines $\{\mathds{1} , \eta , M\}$ generate a non-anomalous symmetry which is equivalent to $\mathrm{Rep}(S_3)$. The boundaries $\ket{M}$ and $\ket{MW}$ are weakly (but not strongly) symmetric  under this $\mathrm{Rep}(S_3)$ fusion category, as can be seen below,
\begin{align}\label{weakreps3}
\begin{split}
   & \;\hat{\eta} \ket{M} = \ket{M}, \quad\quad\quad\quad\quad\quad\;\quad  \quad  \hat{\eta} \ket{MW} = \ket{M W}, \\ 
   & \hat{M} \ket{M} = \ket{\mathds{1}} \oplus \ket{\eta} \oplus \ket{M}, \quad \quad  \quad \  \hat{M} \ket{MW} = \ket{W} \oplus\ket{\eta W} \oplus \ket{M W}. 
\end{split}
\end{align}
Similarly, the boundaries $\ket{W}$, $\ket{\eta W}$, $\ket{MW}$, $\ket{NW}$, and $\ket{\eta N W}$ are weakly but not strongly symmetric under the unitary Fibonacci fusion category generated by $\{\mathds{1},W\}$.

\subsubsection{$SU(2)_k$ WZW models and Their Verlinde Lines}\label{subsubsec:SU2WZW}

Next, we turn to WZW models. For simplicity, we restrict attention to diagonal RCFTs modeled on the $SU(2)_k$ chiral algebras, where $k$ is quantized to be a positive integer. Such theories possess a fusion category symmetry $(A_1,k)$ which commutes with the full $SU(2)$ current algebra. This category possesses $k+1$ simple lines $\mathcal{L}_j$ which we label by an $SU(2)$ spin $j=0,\frac12,\dots,\frac{k}{2}$. Their fusion rules are 
\begin{align}
    \mathcal{L}_{j_1}\otimes \mathcal{L}_{j_2} = \mathcal{L}_{|j_1-j_2|}\oplus \mathcal{L}_{|j_1-j_2|+1}\oplus\cdots\oplus \mathcal{L}_{\mathrm{min}(j_1+j_2,k-j_1-j_2)}\equiv \bigoplus_{j_3}N_{j_1j_2}^{j_3}\mathcal{L}_{j_3}.
\end{align}
The Cardy boundary conditions $|j\rangle$, which by definition are required to preserve (the half of) the full chiral algebra, are also labeled by a spin $j$ and transform in the regular module of $(A_1,k)$, i.e.,
\begin{align}
    \hat{\mathcal{L}}_{j_1}|j_2\rangle \equiv  \bigoplus_{j_3} \widetilde{N}_{j_1j_2}^{j_3}|j_3\rangle
\end{align}
where $\widetilde{N}_{j_1j_2}^{j_3}=N_{j_1j_2}^{j_3}$. 
In particular, the fusion coefficients  are all equal to $0$ or $1$. Thus, for the same reason as in the case of the diagonal minimal models, none of the Cardy states of the diagonal $SU(2)_k$ WZW models are strongly symmetric with respect to the $(A_1,k)$ Verlinde lines.\footnote{For $k=1$, the Verlinde lines generate an invertible $\mathbb{Z}_2$ symmetry, and there is still no symmetric boundary condition as explained later.}

It turns out that the diagonal $SU(2)_k$ WZW models do not have \emph{weakly} symmetric Cardy states with respect to their Verlinde lines either. Indeed, a line $\mathcal{L}_{j_1}$ with half-integer spin $j_1$ can never topologically terminate on any Cardy state $j_2$ because in such cases $\widetilde{N}_{j_1j_2}^{j_2}=0$.

However, we may instead consider the subcategory $(A_1,k)_{\frac12}$ which by definition is generated by the subset of lines $\mathcal{L}_j$ with \emph{integer} spin $j$.  
For small values of $k$, familiar symmetries are recovered, 
\begin{align}\label{eqn:smallk(A1,k)1/2}
    (A_1,k)_{\frac12} = \begin{cases}
        \mathrm{Vec}, & k=1 \\ 
        \mathrm{Vec}_{\mathbb{Z}_2}, & k=2 \\
        \mathrm{Fib}, & k=3 \\
        \mathrm{Rep}(S_3), & k=4 \\ 
        ~~~~\vdots 
    \end{cases}
\end{align}
Because we have disposed of the problematic lines with half-integer spin, there is a chance that there are Cardy states which are weakly symmetric under $(A_1,k)_{\frac12}$. In fact, we claim that when $k$ is even, the Cardy state $|\frac{k}{4}\rangle$ is weakly symmetric, while when $k$ is odd, the two Cardy states $|\frac{k}{4}\pm \frac14\rangle$ are weakly symmetric. For example, when $k$ is even, showing that all the $\mathcal{L}_j$ in $(A_1,k)_{\frac12}$ can topologically terminate on $|\frac{k}{4}\rangle$ amounts to showing that $\frac{k}{4}\geq |j-\frac{k}{4}|$ and $\frac{k}{4}\leq \mathrm{min}(j+\frac{k}{4},\frac{3k}{4}-j)$ for all $j=0,1,\dots,\frac{k}{2}$. We leave this as a straightforward exercise for the reader.

We comment that, on kinematic grounds, weakly symmetric boundaries for $(A_1,k)_{\frac12}$ are the best one could hope for, at least when $k\neq 1,2,4$. Indeed, consider the quantum dimensions of the lines $\mathcal{L}_j$, 
\begin{align}
    \dim(\mathcal{L}_j) = \frac{\sin\big((2j+1)\pi/(k+2)\big)}{\sin\big(\pi/(k+2)\big)}.
\end{align}
The dimension of the $j=1$ line simplifies to the expression
\begin{align}
    \dim(\mathcal{L}_1) = 2\cos(2\pi/(k+2))+1
\end{align}
from which it is straightforward to see that $2<\dim(\mathcal{L}_1)<3$ when $k>4$. In particular, when $k>4$, the dimension is non-integral, and hence the category is kinematically incompatible with the existence of a strongly symmetric boundary.  The same argument can be invoked for $(A_1,3)_{\frac12}=\mathrm{Fib}$. On the other hand, from \eqref{eqn:smallk(A1,k)1/2} we see that the $(A_1,k)_{\frac12}$ categories corresponding to $k=1,2,4$ have all been considered in previous sections, where it was shown that they \emph{do} admit strongly symmetric boundaries.

Finally, we note that the full $(A_1,k)$ fusion category is never kinematically compatible with a strongly symmetric boundary. Because $(A_1,k)_{\frac12}$ is a subcategory of $(A_1,k)$, the previous paragraph demonstrates this fact when $k\neq 1,2,4$. When $k=2$ or $4$, it can be checked by hand that there are lines with non-integer quantum dimensions; when $k=1$, we have $(A_1,1)\cong \mathrm{Vec}_{\mathbb{Z}_2}^\omega$ with $\omega\neq 0$, and hence the category is obstructed from admitting a symmetric boundary by its ordinary 't Hooft anomaly.

\subsubsection{Diagonal RCFTs and Their Verlinde Lines}\label{subsubsec:diagonalRCFTs}

We now consider general diagonal RCFTs. Let $\cal C$ be the MTC associated with such a theory.  
Then the Verlinde lines (i.e., those topological lines that commute with the extended chiral algebra) form a fusion category given by forgetting the braiding and twist structures of $\cal C$. 
Similarly, the Cardy states (i.e., those conformal boundary conditions that respect the extended chiral algebra) form the regular module category of $\cal C$. 

If there is a simple object $X$ in  $\cal C$ that obeys
\begin{align}\label{eqn:weaklysymmetricboundary}
    \mathrm{dim}_{\mathbb{C}}\,\mathrm{Hom}_{\mathcal{C}}(Y\otimes X,X)\geq 1 \text{ for all }Y \text{ in }\mathcal{C},
\end{align}
then the corresponding Cardy state is a weakly symmetric boundary with respect to the Verlinde lines $\cal C$.

The simplest example of this is obtained by taking $\mathcal{C}$ to be either of the two Fibonacci modular categories, both of which have fusion rules
\ie
W \otimes W=\mathds{1}\oplus W.
\fe
They arise, for instance, as the MTCs associated to the affine Kac-Moody algebras $(G_{2})_1$ and $(F_4)_1$, respectively.
They reduce to the same unitary Fibonacci fusion category if we forget the braiding and twist structures. 
In the WZW model obtained as the diagonal RCFT corresponding to, say,  $(G_2)_1$, the Cardy state $|W\rangle$ labeled by the non-vacuum primary is then a weakly symmetric boundary condition with respect to the symmetry lines in $\mathcal{C}$. Indeed, the non-trivial Fibonacci line acts on the two Cardy states $|\mathds{1}\rangle$ and $|W\rangle$ as 
\begin{align}
    \hat{W}|\mathds{1}\rangle = |W\rangle, \ \ \ \ \ \hat{W}|W\rangle = |\mathds{1}\rangle\oplus|W\rangle
\end{align}
which reveals that the topological line $W$ can end topologically on the boundary condition $|W\rangle$. 
 
\begin{table}[]
    \centering
    \begin{tabular}{c|c}
    $\mathcal{C}$ & $\mathcal{V}$ \\\toprule 
        $\text{Fib}$ & $(G_2)_1$ \\
    $\overline{\text{Fib}}$ & $(F_4)_1$  \\ 
    $(A_1,5)_{\frac12}$ & $\mathrm{Ex}[(E_6)_1L_{6/7}]$ \\ 
    $\overline{(A_1,5)}_{\frac12}$ & $\mathrm{Ex}[(A_1)_5(E_7)_1]$\\ 
     $(A_1,7)_{\frac12}$ & $\mathrm{Ex}[(A_1)_1 (A_1)_7]$ \\ 
     $\overline{(A_1,7)}_{\frac12}$ & $(G_2)_2$ \\
    $\text{Fib}\boxtimes \text{Fib}$ & $(G_2)_1^2$\\ 
    $\overline{\text{Fib}}\boxtimes \overline{\text{Fib}}$ & $\mathrm{Ex}[(A_1)_8]$ \\ 
    $\text{Fib}\boxtimes \overline{\text{Fib}}$ & $(G_2)_1(F_4)_1$
    \end{tabular}
    \caption{Examples of RCFTs with low central charge and few primary operators which admit weakly symmetric boundary conditions. More specifically, the diagonal RCFT built out of the chiral algebra $\mathcal{V}$ admits a weakly symmetric boundary condition with respect to its category of Verlinde lines, which is the MTC $\mathcal{C}$ thought of as a fusion category by forgetting its braiding and twist.}
    \label{tab:weaklysymmetrictheories}
\end{table}

 More generally, of the MTCs $\mathcal{C}$ with $\mathrm{rank}(\mathcal{C})\leq 4$, which were classified in \cite{rowell2009classification}, it is straightforward to compute that it is precisely the  ones which appear in Table \ref{tab:weaklysymmetrictheories} that  satisfy  \eqref{eqn:weaklysymmetricboundary}. Here, $\overline{\mathcal{C}}$ is the complex conjugate of the MTC $\mathcal{C}$, obtained by reversing the braiding and conjugating the twist. We comment that complex conjugate MTCs have the same underlying fusion category.

 In \cite{Rayhaun:2023pgc} (see also \cite{Mathur:1988na,Mason:2021xfs,Tener:2016lcn,Mukhi:2022bte} for prior related work) it was shown that, for each MTC $\mathcal{C}$ with $\mathrm{rank}(\mathcal{C})\leq 4$, the lowest value of $c$ which supports a chiral algebra with MTC given by $\mathcal{C}$ actually supports a unique one. For each MTC in Table \ref{tab:weaklysymmetrictheories}, we report this corresponding chiral algebra $\mathcal{V}$, which typically involves products of WZW models (with the exception of the chiral algebra corresponding to $(A_1,5)_{\frac12}$, which involves the Virasoro algebra $L_{6/7}$ with $c=6/7$).  The notation $\mathrm{Ex}[\mathcal{W}]$  denotes a VOA extension of $\mathcal{W}$;\footnote{The precise extensions are recorded in Appendix D.1 of \cite{Rayhaun:2023pgc}.} equivalently, one could just as well apply a non-diagonal modular invariant to $\mathcal{W}$ instead of using the diagonal modular invariant for $\mathrm{Ex}[\mathcal{W}]$. For example, the theory obtained using the diagonal modular invariant on $\mathrm{Ex}[(A_1)_8]$ is the same as the theory obtained using the D-type modular invariant on $(A_1)_8$.
 
 It follows that, for each of these chiral algebras, the corresponding RCFT admits a boundary condition which is weakly symmetric with respect to its Verlinde lines. Appendix E of \cite{Rayhaun:2023pgc} provides many more examples of chiral algebras with these MTCs, to which the same conclusions apply.

\subsubsection{Golden Chain and Fibonacci Category}\label{sec:goldenchain}

\begin{figure}
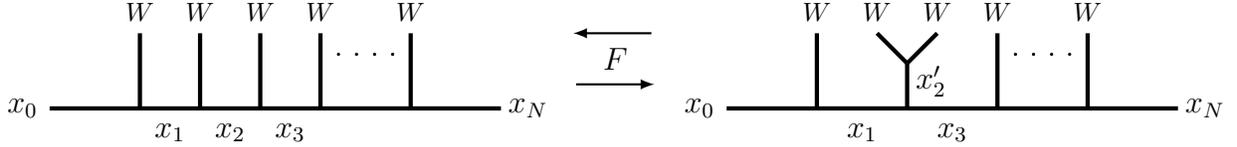

    \centering
    \ctikzfig{figures/goldenchain1}
    \caption{Left: The state $\ket{x_1,x_2,...x_{N-1}}$ in the open golden chain with boundary conditions $x_0,x_N$ on the two ends. Right: The state $\ket{x_1,x'_2,...x_{N-1}}_2$ written in a basis where the 2\textsuperscript{nd} and 3\textsuperscript{rd} $W$ anyons are fused in the $x'_2$ channel. One can go back and forth between these two bases using the $F$-symbols.}
    \label{fig:goldenchain1}
\end{figure}
Examples of weakly symmetric boundaries for non-invertible symmetries are natural in anyonic chains \cite{Feiguin:2006ydp,2013PhRvB..87w5120G,Buican:2017rxc}. As an illustration, we first examine the golden chain \cite{Feiguin:2006ydp}, which is a lattice model with the Fibonacci fusion category symmetry $\mathrm{Fib}=\langle \mathds{1}, W\rangle $. 
An open golden chain can be described by a fusion tree where each link is labeled using the simple lines $\{\mathds{1},W\}$, as shown on the left of Figure \ref{fig:goldenchain1}.
There is an array of $N$ Fibonacci anyons $W$ shown as vertical edges. 
The horizontal edges $x_0,x_1,\cdots, x_N$ take values in $\mathds{1},W$ and are constrained by the fusion rule of the Fibonacci fusion category
\ie
W\otimes W=\mathds{1}\oplus W\,,~~~~\mathds{1}\otimes W=W\otimes \mathds{1}=W.
\fe
The horizontal edges $x_0$ and $x_N$ at the ends define a choice of the boundary conditions. 
The Hilbert space with a particular choice of $x_0,x_N$ consists of states  $\ket{x_1,x_2,...,x_{N-1}}$ compatible with the fusion rule, and it is not a tensor product of local Hilbert spaces.

To define the Hamiltonian, we introduce another basis of states. 
One  fuses the $l^{\mathrm{th}}$ and $(l+1)^{\mathrm{th}}$ $W$ anyons (vertical lines) using the $F$-symbols\footnote{Since the trivalent junction vector spaces consisting of the topological lines $\mathds{1},W$ are at most 1 dimensional, we suppress the indices for the junction vectors in the $F$-symbols, once we pick a basis for them. Specifically, we write $\left[ F^{W}_{x_{l-1}x_{l+1}W} \right]_{(x_l,\delta,\lambda) (x'_l,\rho,\sigma)} \equiv \left[ F^{W}_{x_{l-1}x_{l+1}W} \right]_{x_l x'_l}$ suppressing redundant the labels $\delta$,$\lambda$,$\rho$ and $\sigma$ for the junction vectors.} (see Appendix \ref{app:categories}) as illustrated in Figure \ref{fig:goldenchain1} for $l=2$. This defines a change of basis of the Hilbert space from $\ket{...,x_{l-1},x_{l},x_{l+1},...}$ to $\ket{...,x_{l-1},x'_{l},x_{l+1},...}_l$ as
\begin{equation}
\ket{...,x_{l-1},x_{l},x_{l+1},...} = \sum_{x'_l = \mathds{1},W} \left[ F^{W}_{x_{l-1}x_{l+1}W} \right]_{x_l x'_l} \ket{...,x_{l-1},x'_{l},x_{l+1},...}_l,
\end{equation}
where the subscript $l$ on $\ket{....}_l$ denotes the basis in which the $l^{\mathrm{th}}$ and $(l+1)^{\mathrm{th}}$ anyons are fused. One can go back and forth between these two bases using the $F$-symbols.

There is a pairwise interaction between every two consecutive $W$ anyons governed by a Hamiltonian $H = \sum_{l}  H_l$ where $H_l$ denotes the interaction between the $l^{\mathrm{th}}$ and $(l+1)^{\mathrm{th}}$ $W$ anyons. $H_l$ is easily described in the latter basis $\ket{...,x_{l-1},x'_{l},x_{l+1},...}_l$  as
\begin{equation}\label{Hlaction}
    H_l \ket{...,x_{l-1},x'_{l},x_{l+1},...}_l = -\delta_{x'_l,\mathds{1}}\ket{...,x_{l-1},x'_{l},x_{l+1},...}_l.
\end{equation}
In other words, it assigns an interaction energy $E = -1$ between the $l^{\mathrm{th}}$ and $(l+1)^{\mathrm{th}}$ $W$ anyons when their fusion is in the trivial channel, and $E=0$ when the fusion is in the $W$ channel. Therefore,  $H_l$ is a projector in the $\ket{...,x_{l-1},x'_{l},x_{l+1},...}_l$ basis 
\begin{equation}
    H_l = -\Pi^{\mathds{1}}_{l,l+1},
\end{equation} where $\Pi^{\mathds{1}}_{l,l+1}$ is a projection operator that projects onto the subspace of states in which the fusion of $l^{\mathrm{th}}$ and $(l+1)^{\mathrm{th}}$ $W$ anyons is in the trivial channel $x'_l=\mathds{1}$. 

\begin{figure}
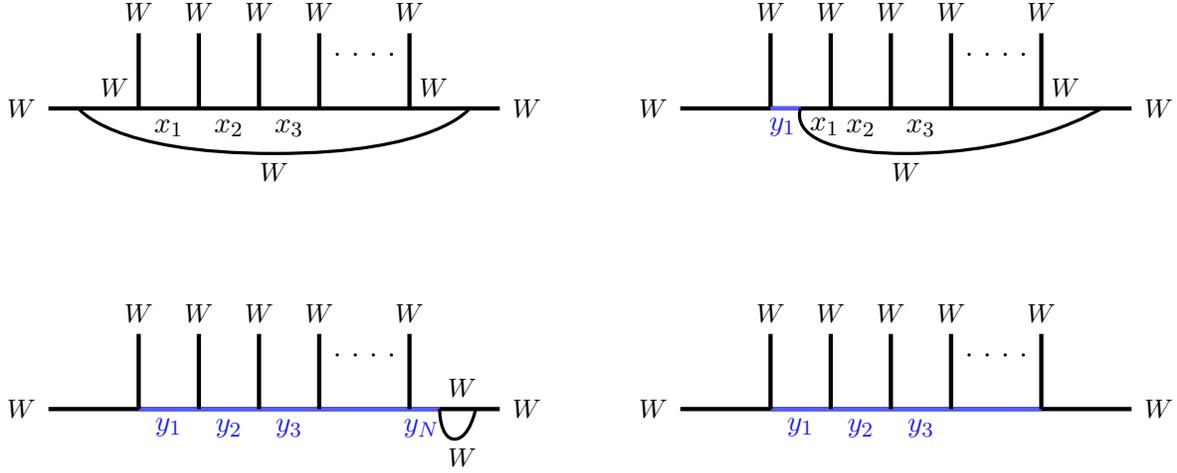

    \centering
   \ctikzfig{figures/goldenchain2}
    \caption{The action of the ``topological symmetry"  operator $\hat{W}$ on the state $\ket{x_1,x_2,...,x_N}$ shown by attaching an additional $W$ line to the fusion tree. On the top left, we use the $F$-symbol to sweep the left end of the $W$ line to obtain the figure on the top right. We continue sweeping it through other edges using the $F$ symbols until we get a loop of $W$ as shown in the bottom left. Shrinking the loop then gives the final state after the $\hat W$ action (bottom right). We have suppressed the $F$-symbols in the figure.}
    \label{fig:goldenchain2}
\end{figure}

Depending on the boundary conditions $x_0$ and $x_N$, the open golden chain may or may not have a conserved operator that commutes with the Hamiltonian $H$. Specifically, we show that there is a non-trivial conserved operator $\hat{W}$ which commutes with the Hamiltonian if (and only if) both of the boundary conditions $x_0$ and $x_N$ are set to $W$.
It is a non-local operator and is commonly called the ``topological symmetry" in the literature. 
Its action on the Hilbert space is most easily described pictorically as in Figure \ref{fig:goldenchain2}. 
We attach an additional $W$ line to the fusion tree that ends at the boundaries $x_0=W$ and $x_N=W$ as shown in the top left of Figure \ref{fig:goldenchain2}. 
Using a sequence of moves shown in the figure, the action of $\hat{W}$ on a state $\ket{x_1,x_2,...,x_{N-1}}$ can be explicitly written using the $F$-symbols  as
\begin{equation}
\hat{W} \ket{x_1,x_2,...,x_{N-1}} =  \langle W \rangle \left[F_{WWW}^{W}\right]_{W \mathds{1}}\sum_{y_1,y_2,..,y_{N-1} = \mathds{1},W} \prod_{l=0}^{N-1}  \left[F_{Wx_{l+1}W}^{y_{l}}\right]_{x_{l}y_{l+1}} \ket{y_1,y_2,...,y_{N-1}}
\end{equation}
where $y_0 \equiv W$ and $y_N\equiv W$, and $\langle W \rangle$ is the quantum dimension of the $W$ line. 
Note that the attachment of this $W$ line does not change the boundary edges because $W\otimes W$ contains a $W$ channel. 

\begin{figure}
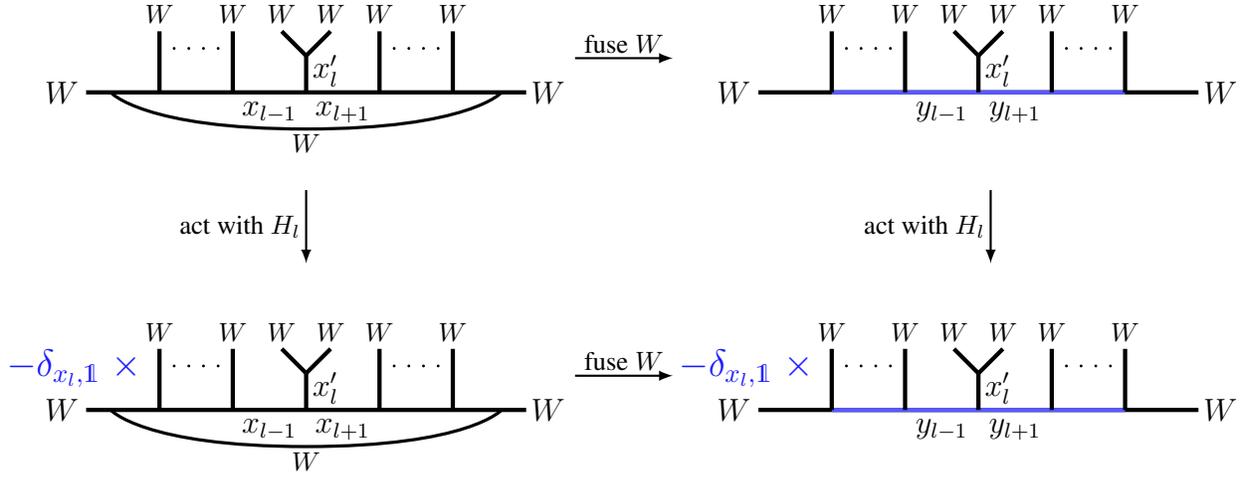

    \centering
 \ctikzfig{figures/goldenchain3}
    \caption{A commutative diagram showing that the topological symmetry operator $\hat{W}$ and the local Hamiltonian   $H_l$ commute for every $l$.}
    \label{fig:goldenchain3}
\end{figure}
To see that $\hat{W}$ commutes with the Hamiltonian, we first define a subspace $\mathcal{H}_{l,x'_l}$ of the Hilbert space $\mathcal{H}$ as the linear span of the basis states $\ket{...,x_{l-1},x'_{l},x_{l+1},...}_l$ for a fixed value of $l$ and $x'_{l}$. The full Hilbert space $\cal H$ can then be decomposed into a direct sum 
\begin{equation}
    \mathcal{H} = \bigoplus_{x'_l=\mathds{1},W} \mathcal{H}_{l,x'_l}
\end{equation}
Now, consider the action of $\hat{W}$ and $H_l$ on the $\mathcal{H}_{l,x'_l}$ for a fixed $l$ and $x'_l$. The action of $\hat{W}$ on the $\ket{...,x_{l-1},x'_{l},x_{l+1},...}_l$ basis corresponds to attaching an additional $W$ line onto the fusion tree, similar to the action on the $\ket{x_1,x_2,...,x_{N-1}}$ basis, as shown in the top left of Figure \ref{fig:goldenchain3}. Such a fusion cannot affect the channel $x'_{l}$ into which the $l^{\mathrm{th}}$ and $(l+1)^{\mathrm{th}}$ anyons fuse; it only changes the values of $\{x_1,..x_{l-1},x_{l+1},...x_{N-1}\}$. This implies that $\mathcal{H}_{l,x'_l}$ is an invariant subspace under the action of $\hat{W}$. On the other hand, $H_l$ acts like a c-number on $\mathcal{H}_{l,x'_l}$ as shown in \eqref{Hlaction}. Therefore, when restricted to the subspace $\mathcal{H}_{l,x'_l}$, $H_l$ commutes with $\hat{W}$. See Figure \ref{fig:goldenchain3}. However, since $x'_l$ and $l$ are arbitrary, we get $[\hat{W},H_l] = 0$ for all $l\in\{1,2,..,N-1\}$ in the full Hilbert space $\mathcal{H}$.  In other words, 
\begin{equation}
    [\hat{W},H] = 0.
\end{equation}

In the continuum limit, the golden chain is described by the tricritical Ising CFT with central charge $c=7/10$. 
The ``topological symmetry" $\hat{W}$ flows to a topological line $W$ in space-time that generates a unitary Fibonacci fusion category. 
 The boundary corresponding to $x_0=W$ (and similarly for $x_N=W$)  on the golden chain, which we denote as ${\cal B}_W$, flows to a conformal boundary in the tricritical Ising CFT. 
Since $\hat W$ is a conserved operator on the open golden chain, this means that in the continuum, the topological line $W$ admits a topological junction on the boundary. In other words, the boundary ${\cal B}_W$, in the continuum limit of the golden chain, is weakly symmetric under the topological line $W$. 
If it is simple, it must therefore be one of the three weakly symmetric boundaries appearing in \eqref{eq:wbc_tri}.

This idea can be generalized to any anyonic chain based on a fusion category $\mathcal{C}$. 
For simplicity, we assume the fusion coefficients $N^k_{ij}$ are no greater than 1, and we refer the readers to \cite{Inamura:2023qzl} for the more general cases.
We first pick a reference anyon  ${\cal L}_0$ in $\cal C$, and 
consider a fusion tree with $N$ vertical ${\cal L}_0$ anyons. 
For the open anyonic chain, we pick two boundary horizontal edges   labeled by some anyons $x_0$ and $x_N$. 
The Hilbert space $\mathcal{H}$ consists of all states  $\ket{x_1,x_2,...,x_{N-1}}$  with the edge variables $x_i$ 
 compatible with the fusion rule of $\cal C$. 
 $F$-symbols are used to fuse two nearby anyons to go to another   basis $\ket{...,x_{l-1},x'_{l},x_{l+1},...}_l$ where the $l^{\mathrm{th}}$ and $(l+1)^{\mathrm{th}}$ anyons are fused in the $x'_l$ channel. 
 To define a Hamiltonian, we choose a second reference anyon ${\cal L}_H$ (which may or may not be the same as ${\cal L}_0$), and define   $H = -\sum_{l} \Pi^{\mathcal{L}_H}_{l,l+1}$  in the $\ket{...,x_{l-1},x'_{l},x_{l+1},...}_l$ basis.

For simplicity, we choose the boundary conditions  on the two ends to be the same, given by a choice of an anyon ${\cal L}_B$, i.e.,  $x_0=x_N=\mathcal{L}_B$. 
If there exists a line $\cal L$ which obeys  ${\cal L}\otimes {\cal L}_B={\cal L}_B\oplus\cdots$,  then we can define a topological symmetry operator $\hat{\mathcal{L}}$ on the Hilbert space $\mathcal{H}$ by attaching an $\cal L$ line to the open chain. 
By a similar argument in the golden chain case, it commutes with the Hamiltonian. 
In the continuum, the boundary conditions corresponding to $x_0=x_N=\mathcal{L}_B$ will be weakly symmetric under the topological line $\cal L$. 

\subsection{Boundary Conditions in Arbitrary Module Categories}\label{subsec:arbitraryModuleCategory}

Suppose that we are given a 1+1d CFT $\mathcal{T}$ with fusion category symmetry $\mathcal{C}$. 
Let $\cal B$ be an object in an arbitrary module category $\mathcal{M}$ of $\mathcal{C}$. The theory $\mathcal{T}$ may not admit a physical boundary condition which transforms like $\mathcal{B}$ under the action of $\mathcal{C}$. However, we show in this subsection that there exists another CFT $\mathcal{T}_{\cal B}$, which also has symmetry category $\mathcal{C}$, and which moreover \emph{does} have a boundary condition transforming like $\cal B$ in $\mathcal{M}$.

Before we provide the derivation of this result, we pause to explain how one might obtain a 1+1d CFT $\mathcal{T}$ with arbitrary fusion category symmetry $\mathcal{C}$ in the first place. In the case that $\mathcal{C}=\mathrm{Vec}_G$ for some finite group $G$, we described in Section \ref{subsubsec:repG} how to obtain such a theory from products of holomorphic VOAs (subject to a particular conjecture in the case that $G$ is not solvable, see Footnote \ref{footnote:regularityconjecture}). What about a more general fusion category? One construction is the following. We assume the widely-believed conjecture that any modular tensor category is realized as the representation category of some VOA. In the present situation, we apply this conjecture to the Drinfeld center $\mathcal{Z}(\mathcal{C})$ to obtain a VOA $\mathcal{V}$. Then, it is known that there exists a Lagrangian algebra in $\mathcal{Z}(\mathcal{C})$ whose fusion category of modules is precisely $\mathcal{C}$. 
By the results of \cite{Huang:2014ixa}, such a Lagrangian algebra defines a holomorphic VOA $\mathcal{T}$ containing $\mathcal{V}$ as a subalgebra. Furthermore, the fusion category of topological lines of $\mathcal{T}$ which commute with the operators in $\mathcal{V}$ is precisely $\mathcal{C}$ (see \cite{Burbano:2021loy,Rayhaun:2023pgc} for further details on this idea).

Now, assume that $\mathcal{T}$ is some CFT with an action of $\mathcal{C}$ by symmetries, constructed either as in the previous paragraph or by some other method. 
In Appendix \ref{app:categories} we explain that, given an object $\cal B$ in a module category $\mathcal{M}$, one can obtain an algebra $\mathcal{A}=\underline{\mathrm{Hom}}({\cal B},{\cal B})$ in $\mathcal{C}$ via the internal-Hom construction such that $\mathcal{M}\cong\mathcal{C}_{\mathcal{A}}$. In other words, $\mathcal{M}$ is equivalent to the category of $\mathcal{A}$-modules in $\mathcal{C}$, and under this equivalence, $\mathcal{A}$ in $\mathcal{C}_{\mathcal{A}}$ is sent to $\cal B$ in $\mathcal{M}$. Let us consider gauging this algebra object $\mathcal{A}$ by inserting a mesh of it in (say, the right) half of spacetime, using the rules of \cite{Bhardwaj:2017xup}. 
With respect to the $\mathcal{C}$ lines in the theory $\mathcal{T}$ to the left, the interface behaves like the object $\cal B$ in $\mathcal{M}$ (or equivalently, $\mathcal{A}$ in $\mathcal{C}_{\mathcal{A}}$). Upon folding, we then obtain a boundary condition in the theory $\mathcal{T}_{\cal B} = \mathcal{T} \otimes \overline{\mathcal{T}\big/\mathcal{A}}$. The theory $\mathcal{T}_{\mathcal{B}}$  also admits an action of $\mathcal{C}$ by symmetries, simply because $\mathcal{T}$ does, and furthermore the boundary condition just described behaves with respect to the lines in $\mathcal{C}$ like the object $\mathcal{B}$ in the module category $\mathcal{M}$. This completes the construction.

One interesting consequence of this line of thinking is the following. Assume the conjecture that any fusion category is realized as the symmetries of some CFT. Then, for any fusion category $\mathcal{C}$, and for any object $\cal B$ in one of its module categories $\mathcal{M}$, there exists a theory $\mathcal{T}_{\cal B}$ with symmetry $\mathcal{C}$ and with a boundary condition which transforms like $\mathcal{B}$. In particular, if $\mathcal{C}$ is a category which kinematically admits a strongly symmetric boundary condition, then there is a theory which dynamically does so as well. Similarly, if $\mathcal{C}$ is a category which kinematically admits a weakly symmetric boundary condition, then there is a theory which dynamically does so as well. This may be viewed as a much weaker version of the conjecture discussed in Section \ref{subsec:symBoundaryConj}.

\section{Applications}\label{sec:app}

In this section, we describe applications of   strongly  and weakly symmetric boundaries to  bulk and boundary RG flows, respectively.

\subsection{Weakly Symmetric Boundaries and Constraints on Boundary RG Flows}

\begin{figure}
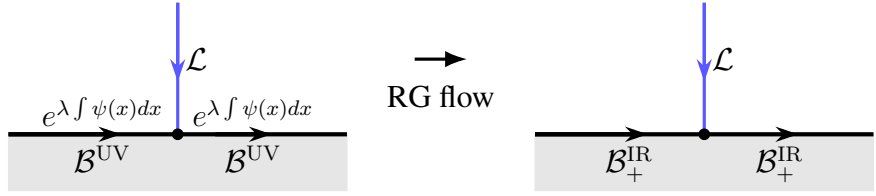

    \centering
    \ctikzfig{figures/boundaryrgcommuting}
    \caption{A relevant boundary deformation $e^{\lambda \int \psi(x) dx}$ (or $e^{- \lambda \int \psi(x) dx}$) is applied on the conformal boundary $\mathcal{B}^{\mathrm{UV}}$ in the presence of a topological line $\mathcal{L}$ which admits a topological junction on $\mathcal{B}^{\mathrm{UV}}$. Here we assume $\psi$ commutes with $\mathcal{L}$. In the IR, $\mathcal{L}$ admits a topological junction on the corresponding boundary $\mathcal{B}^{\mathrm{IR}}_{+}$ (or $\mathcal{B}^{\mathrm{IR}}_{-}$).}
    \label{fig:boundaryrgcommuting}
\end{figure}

We start by reviewing \cite{Konechny:2019wff}, where it was shown that weakly symmetric boundaries can be used to constrain boundary RG flows.
Consider a conformal boundary $\mathcal{B}^{\mathrm{UV}}$ which admits a relevant boundary operator $\psi$. One can trigger a boundary RG flow using $\psi$ by adding $\exp(\pm \lambda \int \psi(x) dx)$ to the action with $\lambda>0$, where the integral is taken along the boundary. 
In the IR, one obtains a conformal boundary $\mathcal{B}^{\mathrm{IR}}_{+}$ or $\mathcal{B}^{\mathrm{IR}}_{-}$ depending on the sign in front of $\lambda$. In general, the conformal boundaries arising from these boundary RG flows might be non-simple. The $g$-theorem \cite{PhysRevLett.67.161 ,Friedan_2004} constrains the $g$ function of the IR conformal boundary to be strictly less than that of the UV conformal boundary, i.e., $g(\mathcal{B}^{\mathrm{UV}}) > g(\mathcal{B}^{\mathrm{IR}}_{\pm}) $.

If $\mathcal{B}^{\mathrm{UV}}$ is weakly symmetric under a topological line $\mathcal{L}$, and if the relevant deformation $\psi$ commutes with $\mathcal{L}$, then the resulting conformal boundary $\mathcal{B}^{\mathrm{IR}}_{\pm}$ will also be weakly symmetric, as shown in Figure \ref{fig:boundaryrgcommuting}. In other words, if $\mathcal{L}$ admits a topological junction on $\mathcal{B}^{\mathrm{UV}}$, then the boundary RG flow triggered by such a relevant boundary operator  results in a boundary $\mathcal{B}^{\mathrm{IR}}_{\pm}$ on which $\mathcal{L}$ also admits a topological junction.

\begin{figure}
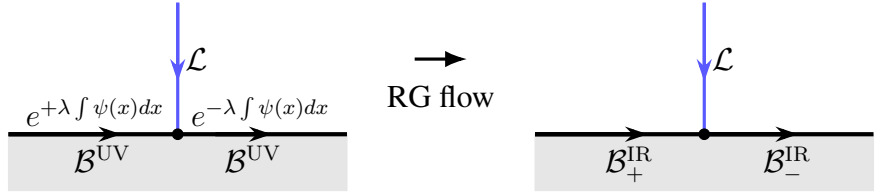

    \centering
    \ctikzfig{figures/boundaryrganticommuting}
    \caption{A relevant boundary deformation $e^{+ \lambda \int \psi(x) dx}$ is applied on the conformal boundary $\mathcal{B}^{\mathrm{UV}}$ to the left of $\mathcal{L}$ and $e^{- \lambda \int \psi(x) dx}$ is applied on $\mathcal{B}^{\mathrm{UV}}$ to the right of $\mathcal{L}$, where $\mathcal{L}$ is a topological line which admits a topological junction on $\mathcal{B}^{\mathrm{UV}}$. Here we assume $\psi$ anticommutes with $\mathcal{L}$. In the IR, this becomes a   topological junction between $\mathcal{L}$, $\mathcal{B}^{\mathrm{IR}}_{+}$ and $\mathcal{B}^{\mathrm{IR}}_{-}$.}
    \label{fig:boundaryrganticommuting}
\end{figure}

On the other hand, if $\psi$ anti-commutes with $\mathcal{L}$, then one can trigger a boundary RG flow in the presence of $\mathcal{L}$ with one sign of the coupling to the left of $\mathcal{L}$, and with the opposite sign of the coupling to the right of $\mathcal{L}$. 
In the IR, $\mathcal{L}$ intersects at a topological junction between $\mathcal{B}^{\mathrm{IR}}_{+}$ and $\mathcal{B}^{\mathrm{IR}}_{-}$, as shown in Figure \ref{fig:boundaryrganticommuting}. 
This, along with the $g$-theorem, can be used to narrow down the possibilities for the IR conformal boundaries. Moreover, if there is more than one line that commutes or anti-commutes with $\psi$ on the boundary $\mathcal{B}^{\mathrm{UV}}$, one can also match the algebra \eqref{projective algebra} for those lines in the UV and in the IR to obtain additional constraints.

As an example, consider the the boundary corresponding to the primary operator labeled by $(r,s) = (3,3)$ in the tetracritical Ising model (see Section \ref{minimal model} for our notation and conventions on minimal models). To make the action of the topological lines more transparent, we label this boundary using its corresponding Verlinde line, $\ket{MW}$. It is then straightforward to see that $\ket{MW}$ is weakly symmetric with respect to $\eta$, $M$, $W$, $\eta W$ and $MW$. This boundary supports a relevant boundary operator $\psi_{2,1}$ with conformal dimension $h = \frac{2}{5}$ that commutes with $\eta$ and anti-commutes with $M$. Upon triggering a boundary RG flow using $\exp(\pm \lambda \int \psi_{2,1}(x) dx)$ with $\lambda > 0$, the resulting IR boundaries $\mathcal{B}^{\mathrm{IR}}_{+}$ and $\mathcal{B}^{\mathrm{IR}}_{-}$ will be weakly symmetric under $\eta$. Moreover, $M$ will have a topological junction between $\mathcal{B}^{\mathrm{IR}}_{+}$ and $\mathcal{B}^{\mathrm{IR}}_{-}$. These constraints, along with the $g$-theorem, narrow down the possibilities to either 
$(\mathcal{B}^{\mathrm{IR}}_{-},\mathcal{B}^{\mathrm{IR}}_{+}) = (\ket{M} , \ket{\mathds{1}} \oplus \ket{\eta})$ or  $(\mathcal{B}^{\mathrm{IR}}_{-},\mathcal{B}^{\mathrm{IR}}_{+}) = (\ket{M} , \ket{M})$ (up to swapping $\mathcal{B}^{\mathrm{IR}}_{-}$ and $\mathcal{B}^{\mathrm{IR}}_{+}$).  Using TCSA technique, it can be shown that the flow in fact goes over to $(\mathcal{B}^{\mathrm{IR}}_{-},\mathcal{B}^{\mathrm{IR}}_{+}) = (\ket{M} , \ket{\mathds{1}} \oplus \ket{\eta})$ \cite{Konechny:2019wff}.

\subsection{Strongly Symmetric Boundaries and Constraints on Bulk RG Flows}
\label{sec:RG}
 
In Section \ref{sec:anomaly}, we discussed the kinematic relation between symmetric boundary conditions and anomalies. 
We now ask if such a boundary condition is dynamically realized in a given CFT. 
More generally, in a given 1+1d CFT with a generalized global symmetry, we discuss the relations between the following three properties: (1) the existence of a (strongly or weakly) symmetric conformal boundary, (2) the existence of a symmetric relevant deformation that trivially gaps the CFT, and (3) the anomaly of the global symmetry.

Using these relations, the analysis of strongly symmetric boundary conditions leads to constraints on bulk renormalization group flows. Specifically,  we find an example among the minimal models where the CFT cannot be trivially gapped  by any relevant deformation  preserving a fusion category symmetry, even though that symmetry is non-anomalous.

\subsubsection{Is There a Symmetric Boundary for Every Non-Anomalous Symmetry?}\label{subsec:symBoundaryConj}

Given a CFT with an invertible global symmetry, is there always a simple, symmetric conformal boundary condition?\footnote{The adjective ``simple'' appearing in this question is important, especially for a finite symmetry. Otherwise, one can always construct a non-simple, symmetric boundary condition by starting with an arbitrary boundary condition, and adding its images under the symmetry action. For instance, starting from the fixed boundary $\ket{\uparrow}$ in the Ising CFT, we can add to it its $\mathbb{Z}_2$ image $\ket{\downarrow}$ to obtain a non-simple, $\mathbb{Z}_2$-symmetric boundary $\ket{\uparrow}\oplus\ket{\downarrow}$.} 
There are obstructions to boundary conditions from anomalies. 
First, the gravitational anomaly (such as a nonzero chiral central charge $c_L-c_R$ in a 1+1d CFT) presents an obstruction to the existence of any conformal boundary condition (see, for example, \cite{Hellerman:2021fla} for a recent discussion).
Second, the 't Hooft anomaly of an internal global symmetry presents a kinematic obstruction to the existence of a symmetry-preserving boundary condition \cite{Han:2017hdv,Jensen:2017eof,Numasawa:2017crf,Thorngren:2020yht,Li:2022drc}.

A natural question is then the following:
given a CFT with an internal global symmetry that is free of gravitational and 't Hooft anomalies, is a simple, symmetric boundary condition always dynamically realized by the theory? 
For invertible symmetries in 1+1d, this question was studied in \cite{Wang:2013yta,Han:2017hdv,Smith:2020rru,Smith:2020nuf,Tong:2021phe,Li:2022drc,Zeng:2022grc,Wang:2022ucy}, where it was found that the answer is positive for a large class of CFTs including the minimal models.

In this subsection, we extend this discussion to non-invertible symmetries. 
We will discuss an example of a CFT with a non-anomalous, internal fusion category symmetry, where there is no strongly symmetric boundary condition, only a weakly symmetric one.

Consider the $c=4/5$ tetracritical Ising CFT. 
As discussed in \cite{Chang:2018iay}, it has a fusion category symmetry $\mathcal{C}=\mathrm{Rep}(S_3)$, one of the simplest non-invertible fusion categories. 
It  has three simple topological lines, $\mathds{1},\eta,M$, corresponding to the trivial, sign, and standard irreducible representations of $S_3$. Their quantum dimensions are $1,1,2$, respectively, which are the dimensions of the corresponding representations. 
They obey the fusion rule of the representation ring of $S_3$,
\begin{equation}\label{reps3fusion}
    \eta\otimes \eta=\mathds{1}\,,~~~~
    \eta\otimes M=M\otimes \eta = M \,,~~~~
    M\otimes M =\mathds{1}\oplus \eta\oplus M\,.
\end{equation}
We see that $\eta$ is a $\mathbb{Z}_2$ line and $M$ is a non-invertible line.

The origin of the non-invertible symmetry $\mathrm{Rep}(S_3)$ in the tetracritical Ising CFT can be seen as follows. 
The tetracritical Ising CFT is a diagonal minimal model which can be obtained from the non-diagonal 3-state Potts CFT by gauging the $S_3$ symmetry.
The quantum symmetry \cite{Vafa:1989ih,Bhardwaj:2017xup,Tachikawa:2017gyf} which arises from this gauging is the  non-invertible $\mathrm{Rep}(S_3)$ symmetry. 
Conversely,  following the generalized orbifold procedure in \cite{Fuchs:2002cm,Brunner:2014lua,Bhardwaj:2017xup}, one can gauge the $\mathrm{Rep}(S_3)$ symmetry of the tetracritical Ising CFT to obtain the 3-state Potts model.

 The fusion category $\mathrm{Rep}(S_3)$ is anomaly-free and is kinematically compatible with a trivially gapped phase (i.e., there is a fiber functor).\footnote{There are two other anomalous fusion categories obeying the same fusion rule as in \eqref{reps3fusion}, but with different $F$-symbols.} 
Following the discussion in Section \ref{sec:anomaly}, this further implies that a strongly symmetric boundary condition $\mathcal{B}$ is \emph{kinematically} allowed for $\mathrm{Rep}(S_3)$. If this boundary were to exist, it would necessarily be acted on by the topological lines as
\begin{equation}
\hat{\mathds{1}} |\mathcal{B}\rangle=|\mathcal{B}\rangle\,,~~~~\hat{\eta} |\mathcal{B}\rangle=|\mathcal{B}\rangle\,,~~~~\hat{M}|\mathcal{B}\rangle=2|\mathcal{B}\rangle\,.
\end{equation}
But is such a strongly symmetric boundary condition \emph{dynamically} realized in the tetracritical Ising CFT?

As reviewed in Section \ref{minimal model}, 
all conformal boundary conditions in the diagonal minimal models are explicitly classified by Cardy \cite{Cardy:1989ir}. The action of the topological lines on these Cardy states is given by (\ref{VonC}), with the coefficients $N_{ij}^k$ equal to the fusion coefficients of the corresponding minimal model. Since each $N^k_{ij}$ is either $0$ or $1$ in the minimal models, we see that they do not realize any simple, strongly symmetric boundary condition for their non-invertible fusion category symmetries.

In particular,  even though a strongly symmetric boundary condition is \emph{kinematically} possible for  the $\mathrm{Rep}(S_3)$ symmetry, it is not \emph{dynamically} realized in the tetracritical Ising CFT. 
On the other hand, in the tetracritical Ising CFT, there is a boundary condition that is weakly symmetric under $\mathrm{Rep}(S_3)$, as shown in  (\ref{weakreps3}).

\subsubsection{RG Boundary Conditions}\label{sec:RGbdy}

 Next, we discuss the relation between  symmetric boundary conditions and    symmetric relevant deformations that trivially gap a CFT. 
 See \cite{Wang:2013yta,Han:2017hdv,Li:2022drc,Wang:2022ucy} for discussions in the case of invertible symmetries.

Consider  a local relevant scalar operator $\phi$ in a 1+1d CFT. 
We assume that the CFT flows to a trivially gapped phase with a non-degenerate vacuum when we activate this relevant deformation everywhere in spacetime, $\exp\left({\lambda \int_{\mathbb{R}^2} d^2x \phi(x)}\right)$ with $\lambda>0$. 
Alternatively, we can  activate $\phi$ just in half of spacetime,
\begin{align}
\exp\left(\lambda \int_{x>0} d^2x\phi(x)\right) \,.
\end{align}
At low energies, the system is then described by the original gapless CFT in the region $x<0$, and is in a trivially gapped phase with a non-degenerate vacuum in the region $x>0$.
At $x=0$, this construction yields a boundary condition for the CFT, known as an RG boundary. We denote the corresponding boundary state as $|\phi\rangle_\text{RG}$. 
Suppose the CFT has a non-anomalous global symmetry and the relevant operator $\phi$ is symmetric under this symmetry. Then, the resulting RG boundary condition $\ket{\phi}_{\mathrm{RG}}$ is also symmetric. 
 See \cite{Gaiotto:2012np,Konechny:2016eek,Cardy:2017ufe,Konechny:2020jym,Cordova:2022lms}  for discussions of RG boundaries in 1+1d.

The converse is more subtle. The existence of a symmetric boundary condition does not guarantee a symmetric relevant deformation that trivially gaps the CFT.  
For instance, consider the tensor product of the chiral and antichiral Monster CFT (so that the gravitational anomaly vanishes). 
It was shown in \cite{Craps:2002rw} that there are  boundary conditions invariant under various centralizers of the diagonal Monster group, and yet there is simply no relevant deformation at all in this CFT. However, it should be stressed that there are other mechanisms to gap a system than merely turning on relevant deformations. For example, one can also add additional massive degrees of freedom, and attempt to gap the combined system in a symmetry-preserving manner. See \cite{nati} for more discussions.

Let us discuss the RG boundaries in the Ising CFT. Turning on the $\mathbb{Z}_2$-odd $\sigma$ deformation in the Ising CFT in half of spacetime leads to the two fixed boundary conditions $\ket{\uparrow},\ket{\downarrow}$, depending on the sign of the relevant coupling $\lambda$. The two fixed boundaries are exchanged by the $\mathbb{Z}_2$ symmetry. 
For one sign of the $\mathbb{Z}_2$-even thermal deformation $\varepsilon$, the corresponding RG boundary is the $\mathbb{Z}_2$-symmetric free boundary $|f\rangle$. (For the other sign, the low energy phase is not trivially gapped, and the half spacetime RG does not lead to a boundary state. We refer readers to \cite{Chang:2018iay} for more discussions.)

\begin{figure}[t]
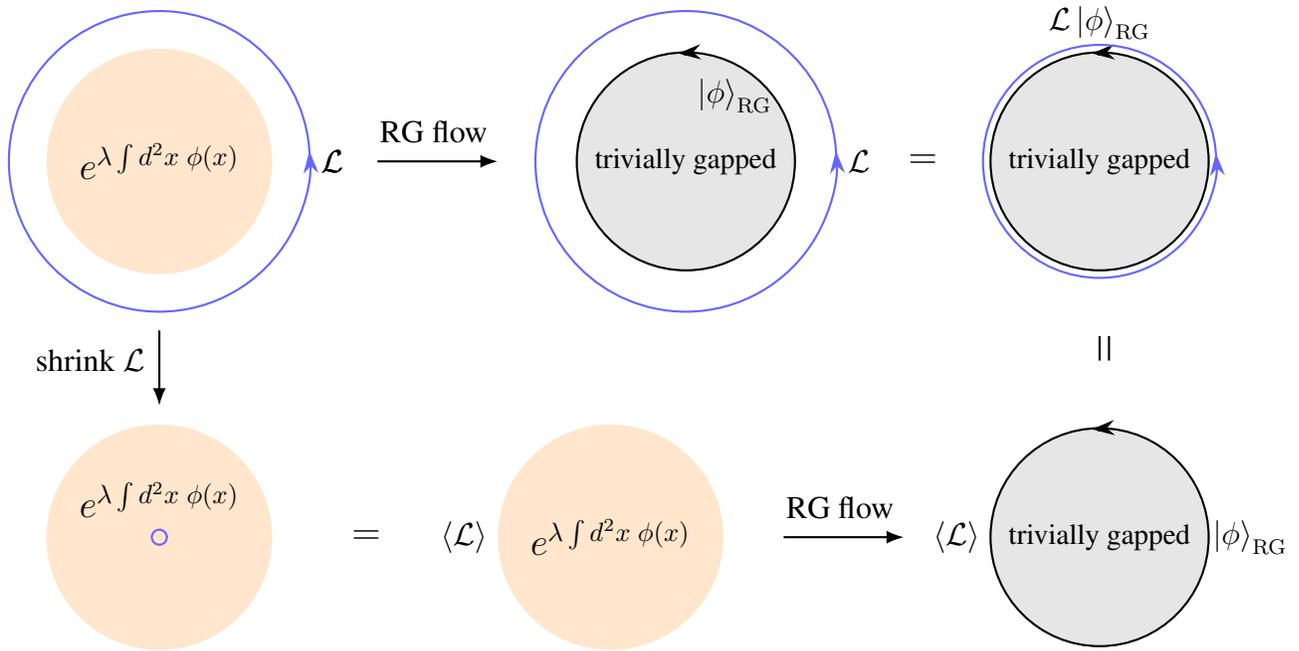

    \centering
 \ctikzfig{triviallygapssbc}
    \caption{
    We turn on a relevant operator $\phi$ which commutes with a line $\mathcal{L}$ and trivially gaps the theory, inside a disk region (light orange) in the bulk. The resulting interface between the trivially gapped theory and the bulk CFT is a strongly symmetric conformal boundary $\ket{\phi}_{\mathrm{RG}}$ under $\mathcal{L}$. On the top left, we first flow to the IR and create a conformal boundary $\ket{\phi}_{\mathrm{RG}}$. We then  shrink $\mathcal{L}$ onto $\ket{\phi}_{\mathrm{RG}}$ to obtain $\mathcal{L}\ket{\phi}_{\mathrm{RG}}$ on the top right. Alternatively, starting from the top left, we can shrink the loop of $\mathcal{L}$ inside the deformed region to get $\langle \mathcal{L} \rangle$. We then   flow to the IR and obtain $\langle \mathcal{L} \rangle \ket{\phi}_{\mathrm{RG}}$. This commutative diagram shows that such a relevant deformation creates a strongly symmetric boundary, i.e., ${\cal L} \ket{\phi}_{\mathrm{RG}} = \langle{\cal L}\rangle \ket{\phi}_{\mathrm{RG}}$. }
    \label{fig:triviallygapssbc}
\end{figure}
How do we extend this discussion to a CFT with a non-invertible global symmetry described by a fusion category $\mathcal{C}$? 
A local operator $\phi$ is said to be symmetric under $\mathcal{C}$ if every line of $\mathcal{C}$ commutes with $\phi$ \cite{Chang:2018iay,Lin:2023uvm}. 
Let $\phi$ be a $\mathcal{C}$-symmetric relevant deformation such that $\exp(\lambda \int_{\mathbb{R}^2}d^2 x \phi(x))$ drives the CFT to a trivially gapped phase, which in particular implies that $\mathcal{C}$ is anomaly-free (i.e., it admits a fiber functor). 
Then we claim that the corresponding RG boundary condition $|\phi\rangle_\text{RG}$ is not only weakly symmetric under $\mathcal{C}$, but also strongly symmetric.\footnote{In all examples we know of, the RG boundary condition from such a relevant deformation is always a simple boundary, meaning that it cannot be written as the superposition of other boundary conditions. It is conceivable that with more than one relevant coupling constant, one can fine-tune the couplings such that the resulting RG boundary is not simple. We will focus on relevant deformation triggered by a single relevant deformation here and assume that the RG boundary is simple.} This can be most easily seen from Figure \ref{fig:triviallygapssbc}.

Turning the argument around, we reach the following statement: \textit{Given an anomaly-free fusion category symmetry $\mathcal{C}$, if there does not exist a simple, strongly symmetric boundary, then there cannot be a symmetric  deformation that trivially gaps the bulk CFT triggered by a  relevant operator.}
In the next subsection, we will use this to constrain bulk RG flows in CFT.

\subsubsection{Non-Trivial RG Flows without Anomalies}

Following the discussion in Section \ref{sec:RGbdy}, the absence of a strongly symmetric boundary in the tetracritical Ising CFT leads to the conclusion that this CFT cannot be trivially gapped by a relevant deformation preserving the non-invertible symmetry $\mathrm{Rep}(S_3)$. 
This is somewhat surprising given that the non-invertible symmetry $\mathrm{Rep}(S_3)$ is anomaly-free and is kinematically compatible with a trivially gapped phase.

Let us verify this claim. 
The tetracritical Ising CFT has a unique $\mathrm{Rep}(S_3)$-symmetric relevant local operator $\phi_{2,1}$ of dimension $(h,\bar h)=(2/5,2/5)$ \cite{Chang:2018iay}.  
Furthermore, $\phi_{2,1}$ breaks all the other internal symmetries (including a Fibonacci category) and only preserves the non-anomalous $\mathrm{Rep}(S_3)$.
The flow triggered by $\phi_{2,1}$ is integrable and has been studied in \cite{Fateev:1997yg}.\footnote{We thank A.\ Zamolodchikov for pointing out this reference to us.} 
Indeed, it was found that the low energy phase is gapped, and has 2 states for one sign and 3 states for the other sign of the relevant coupling constant. 
The gapped phase with 3 states is a 1+1d TQFT with a $\mathrm{Rep}(S_3)$ fusion category symmetry, which can be viewed as a spontaneously broken phase of $\mathrm{Rep}(S_3)$.  (See \cite{Chang:2018iay,Huang:2021zvu} for general discussions on TQFTs with fusion category symmetries).
This massive vacuum structure can also be easily verified using the variational calculation in \cite{Cardy:2017ufe}.

To conclude, we find that the tetracritical Ising CFT cannot be trivially gapped by any relevant deformation while preserving its non-anomalous $\mathrm{Rep}(S_3)$ fusion category symmetry.\footnote{It is still  possible that this minimal model can be trivially gapped while respecting the $\mathrm{Rep}(S_3)$ symmetry by introducing additional degrees of freedom and turning on interactions.} 
This conclusion is confirmed from  the analysis of symmetric boundary conditions and from integrability. 
While this is not a contradiction in any way, it is curious that the nontrivial low energy phases do not seem to admit an explanation just in terms of symmetry. 
(See Section \ref{sec:RGbdy} for related discussions in the case of invertible symmetries.)

One might wonder if there are  mixed anomalies involving the fusion category symmetry and spacetime symmetries, such as time-reversal symmetry.\footnote{We thank Z.\ Komargodski for discussions on this point.}  
For instance,  mixed anomalies between time-reversal symmetry and fusion category symmetries in 1+1d have been previously studied in \cite{2021JHEP...05..204I}.
We leave this interesting possibility for future investigations.

\section{Boundaries and Symmetries in Higher Dimensions} \label{sec:2+1d}

In this section, we make some preliminary comments on  boundary conditions and symmetries in higher spacetime dimensions. We leave a complete analysis for the future.

\subsection{Higher Simplicity}

Consider a QFT in $d$ spacetime dimensions with a set of $d-1$-dimensional boundary conditions.  
Given any such boundary condition $\cal B$, we can stack a decoupled $d-1$-dimensional QFT $\cal Q$ on top of $\mathcal{B}$ to obtain another boundary, denoted as ${\cal Q}\otimes {\cal B}$ or ${\cal Q}|{\cal B}\rangle$ for the corresponding boundary state.  More generally, we can consider a direct sum of boundaries  with ``coefficients" being any $d-1$-dimensional QFTs ${\cal Q}_i$,
\ie\label{bdyhigherd}
\bigoplus_i {\cal Q}_i  |{\cal B}_i\rangle \,.
\fe

In the special case that we restrict ourselves to $d$-dimensional CFTs and their conformal boundary conditions, the ``coefficients" ${\cal Q}_i$ are restricted to be $d-1$-dimensional CFTs. 
In the previous sections, where we considered 1+1d bulk CFTs, the corresponding ${\cal Q}_i$ are then  $0+1$-dimensional CFTs, i.e.\ topological quantum mechanics theories. Such theories can be thought of as free $n_i$-dimensional qunits, and each is  completely specified by a non-negative integer $n_i\in \mathbb{Z}_{\ge 0}$. 
Thus, for $d=2$,  \eqref{bdyhigherd} is a general non-negative integer linear combinations of $0+1$d boundary conditions. In higher spacetime dimensions, the non-negative integers are replaced by $d-1$-dimensional field theories.
Similarly, if we restrict ourselves to $d$-dimensional TQFTs with topological boundaries, then the ${\cal Q}_i$ are restricted to be $d-1$-dimensional TQFTs.

In general spacetime dimensions, we further define a notion of \textit{higher simplicity}. 
A $d-1$-dimensional boundary $\cal B$ (or a topological defect) of a $d$-dimensional QFT is called \textit{$q$-simple} if there is no topological defect on $\cal B$ whose dimension is less than or equal to $q$. 
A 0-simple boundary is simple (or elementary) in the usual sense since  there are no topological local operators on it, which implies that it cannot be written as a direct sum of other boundaries.

\subsection{TQFT Matrix Representation}

As discussed previously, the conformal boundary conditions of a 1+1d CFT transform as NIM-reps of a generalized global symmetry. We now extend this to QFTs in $d$ spacetime dimensions.

\begin{figure}
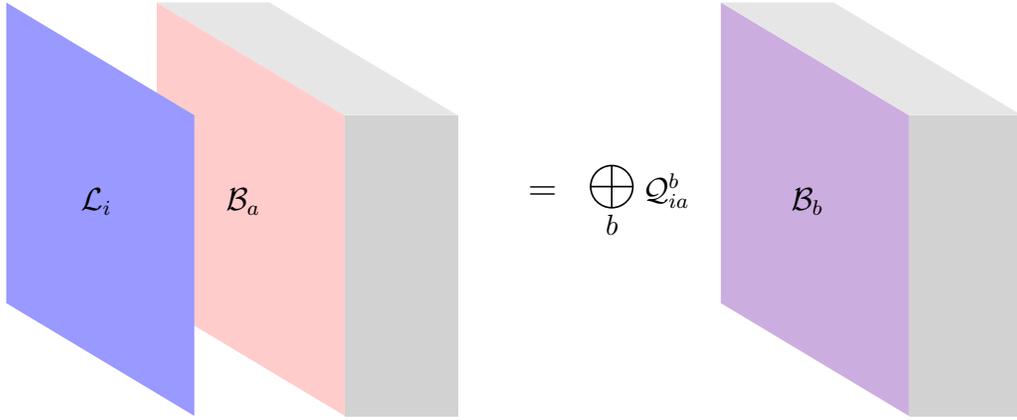

   \centering
   \ctikzfig{figures/NIMhigherd}
   \caption{Parallel fusion of a codimension-1 topological defect $\mathcal{L}_i$ with a boundary $\mathcal{B}_a$ gives a linear combination of boundaries with TQFT coefficients ${\cal Q}^b_{ia}$.}
    \label{fig:NIMhigherd}
\end{figure}

For simplicity, we focus on a 0-form global symmetry, the coarse features of which we distill into a fusion algebra $\mathcal{C}$ generated by codimension-1 topological operators $\mathcal{L}_i$. 
The 0-form symmetry $\cal C$ can be a group algebra corresponding to an invertible finite symmetry, or some more general algebra corresponding to a non-invertible symmetry.  
We start by considering the parallel fusion of codimension-1 topological defects ${\cal L}_i$  with a set of boundary conditions ${\cal B}_a$ as shown in Figure \ref{fig:NIMhigherd}.  

Such an action takes the following form
\ie\label{NIMhigherd}
{\cal L}_i  |{\cal B}_a\rangle = \bigoplus_b {\cal Q}^b_{ia} | {\cal B}_b\rangle 
\fe
where the ``coefficients" ${\cal Q}^b_{ia}$ are now $d-1$-dimensional TQFTs, rather than just non-negative integers for $d=2$. 
These  ${\cal Q}_{ia}^b$s furnish as a matrix representation of the global symmetry $\cal C$ valued in TQFT coefficients. 
 \eqref{NIMhigherd} states that boundary conditions are in   \textit{TQFT matrix representations} of  a generalized global symmetry $\cal C$,  generalizing the NIM-reps for $d=2$. 
 See Figure \ref{fig:NIMhigherd}. 
 Similar observations related to ``TQFT-valued coefficients" arising in the fusion of topological defects have been  made previously in \cite{Roumpedakis:2022aik,Choi:2022zal}.

\subsection{2+1d $U(1)$ Maxwell Theory}

Let us illustrate the  general discussion above in the context of 2+1d free $U(1)$ Maxwell  gauge theory. In Euclidean signature, the Lagrangian is
\ie\label{bulkmaxwell}
{\cal L} =\frac{1}{2g^2}F\wedge \star F\,,
\fe
where $F=dA$ is the gauge-invariant field strength. 
The free Maxwell theory is exactly dual to a free compact scalar $\phi\sim \phi+2\pi$, which is related to the gauge field as $\frac{i}{2\pi}d\phi = {1\over g^2}\star F$.

The global symmetry  of the free Maxwell theory includes $(\tilde{U}(1)^{(0)}\times U(1)^{(1)})\rtimes \mathbb{Z}_2^{(0)}$. 
It has a $U(1)^{(1)}$ 1-form symmetry that shifts the gauge field by a flat connection. 
The 1-form symmetry charge is the electric flux integrated against a curve $\gamma$, or in terms of the dual scalar field, 
it is the winding number ${1\over 2\pi } \oint_\gamma d\phi$.  
The  topological symmetry line is  $\exp\left( {i\alpha\over 2\pi}\oint_\gamma d\phi\right)$.
There is also a $\tilde{U}(1)^{(0)}$ 0-form magnetic global symmetry whose conserved charge is the magnetic flux $\frac{1}{2\pi }\oint_\Sigma F$ through a 2-surface $\Sigma$. The topological symmetry surface is $\exp({i\alpha\over 2\pi}\oint_\Sigma dA)$. 
The $\tilde{U}(1)^{(0)}$ 
 symmetry acts on the scalar field by a shift, i.e., $\phi\to \phi+\alpha$.  
Finally, the charge conjugation symmetry $\mathbb{Z}_2^{(0)}$ flips the sign of $\phi$ and $A$. 

We now discuss some (0-)simple boundary conditions. 
For simplicity, we place the boundary at $x=0$. 
We start with a Neumann boundary condition for the gauge field, which obeys 
\ie
F_{x\mu}\Big|_{x=0} = 0\,.
\fe
Since $A$ is unconstrained on the boundary, we can add a boundary theta angle term
\ie\label{2dbdytheta}
{i\theta\over 2\pi} \oint dA \Big|_{x=0}\,.
\fe
We denote this $S^1$ family of Neumann boundaries as $\widetilde{\cal B}(\theta)$ with $\theta\in[0,2\pi)$. 
It is equivalent to the Dirichlet boundary conditions for the dual compact scalar field $\phi\sim \phi+2\pi$:
\ie
\widetilde{\cal B}(\theta) :~~\phi\Big|_{x=0} = \theta\,.
\fe
The $\tilde{U}(1)^{(0)}$ and $\mathbb{Z}_2^{(0)}$ symmetries act on the Neumann boundary as
\ie
&e^{  {i\alpha \over 2\pi} \oint dA } | \widetilde{\cal B}(\theta)\rangle = |\widetilde{\cal B}(\theta +\alpha)\rangle \,,\\
&\eta |\widetilde {\cal B}(\theta)\rangle =|\widetilde {\cal B}(-\theta)\rangle\,,
\fe
where $\eta$ is a surface operator that generates the $\mathbb{Z}_2^{(0)}$ symmetry.
Finally as we bring the bulk topological line $\exp({i\alpha\over2\pi}\oint d\phi)$ for the $U(1)^{(1)}$ symmetry to the boundary, it becomes trivial on $\widetilde{\cal B}(\theta)$ because $\phi$ is taken to be a fixed value 
 and there is no winding  on this boundary. 
 Hence $\widetilde {\cal B}(\theta)$ is 1-simple, but not 2-simple because of the 2d topological operator, i.e., the boundary theta term \eqref{2dbdytheta}.

There is also  a Dirichlet boundary condition  for the gauge field
\ie
{\cal B} :~~ A\Big|_{x=0} =0\,,
\fe
where $|_{x=0}$ stands for the restriction of the differential form to the boundary. 
The Dirichlet boundary condition has the distinguished feature that a minimally charged Wilson line $\exp(i \int A)$ can terminate on it. 
This is a Neumann boundary condition for the dual field $\phi$. 
The boundary condition ${\cal B}$ is invariant under the $\tilde{U}(1)^{(0)}$ symmetry and the charge conjugation $\mathbb{Z}_2^{(0)}$ symmetry:
\ie
&e^{  {i\alpha \over 2\pi} \oint dA } |{\cal B}\rangle =| {\cal B}\rangle\,,\\
&\eta |{\cal B}\rangle= |{\cal B}\rangle\,.
\fe 
As we bring the $U(1)^{(1)}$ topological line $\exp({i\alpha\over2\pi}\oint_\gamma d\phi)$ to $\cal B$, it becomes a nontrivial line on ${\cal B}$. 
Hence the bulk $U(1)^{(1)}$ 1-form symmetry becomes a boundary $U(1)^{(0)}$ 0-form symmetry.  
 There are two ways to understand this. 
 From the gauge field point of view, since the Wilson line $\exp(i \int A)$ can terminate on $\cal B$, we can link a $U(1)^{(1)}$ line with a Wilson line in the bulk, and push the former to the boundary. Since the linking phase is nontrivial, this process has to give a nontrivial boundary line.  
 Alternatively, from the scalar field point of view, $\phi$ is unconstrained on $\cal B$, and therefore there is a $U(1)^{(0)}$ winding symmetry on the boundary that measures its winding number.\footnote{In contrast, $\widetilde{\cal B}(\theta)$ has a boundary theta angle \eqref{2dbdytheta}, which can be viewed as a boundary $U(1)$ (-1)-form global symmetry.} 
We hence learn that  ${\cal B}$ is not 1-simple, but only 0-simple.

\begin{figure}
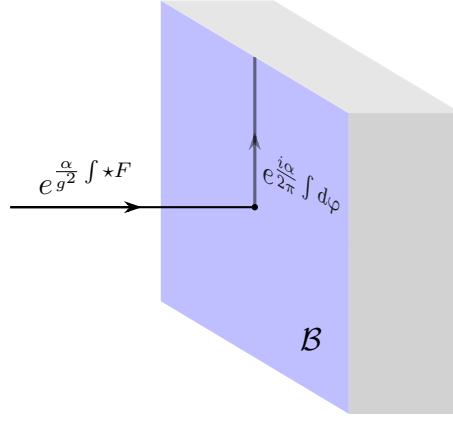

    \centering
    \ctikzfig{figures/maxwell1}
    \caption{The bulk $U(1)^{(1)}$ global symmetry line $\exp({\frac{\alpha}{g^2}\int \star F})$ is connected to a boundary $U(1)^{(0)}$ symmetry line $\exp({\frac{i\alpha}{2\pi}\int d\varphi})$ at a topological junction on the Dirichlet boundary $\mathcal{B}$ for the gauge field $A$. Here $\varphi$ is a compact scalar field living on the boundary.}
    \label{fig:maxwell1}
\end{figure}

The explicit action for $\cal B$ is
\ie\label{bdyaction}
{\cal B}:~~~{i\over 2\pi}\int_{x=0} \varphi dA\,,
\fe
where $\varphi$ is a compact scalar field that lives on the boundary $x=0$ and $A$ is the restriction of the bulk 1-form gauge field to the boundary.\footnote{For the Dirichlet boundary condition $\cal B$, the boundary scalar $\varphi$ is the restriction of the bulk $\phi$ to $\cal B$. Nonetheless we introduce a new symbol $\varphi$ because later when we discuss the partially Dirichlet boundary ${\cal B}_N$, the bulk and boundary compact scalar fields differ by a factor of $N$ \eqref{phivarphi}, i.e., $\phi= N\varphi$.}
The variation of $A$ gives  the  boundary equation  of motion:
\ie
{\cal B}:~~~
{i\over 2\pi} d\phi\Big|_{x=0}=
{1\over g^2} \star F\Big|_{x=0} = {i\over 2\pi} d\varphi
\fe
in addition to the bulk equation of motion. 
This means that the bulk  $U(1)^{(1)}$ symmetry line $\exp( {\alpha\over g^2}\int \star F)=\exp({i\alpha\over2\pi}\int d\phi)$ can be  connected to   a boundary topological line $\exp\left({i\alpha\over 2\pi}\int d\varphi \right)$ at a topological junction on the boundary as shown in Figure \ref{fig:maxwell1}.

 \subsubsection{Partially Dirichlet Boundaries}

We move on to discuss the non-invertible symmetries of the 2+1d Maxwell theory and their action on the above boundaries. 
The non-invertible condensation surface defects $S_N(\Sigma)$ were constructed in \cite{Roumpedakis:2022aik} 
\ie\label{SN}
S_N(\Sigma) = 
\frac{1}{\sqrt{|H_1(\Sigma,\mathbb{Z}_N)|}}\sum_{\gamma\in H_1(\Sigma,\mathbb{Z}_N)} \exp\left(
\frac{i}{N}  \oint_\gamma d\phi
\right)
\fe
where  $N$ is any positive integer and $\Sigma$ is a closed 2-surface. 
They arise from higher gauging the discrete $\mathbb{Z}_N^{(1)}$  subgroup of $U(1)^{(1)}$ on a surface.
The explicit action for this defect inserted along $x=0$ is\footnote{See also \cite{Kapustin:2010if} for the action of related condensation surface defects in Chern-Simons theory.}
\ie\label{SNLag}
& {1\over 2g^2} \int_{x<0}F_L\wedge \star F_L
+ {1\over 2g^2} \int_{x>0}F_R\wedge \star F_R\\
&+ {i N\over 2\pi} \int_{x=0} \Phi (dA_L-dA_R)\,,
\fe
where $A_L,A_R$ are respcetively the gauge fields on the two sides of the defect. 
Here $\Phi\sim \Phi+2\pi$ is an auxiliary compact scalar field living only on the defect $x=0$.

The condensation defect $S_N$ obeys the non-invertible fusion rule 
\ie
S_N \otimes S_{N'} = ({\cal Z}_{\text{gcd}(N,N')})S_{\text{lcm}(N,N')}\,,
\fe
where the fusion ``coefficient" ${\cal Z}_{\text{gcd}(N,N')}$ is a 1+1d $\mathbb{Z}_{\text{gcd}(N,N')}$ gauge theory.

Let us discuss the action of the non-invertible symmetry on the two kinds of boundaries $\cal B$ and $\widetilde{\cal B}(\theta)$ above. 
Since the topological lines $\exp\left( \frac{i}{N} \oint d\phi\right)$ become 1 on $\widetilde{\cal B}(\theta)$, the condensation defect $S_N(\Sigma)$ reduces to the partition function of a decoupled 1+1d $\mathbb{Z}_N$ gauge theory using the expression \eqref{SN}, i.e.,
\ie
S_N | \widetilde{\cal B}(\theta)  \rangle= {\cal Z}_N| \widetilde{\cal B}(\theta)\rangle\,.
\fe
This gives a realization of the general formula \eqref{NIMhigherd} where the ``coefficients" are TQFTs.

Acting with $S_N$ on the other boundary $\cal B$ gives another simple boundary, which we denote by
\ie
|{\cal B}_N\rangle \equiv S_N|{\cal B}\rangle
\fe
with $|{\cal B}_1\rangle = |{\cal B}\rangle$ as $S_1$ is a trivial surface. 
This is equivalent to gauging the $\mathbb{Z}_N^{(0)}$ subgroup of the $U(1)^{(0)}$ symmetry (which arises from the restriction of the bulk $U(1)^{(1)}$ 1-form symmetry) only on the boundary. 
Since $S_N \otimes\exp({2\pi i\over N}\oint d\phi)=S_N$ \cite{Roumpedakis:2022aik}, this new boundary condition $|{\cal B}_N\rangle$ can freely absorb the $\mathbb{Z}_N^{(1)}$ topological lines (but not other lines):
\ie\label{ZNabsorb}
\exp\left( { i \over N}\oint_\gamma  d\phi \right) |{\cal B}_N\rangle = |{\cal B}_N\rangle \,,
\fe
where the curve $\gamma$ is parallel to the boundary. 
This shows that the boundary $|{\cal B}_N\rangle$ is neither a Dirichlet nor a Neumann boundary condition.  
Since the $\mathbb{Z}_N^{(1)}$ topological lines become trivial on ${\cal B}_N$, it follows that only Wilson lines $\exp(i Q \int A)$ with 
\ie
Q\in N\mathbb{Z}
\fe
can terminate on ${\cal B}_N$.
 In this sense the boundary ${\cal B}_N$ is  a \textit{partially Dirichlet boundary condition}.  
Indeed,  $S_N$ annihilates a charge $Q$ Wilson line by encircling it  unless $Q\in N\mathbb{Z}$ \cite{Roumpedakis:2022aik}, implying that the Wilson line can only end on $S_N$ if this condition is satisfied.

Let us describe the partially Dirichlet boundary condition $|{\cal B}_N\rangle$ for the Maxwell theory in detail. 
The fusion between $S_N$ and the Dirichlet boundary $|{\cal B}\rangle$ is represented by the following action
\begin{align}\begin{split}
&{i\over 2\pi} \int_{x=0} \varphi_L dA_L
+{iN\over 2\pi} \int_{x=\epsilon} \Phi (dA_L-dA_R)
\\
&+{1\over 2g^2}\int_{0<x<\epsilon} F_L\wedge\star F_L +\int_{x>\epsilon} F_R\wedge\star F_R
\end{split}\end{align}
where we place the boundary ${\cal B}$ at $x=0$ and the condensation defect $S_N$ at $x=\epsilon$.  
Next, we take $\epsilon\to0$, and $A_L$ becomes a field living only on the boundary. 
Defining $\varphi'= \varphi_L+N\Phi$,  $\varphi=-\Phi$ and renaming $A_R$ as just $A$, the boundary action for $S_N|{\cal B}\rangle$ becomes
\begin{align}\begin{split}
&{i\over 2\pi}\int_{x=0} \varphi'dA_L+ {iN\over2\pi}\int_{x=0} \varphi dA\\
&+{1\over 2g^2}\int_{x>0}F\wedge \star F\,.
\end{split}\end{align} 
Now, $A_L$ is a gauge field that only lives on the boundary. The first term is a trivial 1+1d $\mathbb{Z}_1$ TQFT and can be discarded.

\begin{figure}[h]
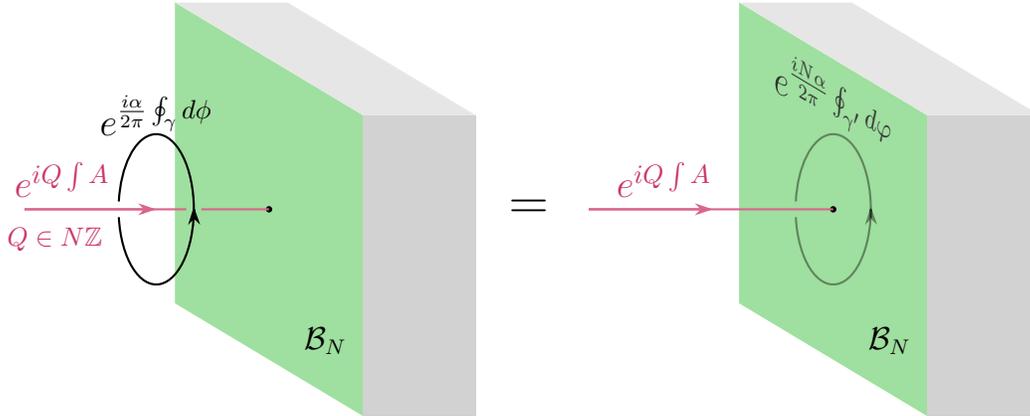

\centering
\ctikzfig{figures/maxwell2b}
\caption{In free Maxwell theory, a Wilson line $\exp(i Q\int A)$   of charge $Q$ can end on the partially Dirichlet boundary ${\cal B}_N$ only if $Q\in N\mathbb{Z}$. The bulk $U(1)^{(1)}$ global symmetry line $\exp({i\alpha\over2\pi}\oint_\gamma d\phi)$ becomes a boundary line $\exp({iN\alpha\over2\pi}\oint_{\gamma'} d\varphi)$ when pushed to ${\cal B}_N$. Here $\phi$ is the bulk compact scalar field dual to the gauge field, and $\varphi$ is a compact scalar field on the boundary.}

\label{fig:maxwell2}
\end{figure}

We therefore conclude that the boundary ${\cal B}_N = S_N\otimes {\cal B}$ is described by the following boundary action:
\ie
{\cal B}_N:~~~{iN\over2\pi} \int_{x=0} \varphi dA\,,
\fe
where $\varphi\sim \varphi+2\pi$ is a compact scalar field that only lives on the boundary $x=0$. 
The $\mathcal{B}_N$ can be viewed as a discrete family of generalizations of the Dirchlet boundary condition \eqref{bdyaction}.
The equation of motion for $A$ gives
\ie
{\cal B}_N:~~~
{i\over 2\pi}d\phi\Big|_{x=0}=
{1\over g^2} \star F\Big|_{x=0} = {iN\over 2\pi} d\varphi.
\fe
Written in terms of the dual scalar field $\phi$, we find the restriction of the bulk $U(1)^{(1)}$ topological lines to the boundary ${\cal B}_N$:
\ie\label{phivarphi}
\exp\left({i\alpha\over 2\pi}\oint_\gamma d\phi\right)\Big|_{x=0}=\exp\left({iN\alpha\over 2\pi}\oint_{\gamma'} d\varphi\right) \,.
\fe
where $\gamma'$ is a curve on ${\cal B}_N$ that is homologous to    $\gamma$. In particular, it means that the lines of the $\mathbb{Z}_N^{(1)}$ subgroup are trivial on the boundary (i.e., those with $\alpha \in 2\pi \mathbb{Z}/N$), 
which is consistent with \eqref{ZNabsorb}. 
See Figure \ref{fig:maxwell2}.

To conclude, ${\cal B}_N=S_N\otimes {\cal B}$ is a discrete generalization of the Dirichlet boundary condition ${\cal B}:A|=0$ obtained from the non-invertible symmetry action on the ordinary Dirichlet boundary. 
Unlike the ordinary Dirichlet boundary, only the charge $N$ Wilson line can terminate on ${\cal B}_N$.

It would be interesting to generalize this discussion to obtain the action of the non-invertible symmetries found in \cite{Choi:2021kmx,Choi:2022zal,Choi:2022jqy,Cordova:2022ieu,Choi:2022rfe,Niro:2022ctq,Choi:2022fgx,Yokokura:2022alv,Yamamoto:2023uzq} on the boundaries of 3+1d $U(1)$ gauge theories. 
See \cite{Damia:2022bcd} for discussions of non-invertible 1-form symmetries in 4+1d gauge theory with boundaries.

\subsection{TQFT Examples}

We further discuss several examples of boundary conditions for 2+1d TQFTs.

\subsubsection{2+1d $\mathbb{Z}_2$ Gauge Theory}

Consider the 2+1d $\mathbb{Z}_2$ gauge theory whose Lagrangian is \cite{Maldacena:2001ss,Banks:2010zn,Kapustin:2014gua}
\ie
{\cal L} = {2i\over 2\pi }Ad\tilde A\,.
\fe
This is the low energy limit of the toric code. 
There are three nontrivial topological lines
\ie
e= e^{i \oint A }\,,~~~~m=e^{i\oint \tilde A}\,,~~~~\psi = e^{i \oint (A+\tilde A)}\,,
\fe
where $e,m$ have spin 0 and $\psi$ is a fermion of spin $1/2$.

Our discussion of the topological boundaries and surfaces follows \cite{Roumpedakis:2022aik}  closely. 
There are two topological boundary conditions \cite{Kapustin:2010hk} (see also \cite{Bravyi:1998sy,2012CMaPh.313..351K,2013PhRvX...3b1009L,2013PhRvB..88x1103B}), given by the Dirichlet boundary conditions of $A$ and $\tilde A$:
\ie
&{\cal B}_e:~~~A|=0\,,\\
&{\cal B}_m:~~~\tilde A|=0\,.
\fe
${\cal B}_e$ is known as the $e$-condensing boundary condition in the sense that the $e$ line becomes trivial when pushed to this boundary, and vice versa for the $m$-condensing boundary ${\cal B}_m$. 
Both ${\cal B}_e$ and ${\cal B}_m$ can be realized by gauging the $\mathbb{Z}_2^{(1)}$ 1-form symmetry generated by $e$ and $m$ in half of the spacetime, respectively \cite{Kaidi:2021gbs}.  
Both of these topological boundary conditions are not 1-simple.

The boundary states obey the following inner products:
\ie\label{inner}
\langle{\cal B}_e |{\cal B}_e \rangle = \langle{\cal B}_m |{\cal B}_m \rangle = {\cal Z}_2\,,~~~~\langle{\cal B}_m |{\cal B}_e \rangle = \langle{\cal B}_e |{\cal B}_m \rangle = 1
\fe
where ${\cal Z}_2$ stands for a 1+1d $\mathbb{Z}_2$ gauge theory.

There are 6 topological surfaces in total, denoted by  $\{\mathds{1},S_\psi,S_e,S_m,S_{em},S_{me} \}$ \cite{Roumpedakis:2022aik} (see also \cite{Lan:2014uaa}).
Here, $\mathds{1}$ is the trivial surface, and $S_\psi$ generates a non-anomalous $\mathbb{Z}_2$ symmetry  that exchanges the $e$ and $m$. We have $S_\psi^2=\mathds{1}$. 
The other four surfaces $S_e,S_m,S_{em},S_{me}$ are non-invertible. 
They can be factorized in terms of the topological boundaries,
\ie
S_e = |{\cal B}_e \rangle\langle{\cal B}_e|\,,~~~~
S_m = |{\cal B}_m\rangle\langle{\cal B}_m|\,,~~~~
S_{em} = |{\cal B}_e \rangle\langle{\cal B}_m|\,,~~~~
S_{me} = |{\cal B}_m\rangle\langle{\cal B}_e|\,.
\fe
The fusion rules between these six surfaces can be found in 
 \cite{Roumpedakis:2022aik}. 
In particular, $S_e$ satisfies  the non-invertible fusion rule $S_e \otimes S_e = ({\cal Z}_2) S_e$. These surfaces act on the two topological boundaries as
\ie
&S_\psi |{\cal B}_e\rangle = |{\cal B}_m\rangle\,,~~~~
S_\psi |{\cal B}_m\rangle = |{\cal B}_e\rangle\,,\\
&S_e|{\cal B}_e \rangle = ({\cal Z}_2 )|{\cal B}_e\rangle\,,~~~~~S_e |{\cal B}_m\rangle = |{\cal B}_e\rangle\,,\\
&S_m|{\cal B}_e \rangle = |{\cal B}_m\rangle\,,~~~~~S_m |{\cal B}_m\rangle = ({\cal Z}_2)|{\cal B}_m\rangle\,,\\
&S_{em}|{\cal B}_e\rangle = |{\cal B}_e\rangle\,,~~~~S_{em}|{\cal B}_m\rangle = ({\cal Z}_2)|{\cal B}_e\rangle\,,\\
&S_{me}|{\cal B}_e\rangle = ({\cal Z}_2)|{\cal B}_m\rangle\,,~~~~S_{me}|{\cal B}_m\rangle = |{\cal B}_m\rangle\,.
\fe

Generally, given a set of topological surfaces, one asks if there is a boundary condition that is an eigenstate under each of the surfaces, with eigenvalue being  the partition function of a decoupled 1+1d TQFT.
By specializing to a spatial torus $T^2$, one finds that a necessary condition for such a boundary condition is the existence of a set of non-negative integers which solves the fusion algebra of the surfaces. 
In the $\mathbb{Z}_2$ gauge theory example, there is no  topological boundary condition that is an eigenstate under all 6 surfaces due to this.\footnote{We thank S.\ Seifnashri for discussions on this point.}

Note that there is no simple topological boundary that is invariant under the $\mathbb{Z}_2$ symmetry $S_\psi$, which is free of 't Hooft anomalies because $H^4(\mathbb{Z}_2,U(1))=0$. 
On the other hand, there is a simple, $\mathbb{Z}_2$-symmetric conformal boundary condition, as we discuss below. 
It can be obtained by gauging the $\mathbb{Z}_2$ symmetry of  the 1+1d critical Ising CFT by the 2+1d $\mathbb{Z}_2$ gauge theory.\footnote{See \cite{Heckman:2022xgu} for a related discussion of boundary conditions for 4+1d discrete 2-form gauge theory.} 
We denote this conformal boundary state   on a spatial slice $\Sigma$ as $\ket{\text{Ising}}$, which takes the form \cite{Gaiotto:2020iye,Kaidi:2022cpf}
\begin{equation}
    \ket{\text{Ising}} = \sum_{a \in H^1(\Sigma,\mathbb{Z}_2)} \mathcal{Z}_\text{Ising}[a] \ket{a} \,,
\end{equation}
where $\mathcal{Z}_\text{Ising}[A]$ denotes the partition function of the 1+1d critical  Ising CFT on $\Sigma$ coupled to the $\mathbb{Z}_2$ gauge field $A$, and $\ket{a}$ is the state of the $\mathbb{Z}_2$ gauge theory in the Hilbert space on $\Sigma$ corresponding to the flat $\mathbb{Z}_2$ connection $a \in H^1(\Sigma,\mathbb{Z}_2)$. 
This gapless boundary of the 2+1d $\mathbb{Z}_2$ gauge theory was also discussed in \cite{Lin:2019kpn,Ji:2019jhk}.

To see that $|\text{Ising}\rangle$ is invariant under the $\mathbb{Z}_2$ symmetry, we note that 
the surface defect $S_\psi$, when acting on the Hilbert space corresponding to a spatial slice $\Sigma$, can be represented as follows \cite{Gaiotto:2020iye,Kaidi:2022cpf}:
\begin{equation} \label{eq:em_surface}
    S_\psi (\Sigma) = \frac{1}{\sqrt{|H^1(\Sigma,\mathbb{Z}_2)|}} \sum_{a,a' \in H^1(\Sigma,\mathbb{Z}_2)} (-1)^{\oint_\Sigma a \cup a'} \ket{a'}\bra{a} \,,
\end{equation}
Using the fact that the Ising CFT is invariant under gauging the $\mathbb{Z}_2$ symmetry, we find
\begin{align*}
    S_\psi \ket{\text{Ising}} &= \sum_{a' \in H^1(\Sigma,\mathbb{Z}_2)}
    \left(    
    \frac{1}{\sqrt{|H^1(\Sigma,\mathbb{Z}_2)|}} \sum_{a \in H^1(\Sigma,\mathbb{Z}_2)} \mathcal{Z}_\text{Ising}[a] (-1)^{\oint_\Sigma a \cup a'}
    \right)
    \ket{a'} \\
    &= \sum_{a' \in H^1(\Sigma,\mathbb{Z}_2)} \mathcal{Z}_\text{Ising}[a'] \ket{a'} \\
    &= \ket{\text{Ising}} \,.
\end{align*}
Indeed, when coupled to the 2+1d $\mathbb{Z}_2$ gauge theory, the non-invertible Kramers-Wannier duality defect of the Ising CFT becomes the end locus of the $\mathbb{Z}_2$ surface on the boundary $|\text{Ising}\rangle$ \cite{Freed:2018cec,Ji:2019ugf}.

\subsubsection{$SU(2)_{16} \times SU(2)_{-16}$ Chern-Simons Theory}

We give an example of \eqref{NIMhigherd} where there is more than one term on the right-hand side. 
Consider first the $SU(2)_{16}$ Chern-Simons theory.
There are two nontrivial surfaces $S_E$ and $S_2$ (corresponding to the $E$ and $D$ modular invariants, respectively), with fusion rule on $T^2$ given by \cite{Fuchs:2002cm,Roumpedakis:2022aik}
\begin{equation} \label{eq:SE}
    S_E \otimes S_E = S_E \oplus S_2 \,.
\end{equation}
On a more general manifold, there can be a relative Euler counterterm for the two terms on the right-hand side.

Now, we consider folding the $SU(2)_{16}$ Chern-Simons theory in the presence of the $S_E$ surface defect inserted at the crease.
This becomes a topological boundary of the doubled $SU(2)_{16} \times SU(2)_{-16}$ Chern-Simons theory which we denote as $\ket{S_E}$.
The fusion rule \eqref{eq:SE} implies
\begin{equation} 
    S_E \ket{ S_E} = \ket{ S_E} \oplus \ket{ S_2} \,,
\end{equation}
where $\ket{S_2}$ is the topological boundary coming from the folding along the $S_2$ surface in $SU(2)_{16}$.
This gives an example of  a non-invertible symmetry action  that produces more than one simple boundary in the outcome.

\subsubsection{Boundary Eigenstate of an Anomalous Symmetry} \label{sec:dsemion}

Finally, we briefly comment on an example of a boundary condition that is an eigenstate under an anomalous,  ordinary global symmetry. 
It was found in \cite{2015PhRvX...5d1013C,Barkeshli:2014cna} that the $U(1)_2$ Chern-Simons theory, also known as the semion model, has a $\mathbb{Z}_2\times \mathbb{Z}_2$ 0-form symmetry with an anomaly valued in $H^4(\mathbb{Z}_2\times \mathbb{Z}_2,U(1))=\mathbb{Z}_2\times \mathbb{Z}_2$. 
While this symmetry acts trivially on the unique local operator, i.e., the identity, it acts projectively on the Wilson lines due to symmetry fractionalization.

Next, consider $U(1)_2\times U(1)_{-2}$, which is the low energy limit of the double semion model. 
Denote the topological lines as $\{\mathds{1}, s, \bar s, s\bar s\}$, with spins $\{0,1/4,-1/4,0\}$. The doubled theory 
$U(1)_2\times U(1)_{-2}$ has a unique topological boundary condition $\cal B$, corresponding to gauging (condensing) the $\mathbb{Z}_2$ 1-form global symmetry generated by $s\bar s$. The associated  Lagrangian algebra is $\mathds{1}\oplus s\bar s$. 
This boundary  is simple, but not 1-simple, since there is a nontrivial $\mathbb{Z}_2$ line on it. 
The anomalous $\mathbb{Z}_2\times \mathbb{Z}_2$ global symmetry extends to a symmetry in $U(1)_2\times U(1)_{-2}$.

Since $\cal B $ is the unique, simple, topological boundary condition in this theory, the 
 $\mathbb{Z}_2\times \mathbb{Z}_2$ 0-form global symmetry   has to act on $|{\cal B}\rangle$ as an eigenstate (whose eigenvalue is generally a 1+1d SPT), even though  this symmetry has an ordinary 't Hooft anomaly valued in $H^4(\mathbb{Z}_2\times \mathbb{Z}_2,U(1))$. 
 This suggests that in general spacetime dimensions, even for invertible symmetries, being a boundary eigenstate is only a necessary, but not sufficient, condition for a symmetry-preserving boundary. 
 We leave a complete treatment of symmetry-preserving boundary conditions in general spacetime dimensions for future investigations.

\section*{Acknowledgements}
We are particularly grateful to Z.\ Komargodski for many enlightening discussions throughout various stages of this project. 
We also thank P.\ Boyle Smith,  Y.-H.\ Lin, C.-T.\ Hsieh, K.\ Ohmori, D.\ Ranard, S.\ Ryu, N.\ Seiberg, S.\ Seifnashri, J.\ Wang, Y.\ Wang,  A.\ Zamolodchikov, C.\ Zhang, and Y.\ Zheng for interesting discussions. 
We are grateful to Y.\ Zheng for comments on a draft. 
The work of SHS was supported in part by an NSF grant PHY-2210182.
SHS thanks Harvard University for its hospitality during the course of this work. 
The authors of this paper were ordered alphabetically.

\appendix

\section{Category Theory Background}\label{app:categories}

As was discussed in the main text, the set of conformal boundary conditions in a 1+1d CFT arrange themselves into non-negative integer-valued matrix representations (NIM-reps) of the fusion algebra of topological lines in the bulk.
Here, we review how the boundary conditions are subject to additional consistency conditions which come from the requirement that they form a module category over the fusion category of bulk topological lines.
This is well-understood for the case of rational CFTs and conformal boundary conditions which preserve (half of) the chiral algebra \cite{Fuchs:2001qc,Fuchs:2002cm}, and is expected to be  true for general boundary conditions in general 1+1d QFTs.
For a mathematically rigorous treatment of the subject, we refer readers to \cite{etingof2016tensor}.

\subsection{Fusion Category} 

\begin{figure}
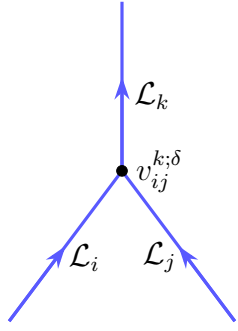

    \centering
    \ctikzfig{figures/trivalentjunctionapp}
    \caption{Topological junction operators $v_{ij}^{k;\delta}$ with $\delta = 1, \cdots, N_{ij}^k$ form a basis for the junction vector space $\mathrm{Hom}_{\mathcal{C}}(\mathcal{L}_i\otimes\mathcal{L}_j,\mathcal{L}_k)$.}
    \label{fig:trivalentjunctionapp}
\end{figure}

We first briefly review the defining data of the fusion category $\mathcal{C}$ formed by the topological line defects of a theory in 1+1d. 
The (finite) set of labels for the simple objects (i.e., topological lines) is denoted as ${\cal I} = \{ i,j,k, \cdots \}$.
Recall, from Section \ref{SWSB}, that the fusion coefficient $N_{ij}^k$  in the fusion of simple lines,
\begin{equation} \label{eq:fusion_lines1}
    \mathcal{L}_i \otimes \mathcal{L}_j = \bigoplus_{k\in\mathcal{I}} N_{ij}^{k} \mathcal{L}_k \,,
\end{equation}
is the dimension of the topological trivalent junction vector space,
\begin{equation}
    N_{ij}^k = \text{dim}_{\mathbb{C}}\,\text{Hom}_{\mathcal{C}}(\mathcal{L}_i \otimes \mathcal{L}_j,\mathcal{L}_k) \,.
\end{equation} 
We use $\delta = 1, \dots, N_{ij}^k$ to label fixed (but arbitrary) basis vectors $v_{ij}^{k;\delta} \in \text{Hom}_{\mathcal{C}}(\mathcal{L}_i \otimes \mathcal{L}_j,\mathcal{L}_k)$ for the trivalent junction vector space, as shown in Figure \ref{fig:trivalentjunctionapp}. 

The tensor product $\otimes$ is associative up to an isomorphism called the \emph{associator}.
For arbitrary (not necessarily simple) objects $X$, $Y$, and $Z$ in $\mathcal{C}$, the associator $\alpha_{XYZ}$ is an isomorphism
\begin{equation}
    \alpha_{XYZ} \in \text{Hom}_{\mathcal{C}}( (X \otimes Y ) \otimes Z  ,  X \otimes ( Y \otimes Z))\,.
\end{equation}
We denote $\alpha_{\mathcal{L}_i \mathcal{L}_j \mathcal{L}_k} \equiv \alpha_{ijk}$ for the simple objects $\mathcal{L}_i$, $\mathcal{L}_j$, and $\mathcal{L}_k$.
The matrix elements of the associator $\alpha_{ijk}$ in a fixed basis (as in Figure \ref{fig:trivalentjunctionapp}) are called the $F$-symbols.
To be more precise, they are defined by the equation
\begin{equation} \label{eq:F}
    v_{ip}^{\ell;\delta} \circ \left( \text{id}_i \otimes v_{jk}^{p;\lambda} \right) \circ \alpha_{ijk} =
    \sum_{q\in \mathcal{I}} \sum_{\rho = 1}^{N_{ij}^q} \sum_{\sigma=1}^{N_{qk}^\ell} 
    \left[ 
        F_{ijk}^{\ell}
    \right]_{(p,\delta,\lambda)(q,\rho,\sigma)}
    \cdot v_{qk}^{\ell;\sigma} \circ (v_{ij}^{q;\rho} \otimes \text{id}_k) \,.
\end{equation}
Here, $\text{id}_i \equiv \text{id}_{\mathcal{L}_i} \in \text{Hom}_{\mathcal{C}}(\mathcal{L}_i,\mathcal{L}_i)$ is the identity morphism, and similarly for $\text{id}_k$.
Note that the two sides of \eqref{eq:F} are both vectors in the four-point junction vector space $\text{Hom}_{\mathcal{C}}( (\mathcal{L}_i \otimes \mathcal{L}_j ) \otimes \mathcal{L}_k  ,  \mathcal{L}_l)$, and the $F$-symbols, $\left[ 
        F_{ijk}^{\ell}
\right]_{(p,\delta,\lambda)(q,\rho,\sigma)}$, are the basis transformation matrix elements for that vector space, relating two inequivalent ways to decompose the four-point junction into two trivalent junctions.
Such a basis transformation is often referred to as an $F$-move or as a crossing relation.
\begin{figure}[h]
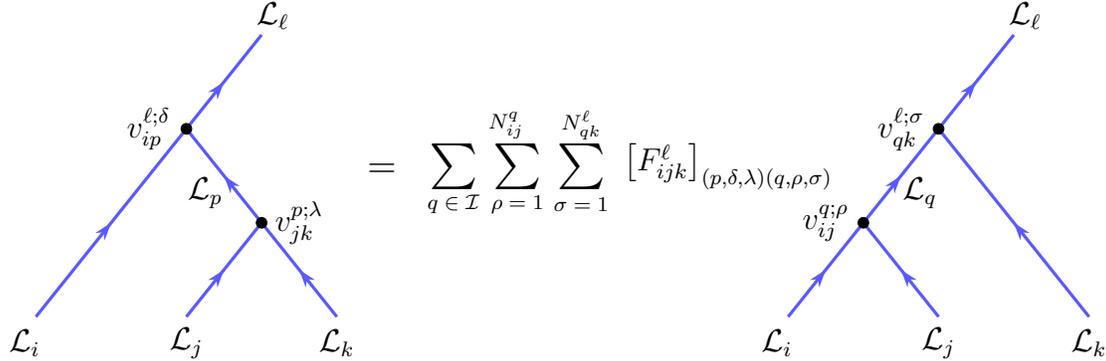

    \centering
    \ctikzfig{figures/associatorapp}
    \caption{The $F$-symbols correspond to the matrix elements of the crossing relation involving four external topological line defects.}
    \label{fig:associatorapp}
\end{figure}
Diagrammatically, \eqref{eq:F} is represented as in Figure \ref{fig:associatorapp}.

\begin{figure}[t]
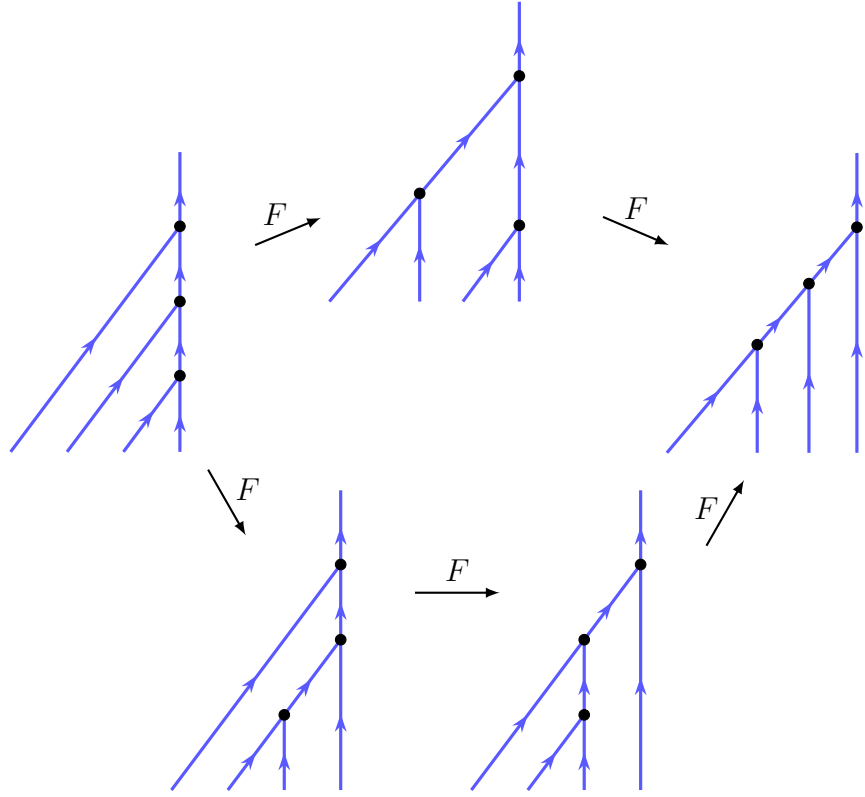

    \centering
    \ctikzfig{figures/pentagon}
    \caption{The $F$-symbols satisfy the pentagon identity, which guarantees that different ways to perform a sequence of $F$-moves involving topological defect lines and junction operators give rise to the same value of the correlation functions.}
    \label{fig:pentagon}
\end{figure}

The associator $\alpha_{XYZ}$ is subject to a consistency condition known as the \emph{pentagon identity}, shown schematically in Figure \ref{fig:pentagon}.
In terms of the $F$-symbols, the pentagon identity can be written as
\begin{align} \label{eq:pentagon}
\begin{split}
    &\sum_{\rho =1}^{N_{qm}^p}
    \left[F_{ijm}^p\right]_{( n, \delta, \lambda)( q, \eta, \rho)}
    \left[F_{qk \ell}^p\right]_{( m, \rho, \sigma)( r, \pi, \xi)} \\
    &=
    \sum_{s \in \mathcal{I} } \sum_{\nu=1}^{N_{jk}^s} \sum_{\tau=1}^{N_{s \ell}^n} \sum_{\omega = 1}^{N_{is}^r} 
    \left[F_{jk \ell}^n \right]_{( m,\lambda, \sigma)( s,\nu , \tau)} 
    \left[F_{i s \ell}^p \right]_{( n,\delta, \tau)(  r,\omega, \xi)}
    \left[F_{ijk}^r \right]_{(s,\omega , \nu)( q,\eta , \pi)} \,.
\end{split}
\end{align}
The data $(\{\mathcal{L}_i\}_{i\in\mathcal{I}},N_{ij}^k,\left[ 
        F_{ijk}^{\ell}
    \right]_{(p,\delta,\lambda)(q,\rho,\sigma)})$, consisting of the set of (isomorphism classes of) simple objects, the fusion coefficients, and the $F$-symbols, determine the fusion category $\mathcal{C}$.
\subsection{Module Category}

We now discuss the definition of a module category $\mathcal{M}$, which is the structure formed by a finite set of boundary conditions which transform into one another under a NIM-rep of the fusion algebra of $\mathcal{C}$.
The set of labels for the simple objects in $\cal M$ (i.e., boundary conditions) is denoted as ${\cal J} = \{a, b,c,\dots \}$.
In Section \ref{SWSB}, we explained that the NIM-rep coefficient $\widetilde{N}_{ia}^b$ in
\begin{equation}
    \mathcal{L}_i \otimes \mathcal{B}_a = \bigoplus_{b\in\mathcal{J}} \widetilde{N}_{ia}^{b} \mathcal{B}_b \,,
\end{equation}
is the dimension of the topological trivalent junction vector space where a bulk topological line meets with two boundaries,
\begin{figure}
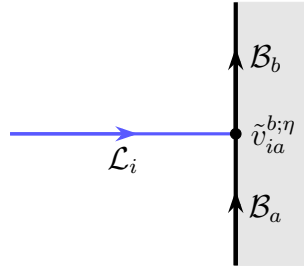

    \centering
    \ctikzfig{figures/boundaryjunctionapp}
    \caption{Topological junction operators $\tilde{v}_{ia}^{b;\eta}$ with $\eta=1, \cdots, \widetilde{N}_{ia}^b$ form a basis for the junction vector space $\text{Hom}_{\mathcal{M}} (\mathcal{L}_i \otimes \mathcal{B}_a, \mathcal{B}_b)$.}
    \label{fig:boundaryjunctionapp}
\end{figure}
\begin{equation}
    \widetilde{N}_{ia}^b = \text{dim}_{\mathbb{C}}\,\text{Hom}_{\mathcal{M}} (\mathcal{L}_i \otimes \mathcal{B}_a, \mathcal{B}_b) \,.
\end{equation}
We use $\eta = 1, \cdots, \widetilde{N}_{ia}^b$ to label fixed (but arbitrary) basis vectors $\tilde{v}_{ia}^{b;\eta} \in \text{Hom}_{\mathcal{M}} (\mathcal{L}_i \otimes \mathcal{B}_a, \mathcal{B}_b)$ for these trivalent junction vector spaces, as shown in Figure \ref{fig:boundaryjunctionapp}.
The action $\otimes$ of $\mathcal{C}$ on the module category $\mathcal{M}$ is again associative up to an isomorphism.
Given two arbitrary objects $X$, $Y$ in $\mathcal{C}$ and an arbitrary object $M$ in $\mathcal{M}$ (all of them not necessarily simple), we have the \emph{module associator}
\begin{equation}
    \widetilde{\alpha}_{XYM} \in \text{Hom}_{\mathcal{M}} ( (X\otimes Y) \otimes M , X \otimes (Y \otimes M )) \,,
\end{equation}
which is an isomorphism.
For simple objects $\mathcal{L}_i$, $\mathcal{L}_j\in {\cal C}$, and $\mathcal{B}_a\in{\cal M}$, we denote $\widetilde{\alpha}_{\mathcal{L}_i \mathcal{L}_j \mathcal{B}_a} \equiv \widetilde{\alpha}_{ija}$.
We call the matrix elements of $\widetilde{\alpha}_{ija}$ in a fixed basis (as in Figure \ref{fig:boundaryjunctionapp}) the $\widetilde{F}$-symbols.
To be more precise, we have
\begin{equation} \label{eq:F_tilde}
    \tilde{v}_{ib}^{c;\eta} \circ \left( \text{id}_i \otimes \tilde{v}_{ja}^{b;\theta} \right) \circ \widetilde{\alpha}_{ija} =
    \sum_{k\in \mathcal{I}} \sum_{\delta = 1}^{N_{ij}^k} \sum_{\tau=1}^{\widetilde{N}_{ka}^c} 
    \left[\widetilde{F}_{ija}^c \right]_{(b,\eta,\theta)(k,\delta,\tau)}
    \cdot \tilde{v}_{ka}^{c;\tau} \circ (v_{ij}^{k;\delta} \otimes \text{id}_a) \,.
\end{equation}
\begin{figure}[h]
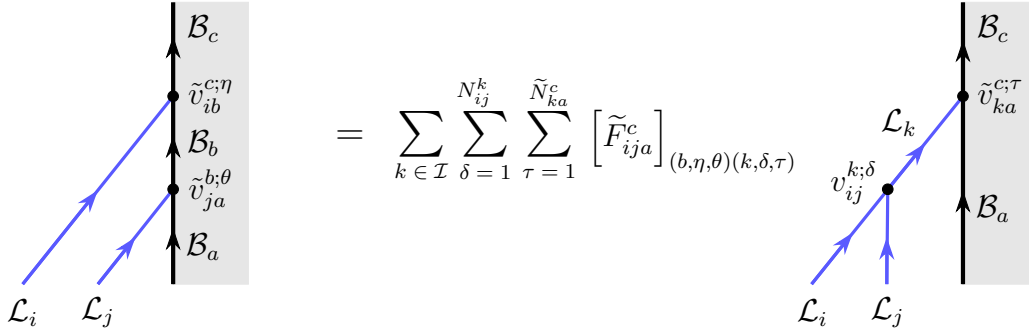

    \centering
    \ctikzfig{figures/tildeFsymbol}
    \caption{The $\widetilde{F}$-symbols are the matrix elements for the boundary crossing relation.}
    \label{fig:tildeFsymbol}
\end{figure}
\begin{figure}[h]
    \centering
    \ctikzfig{figures/modulepentagon}
    \caption{The $\widetilde{F}$-symbols satisfy the module pentagon identity, which guarantees that different ways to perform a sequence of boundary crossing moves involving bulk topological defect lines, boundary conditions, and topological junction operators give rise to the same value of the correlation functions.}
    \label{fig:modulepentagon}
\end{figure}
Here, $\text{id}_a \equiv \text{id}_{\mathcal{B}_a} \in \text{Hom}_{\mathcal{M}} ( \mathcal{B}_a , \mathcal{B}_a) $ is the identity morphism.

Note that the two sides of \eqref{eq:F_tilde} are both vectors in the four-point junction vector space $\text{Hom}_{\mathcal{M}}( (\mathcal{L}_i \otimes \mathcal{L}_j ) \otimes \mathcal{B}_a  ,  \mathcal{B}_c)$.
The $\widetilde{F}$-symbols, $\left[\widetilde{F}_{ija}^c \right]_{(b,\eta,\theta)(k,\delta,\tau)}$, are again basis transformation matrix elements relating different ways to decompose the four-point junction into two trivalent junctions.
We call such a basis transformation  an $\widetilde{F}$-move, or  a boundary crossing relation.
Diagrammatically, \eqref{eq:F_tilde} is represented as in Figure \ref{fig:tildeFsymbol}.
Similar to the fusion category case, the module associator $\widetilde{\alpha}_{XYM}$ is subject to a consistency condition which we call the \emph{module pentagon identity}, shown in Figure \ref{fig:modulepentagon}.
In terms of the $\widetilde{F}$-symbols, the module pentagon identity becomes 
\begin{align} \label{eq:module_pentagon}
    \begin{split}
        &\sum_{\pi =1}^{\widetilde{N}_{nb}^d}
    \left[\widetilde{F}_{ijb}^d \right]_{( c, \eta, \theta)( n,\sigma ,\pi)}
    \left[\widetilde{F}_{nka}^d \right]_{( b,\pi, \tau)( m,\lambda, \xi)} \\
    &=
    \sum_{\ell \in \mathcal{I} } \sum_{\nu=1}^{N_{jk}^\ell} \sum_{\rho=1}^{N_{i \ell}^m} \sum_{\kappa = 1}^{\widetilde{N}_{\ell a}^c} 
    \left[\widetilde{F}_{jka}^c\right]_{( b,\theta, \tau)(  \ell, \nu, \kappa)} 
    \left[\widetilde{F}_{i \ell a}^d\right]_{( c,\eta, \kappa)( m, \rho, \xi)}
    \left[F_{ijk}^m \right]_{( \ell, \rho, \nu)( n,\sigma, \lambda)} \,.
    \end{split}
\end{align}
The data $(\{\mathcal{B}_a \}_{a\in\mathcal{J}}, \widetilde{N}_{ia}^b, \left[\widetilde{F}_{ija}^c \right]_{(b,\eta,\theta)(k,\delta,\tau)})$, consisting of the set of (isomorphism classes of) simple objects, the NIM-rep coefficients, and the $\widetilde{F}$-symbols, determine a (left) module category $\mathcal{M}$ over the fusion category $\mathcal{C}$.

A simple example of a module category over an arbitrary fusion category $\mathcal{C}$ which always exists is $\mathcal{C}$ itself viewed as a module category over $\mathcal{C}$.
This corresponds to simply setting $\widetilde{N} = N$ and $\widetilde{F} = F$.
This module category is often called the \emph{regular} module category.
For instance, the Cardy boundary conditions in the A-series minimal models (or more generally in diagonal RCFTs with extended chiral algebras) form the regular module category over the fusion category of Verlinde lines.
As a check, recall that both the Cardy boundary conditions as well as the Verlinde lines are labeled by the bulk primary operators in these RCFTs \cite{Cardy:1989ir,Petkova:2000ip}.

\subsection{Algebra Object and Internal Hom} \label{app:algebra}

There is a close relation between the module categories over a fusion category $\mathcal{C}$ and the algebra objects in $\mathcal{C}$.
Namely, given a semisimple, indecomposable (left) module category $\mathcal{M}$ over $\mathcal{C}$, one can find an algebra object $\mathcal{A}$ in $\mathcal{C}$ such that $\mathcal{M}$ is equivalent to the category of (right) $\mathcal{A}$-modules in $\mathcal{C}$, denoted as $\mathcal{C}_{\mathcal{A}}$, and vice versa \cite[Theorem~1]{ostrik2003module}.
Physically, this corresponds to a relation between symmetric boundary conditions and gauging, as discussed in Section \ref{sec:anomaly}.
Here we review the definition of an algebra object in more detail.

As was briefly introduced in Section \ref{sec:anomaly}, an algebra object in a fusion category $\mathcal{C}$ is a triple $(\mathcal{A},\mu,u)$.
Here $\mathcal{A}$ is an object in $\mathcal{C}$ which we write as
\begin{equation} \label{eq:algebra2}
\mathcal{A} = \bigoplus_{i\in \mathcal{I}} \langle \mathcal{L}_i , \mathcal{A} \rangle  \mathcal{L}_i \,.
\end{equation}
Moreover, $\mu \in \text{Hom}_{\mathcal{C}}(\mathcal{A} \otimes \mathcal{A}, \mathcal{A})$ is the \emph{multiplication} morphism, and $u \in \text{Hom}_{\mathcal{C}}(\mathds{1},\mathcal{A})$ is the \emph{unit} morphism.
To be explicit, it is convenient to pick a \emph{basis} for the algebra object $\mathcal{A}$, following \cite[Section~3]{Fuchs:2002cm}.
\begin{figure}[t!]
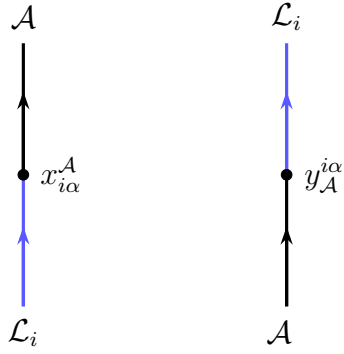
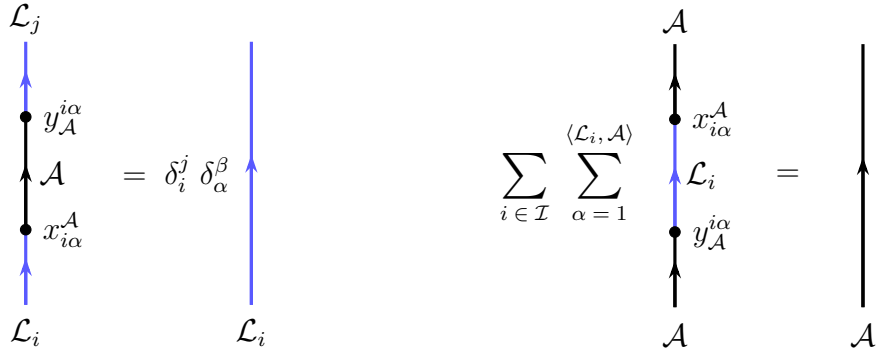


    \centering
    
    \begin{subfigure}{1\textwidth}
    \centering
    \ctikzfig{figures/basisdefnA}
    \caption{Choosing a basis (left) and a dual basis (right) for the algebra object.}
    \label{fig:basisdefnA}
    \end{subfigure}

    \begin{subfigure}{0.45\textwidth}
    \centering
    \ctikzfig{figures/orthogonalA}
    \caption{Orthogonality of the basis.}
    \label{fig:orthogonalA}
    \end{subfigure}
\hspace{0.1em}%
    \begin{subfigure}{0.45\textwidth}
    \centering
    \ctikzfig{figures/completeA}
    \caption{Completeness of the basis.}
    \label{fig:completeA}
    \end{subfigure}

    \caption{An orthogonal and complete basis for the algebra object.}
    \label{fig:basisA}
  
\end{figure}
\noindent We fix the basis vectors $x_{i\alpha}^{\mathcal{A}} \in \text{Hom}_{\mathcal{C}}(\mathcal{L}_i,\mathcal{A})$ and the dual basis vectors $y^{i\alpha}_{\mathcal{A}} \in \text{Hom}_{\mathcal{C}}( 
\mathcal{A} , \mathcal{L}_i)$ with $\alpha = 1, \cdots,  \langle \mathcal{L}_i , \mathcal{A} \rangle$ for each $i \in \mathcal{I}$.
We choose them to be orthonormal, that is,
\begin{equation} 
    y^{j\beta}_{\mathcal{A}} \circ x_{i\alpha}^{\mathcal{A}} = \delta_i^j \delta_{\alpha}^{\beta} \cdot \text{id}_i \,.
\end{equation}
Completeness of the basis implies
\begin{equation}
    \sum_{i\in \mathcal{I}} \sum_{\alpha=1}^{ \langle \mathcal{L}_i , \mathcal{A} \rangle} x_{i\alpha}^{\mathcal{A}} \circ y^{i\alpha}_{\mathcal{A}} = \text{id}_{\mathcal{A}} \,.
\end{equation}
\noindent See Figure \ref{fig:basisA}.
In other words, the $x_{i\alpha}^{\mathcal{A}} : \mathcal{L}_i \rightarrow \mathcal{A}$ morphisms are inclusion maps, and the $y_{\mathcal{A}}^{i\alpha} : \mathcal{A} \rightarrow \mathcal{L}_i$ morphisms are projection maps.

We focus on the case of a haploid (or connected) algebra object, that is, $ \langle \mathds{1} , \mathcal{A} \rangle = 1$.
In such a case, the choice of the unit morphism $u \in \text{Hom}_{\mathcal{C}}(\mathds{1},\mathcal{A})$ is unique up to an overall scale which is not physical, and we set $x_{\mathds{1}}^{\mathcal{A}} = u$.  

\begin{figure}[t]
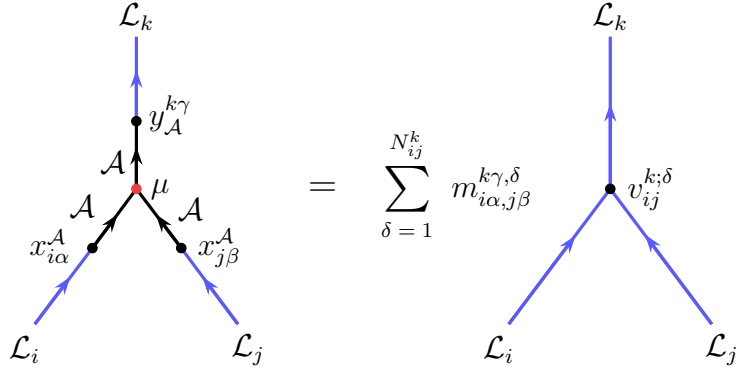

    \centering
    \ctikzfig{figures/basisofmu}
    \caption{Matrix elements $m_{i \alpha, j\beta}^{k \gamma, \delta}$ of the multiplication morphism $\mu$, written in the basis $x_{i\alpha}^{\cal A}$ and $y^{i\alpha}_{\cal A}$ for the algebra object and $v_{ij}^{k;\delta}$ for the trivalent junction of topological lines.}
    \label{fig:basisofmu}
\end{figure}

Given a fixed basis, we can write down the matrix elements of the multiplication morphism $\mu \in \text{Hom}_{\mathcal{C}}(\mathcal{A} \otimes \mathcal{A}, \mathcal{A})$.
To be more precise, the multiplication morphism is characterized by a set of (basis-dependent) complex numbers $\{ m_{i\alpha, j\beta}^{k\gamma;\delta} \}$ where
\begin{equation} \label{eq:multiplication}
    y^{k\gamma}_{\mathcal{A}} \circ \mu \circ \left(
        x_{i\alpha}^{\mathcal{A}} \otimes x_{j\beta}^{\mathcal{A}}
    \right) = \sum_{\delta=1}^{N_{ij}^k} m_{i\alpha, j\beta}^{k\gamma;\delta} \cdot v_{ij}^{k;\delta} \,.
\end{equation}
Note that the two sides of \eqref{eq:multiplication} are both valued in $\text{Hom}_{\mathcal{C}}(\mathcal{L}_i \otimes \mathcal{L}_j, \mathcal{L}_k)$, and recall that $\{ v_{ij}^{k;\delta} \}_{\delta=1, \cdots, N_{ij}^k}$ is the set of fixed basis vectors for the junction vector space $\text{Hom}_{\mathcal{C}}(\mathcal{L}_i \otimes \mathcal{L}_j, \mathcal{L}_k)$.
This is shown diagrammatically in Figure \ref{fig:basisofmu}.

\begin{figure}[h]
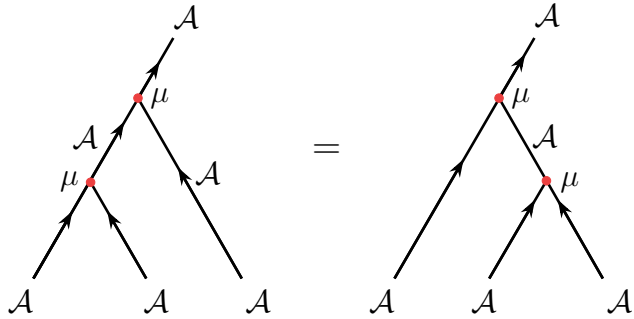

    \centering
    \ctikzfig{figures/associativityAappa}
    \caption{Associativity condition for an algebra object.}
    \label{fig:associativityAappa}
\end{figure}

The multiplication morphism $\mu$ is subject to the \emph{associativity constraint}, which reads
\begin{equation} \label{eq:associativity}
     \mu \circ \left(  \mu  \otimes \text{id}_{\mathcal{A}} \right)
    =
    \mu \circ \left( \text{id}_{\mathcal{A}} \otimes \mu \right) \circ \alpha_{\mathcal{A} \mathcal{A} \mathcal{A}}
     \,.
\end{equation}
Here $\alpha_{\mathcal{A} \mathcal{A} \mathcal{A}} \in \text{Hom}_{\mathcal{C}}((\mathcal{A} \otimes \mathcal{A}) \otimes \mathcal{A}, \mathcal{A} \otimes (\mathcal{A} \otimes \mathcal{A}) )$ is the associator.
Diagramatically, \eqref{eq:associativity} is represented as in Figure \ref{fig:associativityAappa}.
Note that both sides of \eqref{eq:associativity} are vectors in $ \text{Hom}_{\mathcal{C}}((\mathcal{A} \otimes \mathcal{A}) \otimes \mathcal{A}, \mathcal{A} )$.
In a fixed basis, the associativity constraint \eqref{eq:associativity} is explicitly given in terms of the matrix elements $m_{i\alpha, j\beta}^{k\gamma;\delta}$ as \cite{Fuchs:2002cm}
\begin{equation} \label{eq:associativity_components}
    \sum_{\omega = 1}^{\langle \mathcal{L}_q , {\cal A} \rangle } m_{i\alpha, j\beta}^{q \omega ;\rho} m_{q\omega, k\gamma}^{\ell \epsilon;\sigma}
    =
    \sum_{p \in {\cal I}}\sum_{\xi = 1}^{\langle \mathcal{L}_p , {\cal A} \rangle}
    \sum_{\delta =1}^{N_{ip}^\ell} \sum_{\lambda = 1}^{N_{jk}^p}
    m_{i\alpha, p\xi}^{\ell \epsilon ;\delta} m_{j\beta, k\gamma}^{p \xi;\lambda}
    \left[F_{ijk}^\ell \right]_{( p,\delta, \lambda)(  q,\rho, \sigma)} \,.
\end{equation}

\begin{figure}[t]
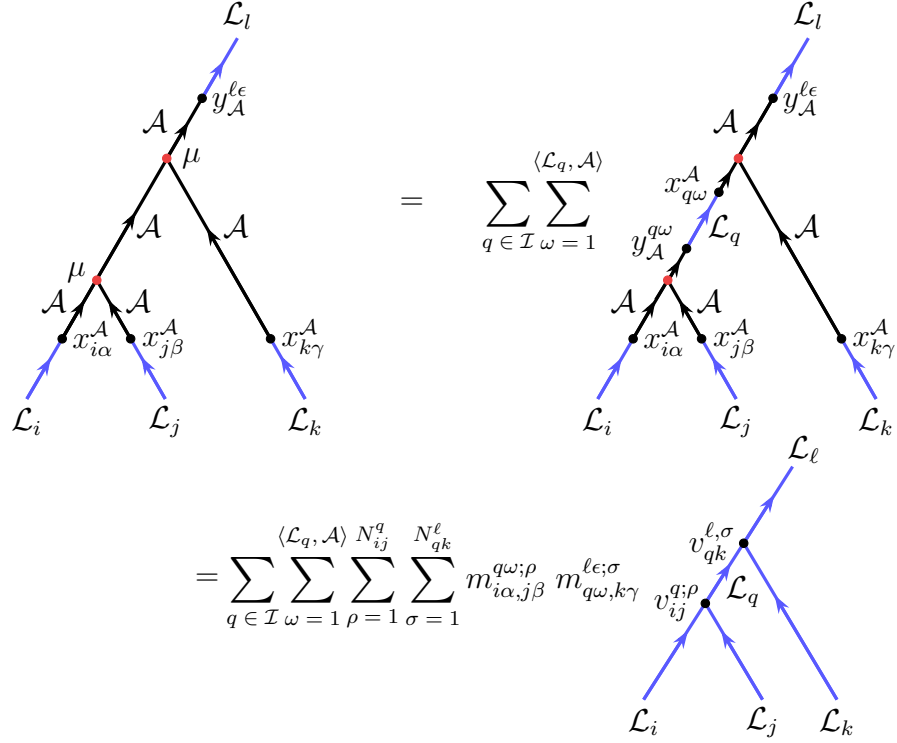

    \centering
    \ctikzfig{figures/associativityAapp0}
    \caption{LHS of Figure \ref{fig:associativityAappa} evaluated in a fixed basis. In the first equality, we insert a complete set of the basis vectors for $\cal A$. Then we use \eqref{eq:multiplication} (see Figure \ref{fig:basisofmu}) to write the multiplication morphism $\mu$ in terms of its matrix elements.}
    \label{fig:LHSassociativityA}
\end{figure}

\begin{figure}[h!]
    \centering
    \ctikzfig{figures/associativityAapp1}
    \ctikzfig{figures/associativityAapp2}

    \caption{RHS of Figure \ref{fig:associativityAappa} evaluated in a fixed basis. In the first equality, we insert a complete set of the basis vectors for $\cal A$. Then we use \eqref{eq:multiplication} (see Figure \ref{fig:basisofmu}) to write the multiplication morphism $\mu$ in terms of its matrix elements. Finally, we perform an $F$-move to convert it into the same form as that in Figure \ref{fig:LHSassociativityA}.
    Comparing the two, we arrive at \eqref{eq:associativity_components}.}

    \label{fig:associativityAapp}
\end{figure}
\noindent See Figure \ref{fig:LHSassociativityA} and \ref{fig:associativityAapp} for the derivation of \eqref{eq:associativity_components}. 
In addition, the $u$ and $\mu$ morphisms must satisfy the \emph{unit constraint} (see Figure \ref{fig:consistencyA}(b)) which reads
\begin{equation} \label{eq:unit_constraint}
   \mu \circ \left( u \otimes \text{id}_{\mathcal{A}} \right) = \text{id}_{\mathcal{A}} = \mu \circ \left( \text{id}_{\mathcal{A}} \otimes u \right) \,.
\end{equation}
In terms of the matrix elements $ m_{i\alpha, j\beta}^{k\gamma;\delta} $, and with the choice $x_{\mathds{1}}^{\mathcal{A}} = u$, the unit constraint \eqref{eq:unit_constraint} becomes
\begin{equation}
    m_{\mathds{1}, i\alpha}^{j\beta} = m_{ i\alpha, \mathds{1}}^{j\beta} = \delta_i^j \delta_\alpha^\beta \,.
\end{equation}
The data $(\{ \langle \mathcal{L}_i , \mathcal{A} \rangle \}_{i\in \mathcal{I}},  m_{i\alpha, j\beta}^{k\gamma;\delta}  )$ satisfying above consistency conditions determine the haploid algebra object $\mathcal{A}$ in $\mathcal{C}$.
When $\mathcal{C}$ is the category of vector spaces, the definition of an algebra object in $\cal C$ reduces to the usual definition of an algebra.

In establishing the relation between  algebra objects and module categories, an essential role is played by the concept of an internal Hom \cite{ostrik2003module}.
To be more precise, given a simple object $\mathcal{B}_a$ in a semisimple, indecomposable (left) module category $\mathcal{M}$ over the fusion category $\mathcal{C}$, we can construct a haploid, semisimple, indecomposable algebra object $\mathcal{A} = \underline{\text{Hom}}(\mathcal{B}_a,\mathcal{B}_a)$.
Explicitly, this is done by setting
\begin{align}
\begin{split}
    \langle \mathcal{L}_i , \mathcal{A} \rangle 
    &= \widetilde{N}_{ia}^a
    \,,\\
    m_{i\alpha, j\beta}^{k\gamma ;\delta} &= \left[\widetilde{F}_{ija}^a \right]_{( a, \alpha, \beta)( k,\delta, \gamma)} \,.
\end{split}
\end{align}
One can verify that the module pentagon identity \eqref{eq:module_pentagon} for the $\widetilde{F}$-symbols guarantees that the associativity constraint \eqref{eq:associativity_components} for the multiplication morphsim is satisfied.
The unit constraint \eqref{eq:unit_constraint} is satisfied due to the fact that the $\widetilde{F}$-symbol $\left[\widetilde{F}_{ija}^a \right]_{( a, \alpha, \beta)( k,\delta, \gamma)}$ is trivial if $\mathcal{L}_i = \mathds{1}$ or $\mathcal{L}_j = \mathds{1}$.

The algebra object $\mathcal{A} = \underline{\text{Hom}}(\mathcal{B}_a,\mathcal{B}_a)$ arising from the internal Hom construction has the property that the category of (right) $\mathcal{A}$-modules in $\cal C$ becomes $\mathcal{C}_{\mathcal{A}} \cong \cal M$ as a (left) module category over $\cal C$.
This implies that, by definition, different algebra objects obtained from the internal Hom of different simple objects in a given module category $\cal M$ are all Morita equivalent.
In particular, if $\cal M$ is the regular module category, then all the algebra objects that are produced from the internal Hom of a simple object in $\cal M$ are Morita trivial, as $\mathds{1} \in \cal M$ and $\underline{\text{Hom}}(\mathds{1},\mathds{1}) = \mathds{1}$ is the trivial algebra object.
For more details on the relation between algebra objects and module categories, including a mathematically rigorous definition of the internal Hom, see \cite{ostrik2003module, etingof2016tensor}.

Every haploid, semisimple, indecomposable algebra object $\mathcal{A}$ in $\cal C$ can be obtained from the internal Hom construction.
This is due to the fact that such $\mathcal{A}$ defines a simple object in $\mathcal{C}_{\mathcal{A}}$, which is equipped with the structure of a (left) module category  over $\cal C$, and $\mathcal{A} = \underline{\text{Hom}}(\mathcal{A},\mathcal{A})$ \cite{ostrik2003module,etingof2016tensor}.
In particular, this tells us that $\langle \mathcal{L}_i , \mathcal{A} \rangle = \widetilde{N}_{i \mathcal{A}}^{\mathcal{A}}$, which cannot be greater than the quantum dimension $\langle \mathcal{L}_i \rangle$ \cite{Gannon:2001ki}.
This proves the inequality \cite{Fuchs:2004dz}
\begin{equation}
    \langle \mathcal{L}_i , \mathcal{A} \rangle \leq \langle \mathcal{L}_i \rangle
\end{equation}
for all $i \in \mathcal{I}$.

\section{NIM Representations of Groups}\label{NIMrepG}

In this section, we review an elementary  fact that the finite-dimensional non-negative integer matrix (NIM) representations of a group are always permutation matrices. 
More precisely, let ${\cal G}$ be the  group of  $n \times n$ invertible matrices over the non-negative integers, which is sometimes denoted as $GL(n,\mathbb{Z}_{\ge0})$. 
We will show that $\cal G$ is the symmetric group $S_n$ of $n \times n$ permutation matrices.\footnote{We thank Qiaochu Yuan for helpful discussions.} See \cite{dolzan2008invertible} for instance.

The proof proceeds as follows. Let $P$ be a non-negative integer matrix in $\mathcal{G}$. 
 Define the function $\mathcal{F}: \mathcal{G} \to  \mathbb{Z}_{\geq0}$  as
\begin{equation}
    \mathcal{F}(P)\equiv\sum_{i,j} P_{ij} \,.
\end{equation}
We first show that   $\mathcal{F}$  obeys 
\begin{equation}\label{nim0}
    \mathcal{F}(PQ)\geq \mathcal{F}(P) \,,\;\; \forall \; P,Q\in \mathcal{G} \,.
\end{equation}
This can be seen from the following:
\begin{equation}\label{nim1}
    \mathcal{F}(PQ) = \sum_{i,j,k} P_{ij}Q_{jk} = \sum_{i,j} \left(P_{ij} \sum_{k} Q_{jk}\right) \,.
\end{equation}
Since $Q \in \mathcal{G}$, it is an invertible matrix and has a non-zero determinant. This implies every row and column is non-zero, i.e., each contains at least one positive integer.  Hence, for every $ Q \in \mathcal{G}$, we have $\sum_{k} Q_{jk} \geq 1$ for all $j$. Using this in (\ref{nim1}), we get
\begin{equation}
    \mathcal{F}(PQ) \geq \sum_{i,j}P_{ij} = \mathcal{F}(P) \,.
\end{equation}
This proves (\ref{nim0}). 

Now we plug $Q = P^{-1}$ into (\ref{nim0}) to obtain
\begin{equation}\label{nim2}
n \geq \mathcal{F}(P) \,,\;\; \forall \; P\in \mathcal{G} \,.
\end{equation}
This says that the sum of all the matrix elements of $P$ is always less than or equal to $n$. 
On the other hand, since each row (and each column) of $P$ contains at least one positive integer, the sum over all matrix elements of $P$ must be greater than or equal to $n$, i.e.,
\begin{equation}\label{nim3}
 n\leq \mathcal{F}(P) \,,\;\;\forall \; P\in \mathcal{G}\,.
\end{equation}
Comparing (\ref{nim2}) and (\ref{nim3}), we find
\begin{equation}
  \mathcal{F}(P) = n \,,\;\;\forall \; P\in \mathcal{G}\,.
\end{equation}
That is,  the sum over all the matrix elements of any $P \in \mathcal{G}$ must be exactly equal to $n$. Since every row and every column of $P$ has to contain at least one positive integer in order for $P$ to have a non-zero determinant, every row and every column must contain exactly a single entry equal to $1$; All other elements should vanish in order for the sum of all entries to be equal $n$. In other words, every matrix $P \in \mathcal{G}$ is a permutation matrix. 
This shows the elementary fact that
\begin{equation}
    \mathcal{G} \cong S_n \,.
\end{equation}
Hence,  given  a group $H$, its finite-dimensional NIM representation, which is a homomorphism $\rho: H \to \mathcal{G}$, consists only of permutation matrices. 

In the context of 1+1d CFTs, this implies that finite   invertible symmetries  act on simple conformal boundaries by permutations. 
For non-invertible symmetries, NIM-reps are not necessarily permutation matrices.

\bibliographystyle{JHEP}
\bibliography{ref}

\end{document}